\documentclass[12pt]{iopart}

\usepackage{subcaption}
\usepackage{graphicx}
\begin{document}

\title[]{Saturation of fishbone instability through zonal flows driven by energetic particle transport in tokamak plasmas}

\author{G. Brochard$^{*,1,2}$, C. Liu$^{3}$, X. Wei$^{2}$, W. Heidbrink$^{2}$, Z. Lin$^{*,2}$, M.V. Falessi$^{4,5}$, F. Zonca$^{4,6}$, Z. Qiu$^{4,6}$, N. Gorelenkov$^{3}$, C. Chrystal$^{7}$, X. Du$^{7}$, J. Bao$^{8}$, A. R. Polevoi$^{1}$, M. Schneider$^{1}$, S. H. Kim$^{1}$, S. D. Pinches$^{1}$, P. Liu$^{2}$, J. H. Nicolau$^{2}$, H. L\"utjens$^{9}$ and the ISEP group}

\address{$^1$ITER organisation, Route de Vinon-sur-Verdon, CS 90 046 13067 St., Paul Lez Durance, France\\
	      $^2$Department of Physics and Astronomy, University of California, Irvine, California 92697, USA\\
	      $^3$Princeton Plasma Physics Laboratory, Princeton University, PO Box 451, Princeton, New Jersey 08543,USA\\
	      $^4$Center for Nonlinear Plasma Science and C.R. ENEA Frascati, C.P. 65, 00044 Frascati, Italy \\
	      $^5$Istituto Nazionale di Fisica Nucleare (INFN), Sezione di Roma, Piazzale Aldo Moro 2, Roma, Italy\\
	      $^6$Institute for Fusion Theory and Simulation, School of Physics, Zhejiang University, Hangzhou, People's Republic of China\\
	      $^7$General Atomics, PO Box 85608, San Diego, CA 92186-5608, United States of America\\
	      $^8$Institute of Physics, Chinese Academy of Sciences, Beijing 100190, China\\
	      $^9$CPHT, CNRS, \'Ecole Polytechnique, Institut Polytechnique de Paris, Route de Saclay, 91128 Palaiseau, France}
\ead{guillaume.brochard@iter.org, zhihongl@uci.edu}
\vspace{10pt}

\begin{abstract}
Gyrokinetic and kinetic-MHD simulations are performed for the fishbone instability in the DIII-D discharge \#178631, chosen for validation of first-principles simulations to predict the energetic particle (EP) transport in an ITER prefusion baseline scenario. Fishbone modes are found to generate zonal flows, which dominate the fishbone saturation. The underlying mechanisms of the two-way fishbone-zonal flows nonlinear interplay are discussed in details. Numerical and analytical analyses identify the fishbone-induced EP redistribution as the dominant generation mechanism for zonal flows. The zonal flows modify the nonlinear dynamics of phase space zonal structures, which reduces the amount of EPs able to resonate with the mode, leading to an early fishbone saturation. Simulation results including zonal flows agree quantitatively with DIII-D experimental measurements of the fishbone saturation amplitude and EP transport, supporting this novel saturation mechanism by self-generated zonal flows. Moreover, the wave-particle mode-locking mechanism is shown to determine quantitatively the fishbone frequency down-chirping, as evident in GTC simulation results in agreement with predictions from analytical theory. Finally, the fishbone-induced zonal flows are possibly responsible for the formation of an ion-ITB in the DIII-D discharge. Based on the low EP transport and the large zonal flow shearing rates associated with the fishbone instability in gyrokinetic simulations of the ITER scenario, it is conjectured that high performance scenarios could be designed in ITER burning plasmas through fishbone-induced ITBs.
\end{abstract}

\section{Introduction}
Energetic Particles (EPs) play a critical role in burning plasmas such as those of ITER \cite{ITER}, by providing plasma self-heating through the thermalization of fusion-born alpha particles on the thermal bulk plasma  which compensates for the power losses associated with collisional and turbulent transport. EPs however tend to destabilize plasma instabilities that arise at various spatial scales such as meso-scale  Alfv\'en eigenmodes \cite{Heidbrink2008} and global kinetic-MHD modes, that degrade EP confinement in the core plasma. Understanding and predicting quantitatively the EP transport in burning plasmas is therefore essential to design plasma scenarios that can achieve high fusion performance, ITER operations aiming in particular at a fusion gain of $Q\geq10$. \\ 
The instability named fishbone \cite{McGuire1983}\cite{Chen1984} is one these global kinetic-MHD modes that could induce a large EP transport, due to their macroscopic extent that can cover up to half of the tokamak minor radius. The EP transport associated with this instability is mostly determined by the saturation amplitude of the fishbone mode. The saturation mechanism typically associated with fishbones is the flattening of the EP distribution gradients in phase space through wave-particle resonant interaction \cite{Fu2006}\cite{Brochard2020b}. However other nonlinear processes could participate to the saturation of fishbones, such as $n=0$ zonal flows \cite{Diamond2005}. Zonal flows are indeed known to play a critical role in the nonlinear saturation of both drift-waves \cite{Lin1998} and Alfv\'en eigenmodes \cite{Spong1994}\cite{Bass2010}\cite{Todo2012}\cite{Zhang2013}. As fishbones were observed in multiple kinetic-MHD simulations to nonlinearly generate zonal flows \cite{Brochard2020b}\cite{Shen2015}\cite{Ge2022}, such flows could have a significant impact on their saturation. However to quantify self-consistently this impact, the kinetic contribution of thermal ions needs to be retained, to account for the zonal flows collisionless damping \cite{Rosenbluth1998}. Moreover, the generation mechanism for fishbone-induced zonal flows has not been clearly established, even though previous works conjectured that zonal electric field could be produced through the fishbone EP redistribution \cite{Guenter2001}\cite{Pinches2001}\cite{Liu2023}. Gyrokinetic simulations are required to confirm such a mechanism for the fishbone mode. The generation of strongly sheared zonal flows by fishbones could also have a significant impact of turbulent transport \cite{Hahm1995}, through cross-scale interactions common in both fusion \cite{Liu2022a} and astrophysical plasmas \cite{Giacalone1999}. Furthermore, the non-adiabatic frequency down-chirping of plasma waves, common in both laboratory \cite{Brochard2020b}\cite{Wang2016}\cite{Vlad2016}\cite{Wang2023} and astrophysical plasmas \cite{Tao2021}\cite{Zonca2022}, plays an important role in the fishbone-induced EP transport. Theoretical studies \cite{Zonca2015}\cite{Chen2016} attribute this phenomenon to a mode-locking occurring between the fishbone mode and resonant EPs, which maximizes wave-particle power transfer and leads to EP transport through avalanche processes. Illustrating self-consistently such a mechanism would therefore provide a better understanding of these nonlinear processes, and help identify actuators that can reduce EP transport in tokamak plasmas. \\
In this work, first-principles simulations are performed to study the two-way nonlinear interplay between fishbone modes and zonal flows. A DIII-D experiment \cite{Heidbrink2020} is chosen for experimental validation of first-principles simulations to predict the EP transport in an ITER baseline Augmented First Plasma (AFP) scenario \cite{Polevoi2021}. Global gyrokinetic GTC \cite{Lin1998}\cite{Deng2012}\cite{Dong2017} simulations self-consistently show that fishbone modes can destabilize zonal flows that dominate their saturation. This novel mechanism for fishbone saturation is supported by quantitative agreements between the simulations with zonal flows and DIII-D measurements for the fishbone amplitude and the drop in neutron emissivity associated with EP transport. Successful comparisons between GTC simulations and analytical models \cite{Pinches2001}\cite{Liu2023} demonstrate that fishbones indeed generate zonal flows through radial currents linked to the EP redistribution. Phase space analysis illustrates that zonal flows are able to induce a Doppler-shift on the fishbone resonances, thereby restricting the amount of particle able to interact resonantly with the mode through avalanche processes, which leads to its lower saturation. The chirping rate of fishbone modes in gyrokinetic simulations is found to agree quantitatively with theoretical predictions \cite{Zonca2015}\cite{Chen2016} based on mode-locking, confirming that mode-locking is the underlying mechanism for the non-adiabatic chirping of fishbone modes. Moreover, the shearing rate of fishbone-induced zonal flows is found to exceed the linear growth rate of the most unstable drift-wave modes obtained from GTC electrostatic simulations. The potential suppression of turbulent transport is consistent with the formation of an ion-ITB (Internal Transport Barrier) in this DIII-D experiment shortly after the onsets of fishbones, similarly to recent numerical/experimental analysis on the EAST tokamak \cite{Ge2022}. These results further highlight the correlation that has long been suspected between fishbone and ITB formations \cite{Pinches2001}, as fishbone modes have often been observed to precede the formation of transport barriers in a large number of tokamak experiments \cite{Guenter2001}\cite{Field2011}\cite{Chen2016a}\cite{Gao2018}. Lastly, fishbone simulations for the ITER scenario recover a marginal EP redistribution and zonal flows shearing rate levels that are sufficient to mitigate microturbulence. The intentional destabilization of fishbone instabilities is therefore proposed as a way to enhance fusion performances in burning plasmas. \\
The rest of the paper is organized as follows. In section 2, the DIII-D discharge chosen as a matching case for the ITER scenario is discussed. The numerical models and tools used to performed first-principle fishbone simulations are presented in section 3. Results from nonlinear simulations with and without zonal flows,and their comparison with DIII-D measurements are reported in section 4. The underlying mechanisms of the two-way nonlinear interaction between fishbone and zonal flows are detailed in section 5. In section 6, comparisons between numerical and theoretical predictions identify mode-locking to be the key mechanism leading to fishbone down-chirping. Experimental observations of an ion-ITB in the DIII-D experiment are presented in section 7, together with GTC electrostatic simulations highlighting the potential role of fishbone-induced zonal flows in microturbulence suppression. Section 8 discusses results obtained from GTC fishbone simulations for the ITER scenario. Finally, our key results and their perspectives are summarized in section 9.
\section{Description of the DIII-D experimental discharge and the ITER scenario}
\begin{figure}[h!]
   \centering
   \includegraphics[scale=0.8]{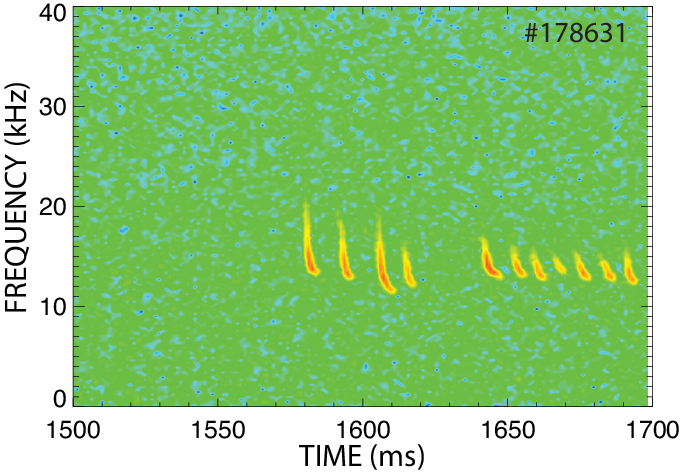}
\caption{Spectogram of the cross power between two magnetic probes separated toroidally by a $\sim\pi/6$ angle, highlighting eleven n=1 fishbone bursts over t$\in[1580,1700]$ms.}
\label{specto}
\end{figure}

The DIII-D discharge \#178631 analysed in this work is a L-mode plasma heated by 3.8MW of 81keV deuterium beams in the co-current direction at the midplane, and by 1.0MW of $2^{nd}$ harmonic central electron cyclotron heating (ECH). The plasma has a near-circular shape, with an elongation of $\kappa=1.17$, a triangularity of $\delta=0.07$, and is limited on the carbon inner wall. The major radius is $R_0=1.74$m, the minor radius $a=0.64$m, the toroidal field is $B_0=2.0T$, the plasma current $I_p=0.88$ MA, and the line-average electron density is $n_{e,0}=2.0\times10^{19}$m$^{-3}$. This discharge has been chosen for validation of first-principles simulations, in order to predict the dynamics of EP-driven instabilities dynamics in a ITER pre-fusion baseline plasma during the AFP phase. This work is itself part of a larger collaboration between the SciDAC Integrated Simulation of Energetic Particles (ISEP) group and the ITPA-EP activities \cite{Lin2023}. This collaboration analyses the EP transport in ITER baseline and steady-state plasmas from every relevant spatial scale, ranging from the microscopic scale ($L\sim\rho_i$, with $\rho_i$ the thermal ion Larmor radius) with microturbulence, mesoscopic scale ($L\sim\rho_f$, with $\rho_f$ the fast ion Larmor radius) with Alfv\'en Eigenmodes (AEs) up to the macroscopic scale ($L\sim a$) with MHD modes such as internal kink and fishbone modes. This work focuses on the macroscopic spatial scale with fishbone modes. \\
The selected ITER baseline scenario is an hydrogen H-mode plasma \cite{Polevoi2021} with $I_p=7.5$MA and $B_0=2.65$T, heated by 33MW of co-passing tangential beams and 20MW of ECH. Several criteria were used in selecting a DIII-D experiment matching this ITER scenario. The chosen DIII-D discharge needs to have EP-driven instabilities, preferably weakly driven, to extrapolate the EP transport in ITER. The DIII-D pulse should also have a similar q profile, $T_e/T_i$ ratio, and normalized beta. The DIII-D discharge \#178631 \cite{Heidbrink2020} at t=1580ms has been chosen for this analysis, primarily because it features eleven n=1 fishbone bursts over $t\in[1580, 1700]$ms as can be observed in Fig. \ref{specto}, and because the weakly reversed shear q profile of this configuration is known accurately and matches very well with the one from the ITER baseline scenario (see Fig. \ref{DIIID_eq}b). The normalized beta and temperature ratio are $\beta_N=2.2$ and $T_e/T_i$=1.67 in ITER and $\beta_N=1.3$ and $T_e/T_i=1.34$ in DIII-D. The time chosen to carry out first-principles based simulations in DIII-D is t=1580ms, just before the first fishbone burst, as the EP distribution evolves classically before the MHD activity and can therefore be reconstructed accurately with the NUBEAM code \cite{Pankin2004}. The q profile is reconstructed using the EFIT code \cite{Lao1985} with both Motional Stark Effects (MSE) \cite{Rice1995} and external magnetics constraints.  It agrees very well with the ECE data, tracking temporally the $q_{min}$ value when reversed shear Alfv\'en eigenmodes (RSAE) and low frequency modes (LFM)  are destabilized prior to the fishbone burst over $t\in[800,1400]$ms \cite{Heidbrink2020}.
\begin{figure}[h!]
   \centering
   \includegraphics[scale=0.8]{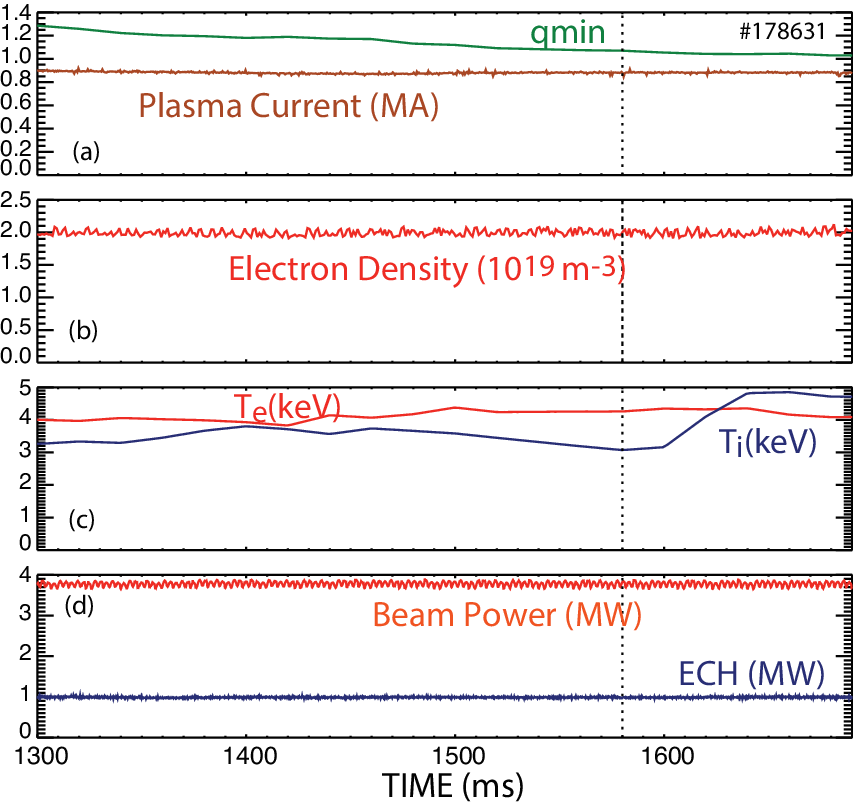}
\caption{Time evolution of experimental signals from DIII-D discharge \#178631 (a) Plasma current and minimum safety factor value (b) On-axis electron density (c) On-axis electron and ion temperature (d) Beam and Electron Cylclotron Heating (ECH) power.}
\label{DIIID_signals}
\end{figure}
The time evolution of the $q_{min}$ value, on-axis quantities and non-inductive heating is displayed on Fig.\ref{DIIID_signals}. The time slice chosen for this analysis is marked by a dashed line. It can be observed that the heating power is constant for multiple slowing-down times before the onset of fishbone modes, which implies that the mode becomes unstable solely due to the drop of $q_{min}$ towards $q_{min}\sim1$, the kinetic drive from EPs varying weakly. This drop in $q_{min}$ is due to the decrease of the toroidal field over the discharge, at constant plasma current. A particularly interesting observation is that the core ion temperature increases by 50\% after the onset of fishbones while the electron temperature barely changes, which implies that the fishbone modes potentially trigger an ion-ITB in this DIII-D plasma, as there is no additional heating power at that time. Such an observation is reminiscent of similar findings in ASDEX \cite{Guenter2001}, MAST \cite{Field2011}, HL-2A \cite{Chen2016} and EAST \cite{Gao2018} plasmas. Therefore, beyond being a good match for validation purposes, this discharge is also particularly well suited to study the nonlinear interplay between fishbones, zonal flows and microturbulence, as fishbones have been identified as the potential cause for increased performances in a large number of tokamak experiments.
\section{Simulation setups}
\begin{figure}[h!]
\begin{subfigure}{.5\textwidth} 
   \centering
   \includegraphics[scale=0.4]{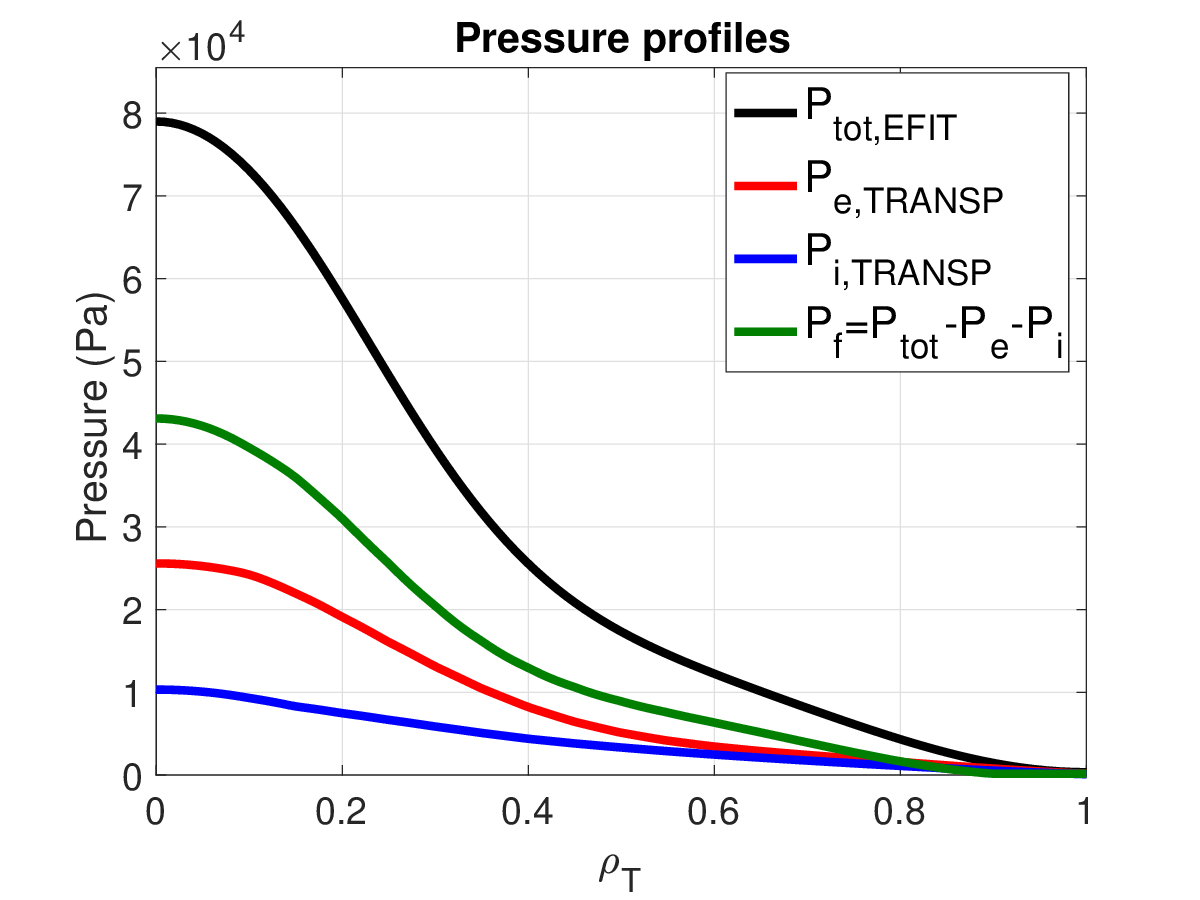}
   \caption{}
\end{subfigure}     
\begin{subfigure}{.5\textwidth} 
   \centering
   \includegraphics[scale=0.4]{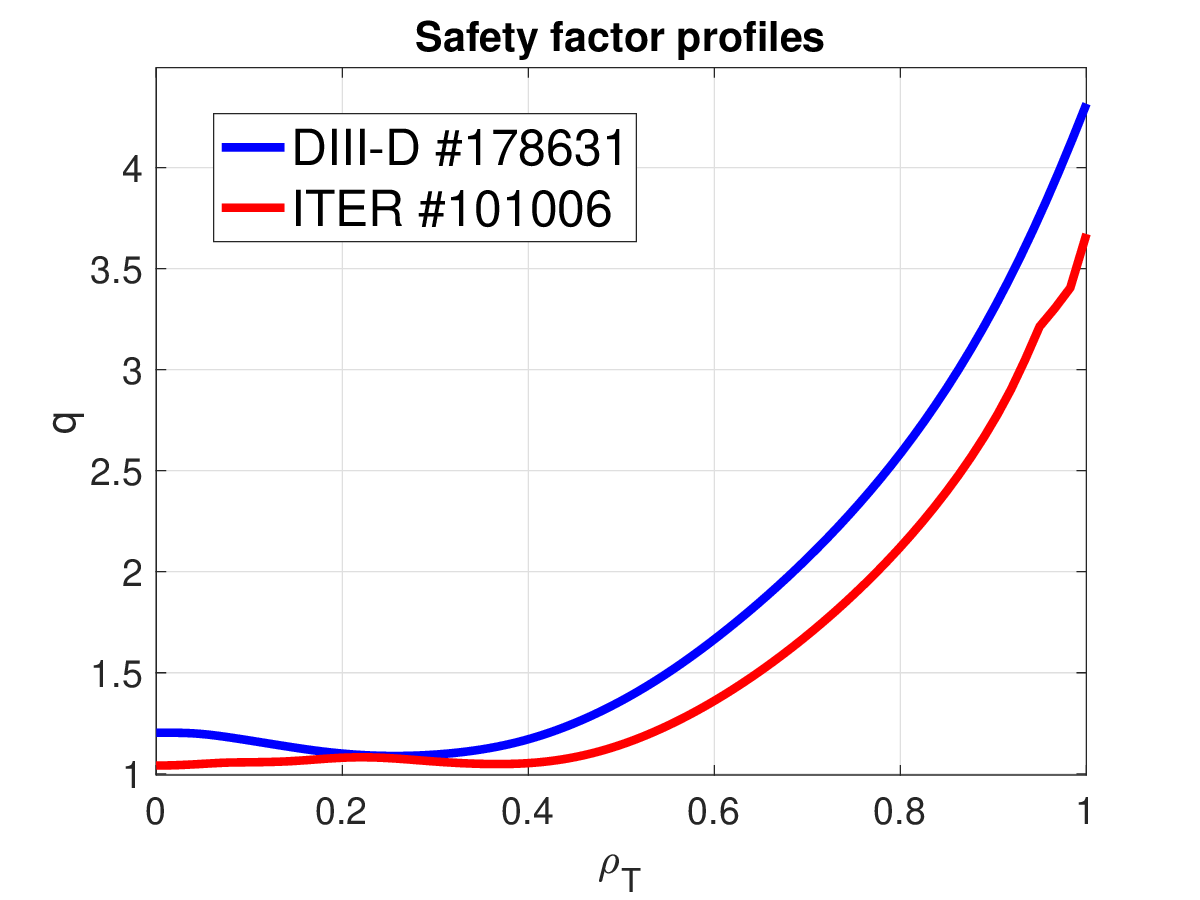}
   \caption{}
\end{subfigure}
\begin{subfigure}{.5\textwidth} 
   \centering
      \includegraphics[scale=0.4]{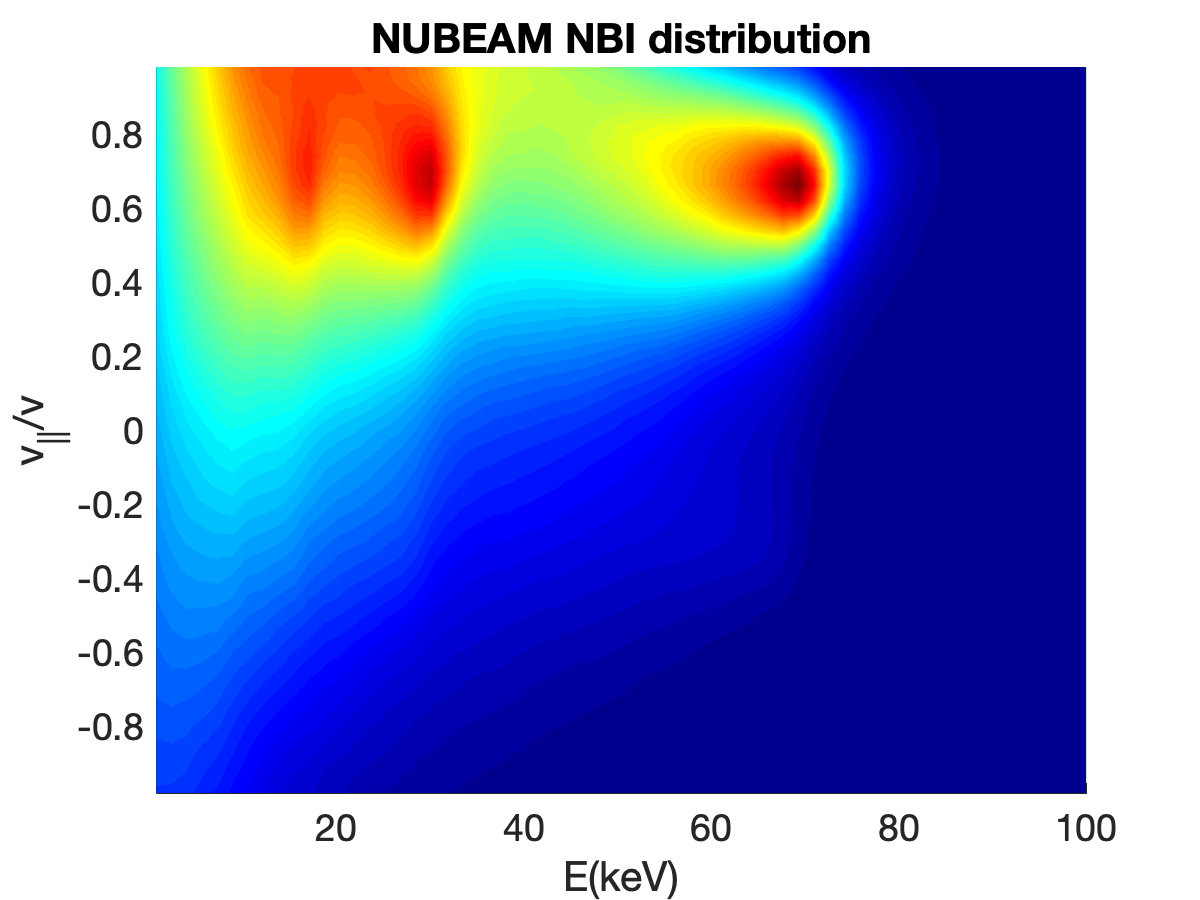}
   \caption{}
\end{subfigure}  
\begin{subfigure}{.5\textwidth} 
   \centering
      \includegraphics[scale=0.4]{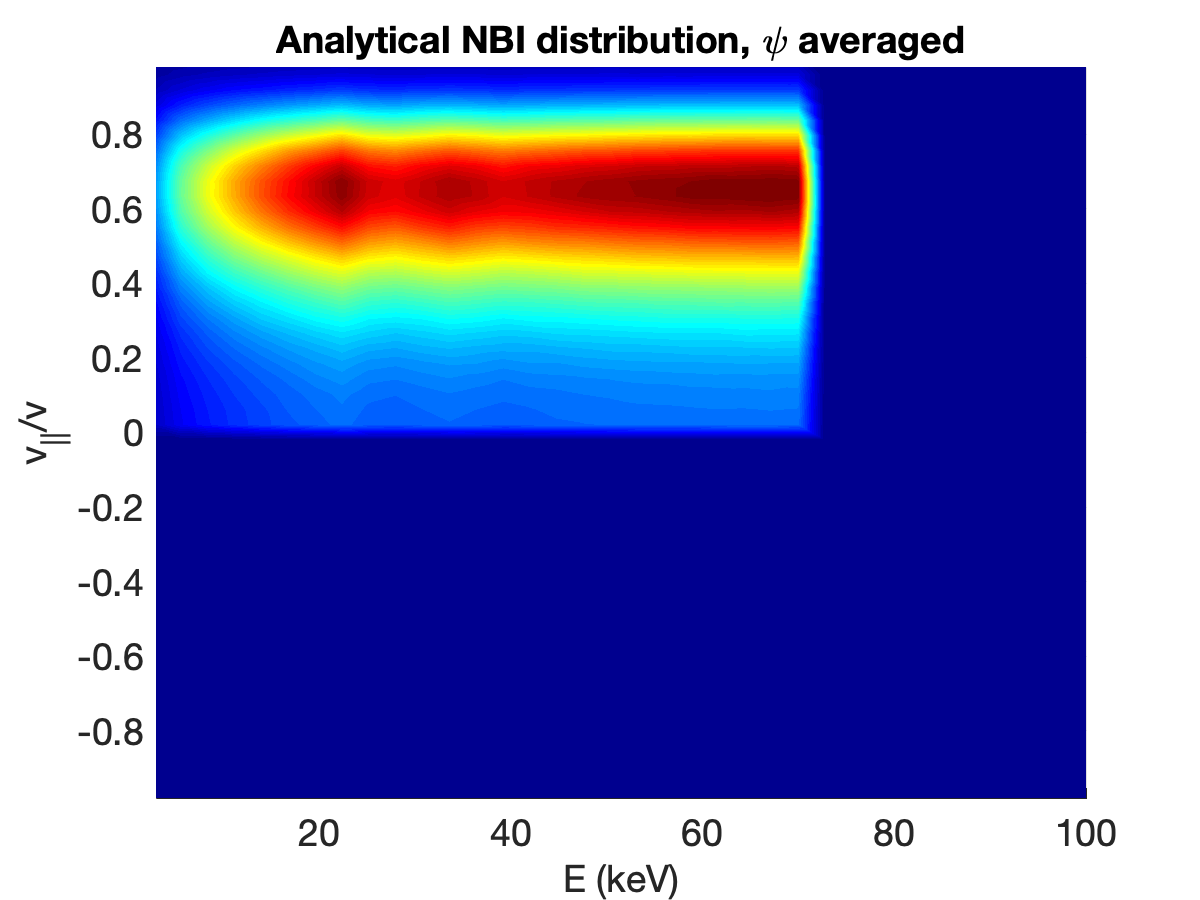}
   \caption{}
\end{subfigure}  
\caption{DIII-D \#178631 numerical equilibrium at t=1580ms. (a) Partial pressure profiles (b) Safety factor profiles, for both DIII-D and ITER scenarios. (c) NUBEAM and (d) analytical NBI distributions in the $(E,v_{\parallel}/v)$ phase space diagram.}
\label{DIIID_eq}
\end{figure}

The global gyrokinetic code GTC was the primary code used for the nonlinear modelling of $n=1$ fishbone modes in the DIII-D discharge and the ITER scenario. Two other first-principles codes, the kinetic-MHD codes M3D-C1 \cite{Jardin2012}\cite{Liu2022} and XTOR-K \cite{Luetjens2010}\cite{Brochard2020a} were also utilised, to provide nonlinear comparisons with GTC. The capability of a gyrokinetic electromagnetic code such as GTC to simulate low-n global MHD modes was recently demonstrated in a verification and validation (V\&V) work \cite{Brochard2022}. This V\&V study was conducted for n=1 internal kink modes in another DIII-D plasma, the verification involving in particular a benchmark between GTC, M3D-C1, XTOR-K and two other MHD codes in the ideal MHD limit. The usual gyrokinetic ordering $k_{\parallel}/k_{\perp}\ll 1$ is respected for n=1 global MHD modes, as $k_{\parallel}\sim0$ and $k_{\perp}^{-1}\sim r_{res}$, with $r_{res}$ the resonant surface at which $\textbf{k}\cdot\textbf{B}_0=0$, $\textbf{k}$ being the mode wave vector. \\ The MHD equilibrium, the plasma profiles and the EP distribution of the DIII-D discharge at t=1580ms were respectively reconstructed by the EFIT, TRANSP \cite{Grierson2018}, and NUBEAM \cite{Pankin2004} codes. The TRANSP plasma profiles were partly modified to enforce the pressure balance of the MHD equilibrium computed by EFIT, the sum of the partial pressures from TRANSP being larger than the total MHD pressure in EFIT. Given that the EP scalar pressure profile has the largest uncertainty in TRANSP simulations, it is redefined as $p_f=p_{tot,EFIT}-p_{i,TRANSP}-p_{e,TRANSP}$, which still yields a large fast ion beta of $\beta_f/\beta_{tot}=54\%$. This modification is crucial to study global MHD modes, as a magnetic configuration that does not have a self-consistent pressure balance $\nabla p = (\nabla\times\textbf{B}_0)\times\textbf{B}_0/\mu_0$ strongly modifies their linear stability \cite{Bussac1975}. The partial pressure profiles of the DIII-D configuration are displayed in Fig.\ref{DIIID_eq}a.\\\\
\begin{figure}[h!]
\begin{subfigure}{.333\textwidth} 
   \centering
   \includegraphics[scale=0.25]{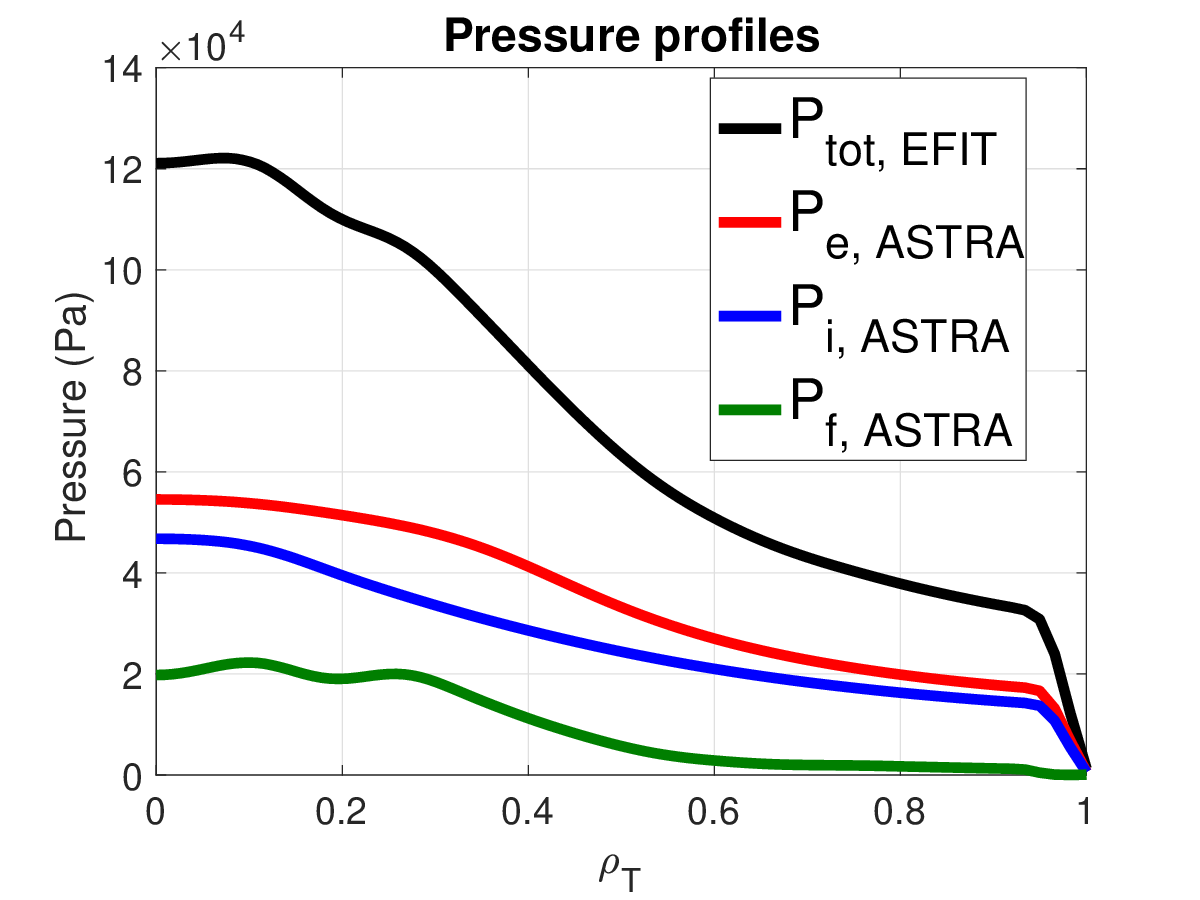}
   \caption{}
\end{subfigure}     
\begin{subfigure}{.333\textwidth} 
   \centering
   \includegraphics[scale=0.25]{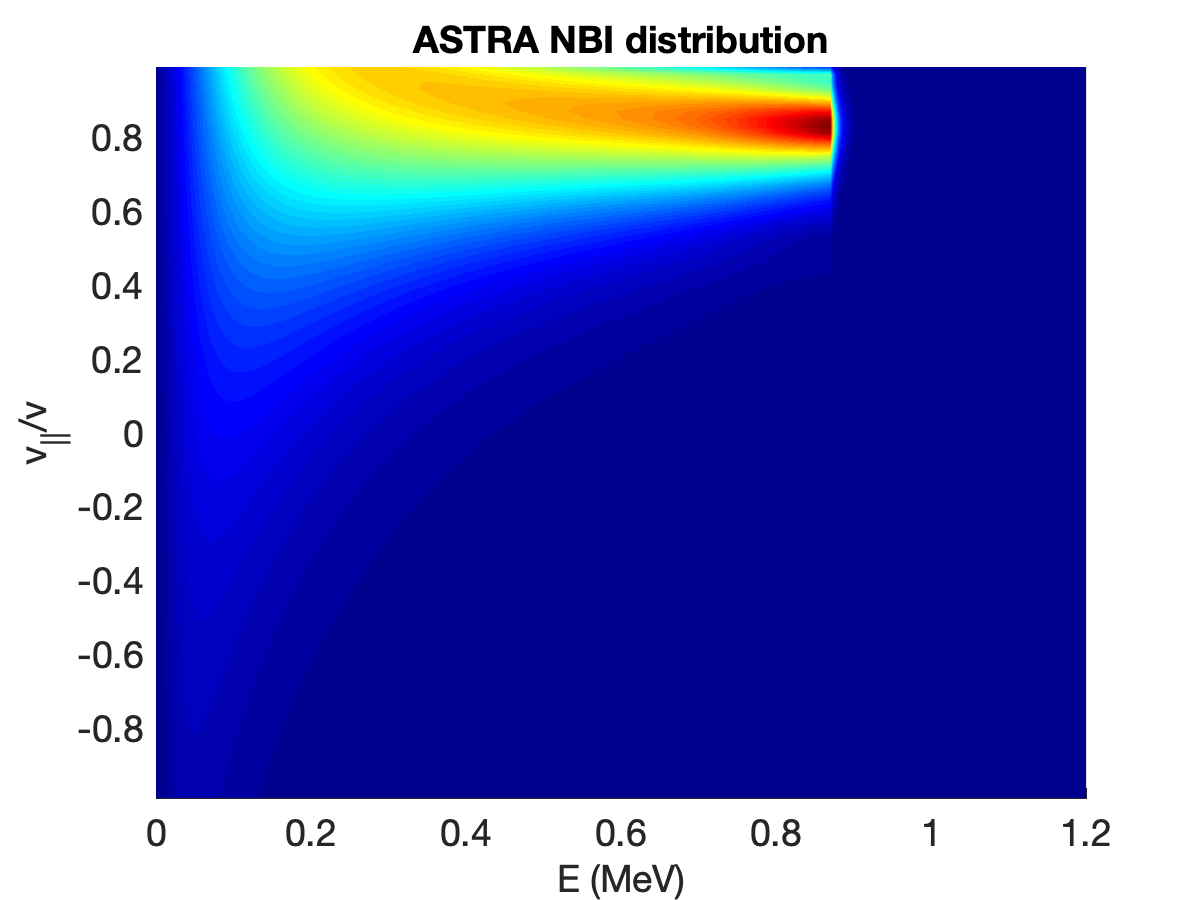}
   \caption{}
\end{subfigure}
\begin{subfigure}{.333\textwidth} 
   \centering
      \includegraphics[scale=0.25]{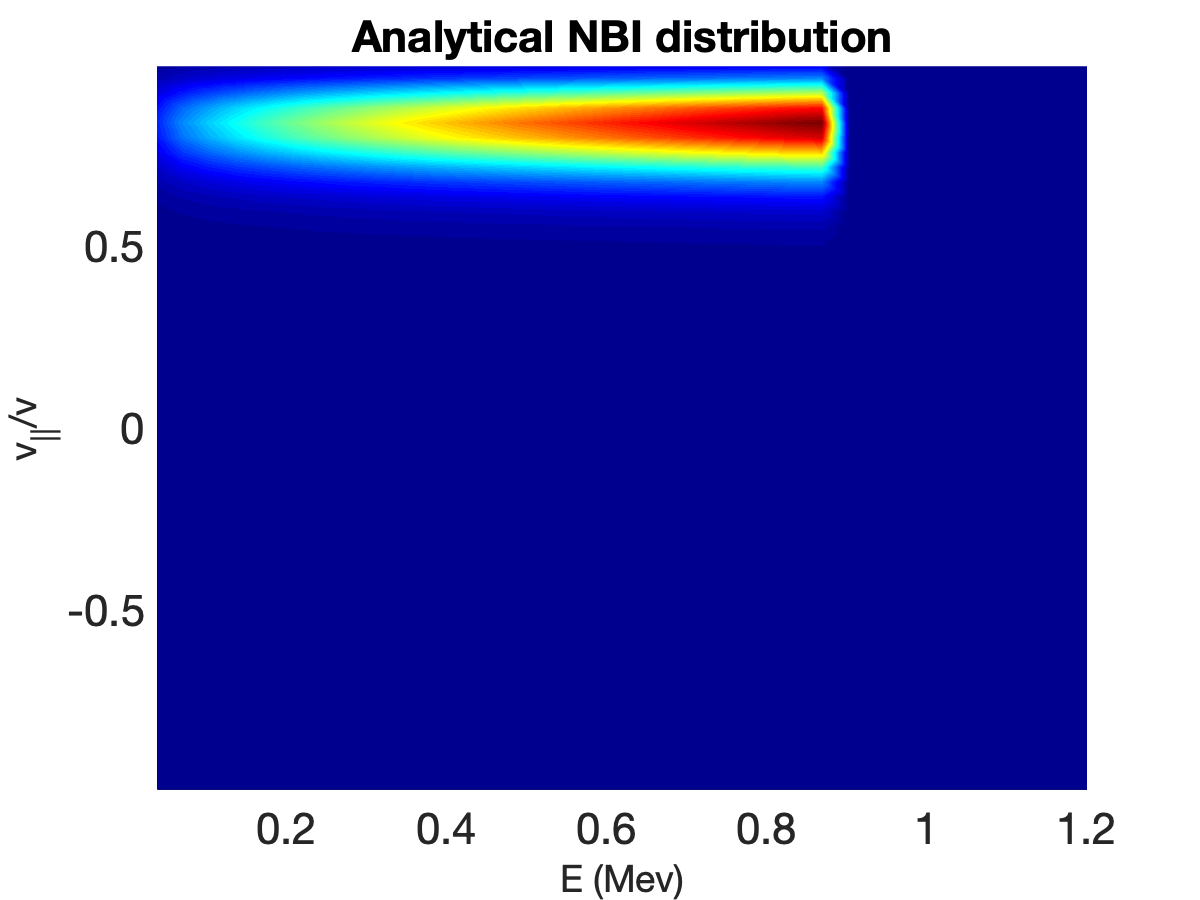}
   \caption{}
\end{subfigure}  
\caption{ITER \#1001006 numerical equilibrium at t=1580ms. (a) Partial pressure profiles (b) ASTRA and (c) analytical NBI distributions in the $(E,v_{\parallel}/v)$ phase space diagram.}
\label{ITER_eq}
\end{figure}
The numerical EP distribution from the NUBEAM code is computed in the 4D phase space $(E,v_{\parallel},R,Z)$, $E$ being the particles' kinetic energy, $R$ and $Z$ their cartesian position in the poloidal plane. This distribution is shown in Fig.\ref{DIIID_eq}c in the $(E,v_{\parallel}/v$) diagram, summing over the different $(R,Z)$ contributions. Three injection energies can be observed at $E_0\sim 70$keV, $E_0/2$ and $E_0/3$, which is characteristic of EP distributions resulting from beams with positive-ion sources \cite{Moseev2019}, as used in DIII-D. Such a complex numerical distribution cannot however be used directly in GTC yet. The code using a PIC $\delta f$ method to describe fast ions, representing an arbitrary numerical distribution $F_{arb}$ would require a precise computation of its first order derivatives $\nabla F_{arb}|_{\mu,v_{\parallel}}$ and $\partial_{v_{\parallel}}F_{arb}|_{\mu,\textbf{X}}$ in the phase-space vicinity of each marker to iterate the weight equation \cite{Dong2017}, which is not straightforward. A general method that can provide smooth $C^2$ inputs for $\delta f$ and full-F codes from numerical distributions, obtained with Fokker-Planck codes \cite{Pankin2004}\cite{Hirvijoki2014} or experimental measurements \cite{Schmidt2023}, will soon be reported in another publication, expanding on a previous work \cite{Bierwage2022} performed for the ITER IMAS (Integrated Modeling \& Analysis Suite) platform (ref). Therefore, to circumvent this issue, the NUBEAM distribution is fitted analytically in GTC, M3D-C1 and XTOR-K, by employing a set of three anisotropic slowing-down distributions with different injection energies. This is an important step for the simulations of fishbone modes, as their drive is significantly modified when using realistic distributions such as slowing-down distributions, compared to equivalent maxwellian distributions. The analytical distribution implemented in GTC reads
\begin{equation}\label{SDani}
F_{SD,ani}(v,\lambda,\psi)=\frac{1}{C}\frac{n_f(\psi)}{v^3+v_c^3}e^{-\big(\frac{\lambda-\lambda_0}{\Delta\lambda}\big)^2}\sum_{i=1}^{3}\alpha_iH(v_0/\sqrt{i}-v)
\end{equation}
with $v$ the particles velocity, $\lambda=\mu B_0/E$ the pitch angle, $\psi$ the poloidal flux, $n_f$ the EP density profile, $v_c$ the critical velocity, $v_0$ the injection velocity, $C$ a normalisation constant and $H$ the Heaviside function. The anisotropy of the distribution is described by a Gaussian in the $\lambda$ direction, $\lambda_0$ being the pitch angle peak and $\Delta\lambda$ the pitch angle width. The $\alpha_i$ factors describe the strength of each injection energy peak, their value is within $[0,1]$ and their sum is equal to one. An analytical fit of the NUBEAM distribution using Eq.(\ref{SDani}) is displayed on Fig.\ref{DIIID_eq}d, where the following parameters have been chosen : $E_0=70$keV, $v_0=2.59\times10^6$m.s$^{-1}$, $\lambda_0=0.6$, $\Delta\lambda=0.3$, $\alpha_1=0.9$, $\alpha_2=0.06$, $\alpha_3=0.04$. The critical velocity is also chosen to be constant to best fit the NUBEAM distribution, using $v_c=1.29v_0$. Due to the beams alignment, the experimental distribution is mostly composed of co-going particles with $v_{\parallel}/v>0$. The pitch angle Gaussian in Eq.(\ref{SDani}) not discriminating against the $v_{\parallel}$ direction, this feature of the NUBEAM distribution is enforced by only loading markers with $v_{\parallel}>0$. The EP density profile is not directly taken from TRANSP, as TRANSP is using a maxwellian approximation for EPs. Instead, the EP density profile is imposed as to maintain the pressure balance of the MHD equilibrium following $n_f=p_f/T_f$, with $T_f=\int d^3\textbf{v} \ mv^2(F_{SD,ani}/n_f)$ the equivalent maxwellian temperature profile for slowing-down distributions. It should be noted here that the EFIT reconstruction considered in this work is isotropic, and does not take into account the anisotropic contribution from the beam ions. Retaining such a contribution in equilibrium codes is possible \cite{Fitzgerald2013}\cite{Qu2014}, however the GTC code is not currently capable of handling anisotropic equilibria, which require describing 1D equilibrium quantities such as density profiles as 2D functions. A generalisation of the GTC code towards anisotropic equilibria will be considered in a future work, as the EP beta in present-day tokamaks is significant, making therefore the MHD equilibrium non-negligibly anisotropic, which can affect the stability of global EP-driven modes. Such a limitation is however less stringent for future burning plasmas experiments, as the $\beta_f/\beta_{tot}$ ratio will be of order 10-20\%, i.e. the ratio of the slowing down time to the energy confinement time.\\
The plasma profiles and the EP distribution for the selected ITER scenario have been computed with the ASTRA code \cite{Pereverzev2002}\cite{Polevoi2019}. The partial pressure profiles and the EP distribution in the $(E,v_{\parallel}/v)$ phase space diagram are shown on Fig.\ref{ITER_eq}a-b. The beam distribution is also mostly co-passing, and is fitted analytically in Fig.\ref{ITER_eq}c with Eq. (\ref{SDani}) using $E_0=870$keV, $v_c=7.07v_0$, $\lambda_0=0.34$, $\Delta\lambda=0.18$, $\alpha_1=1$.\\\\
In all nonlinear simulations, the simulation domains contain the whole plasma volume. GTC uses an outer edge buffer past $\rho_T\sim 0.8$, with $\rho_T$ the normalized square root of the toroidal flux $\psi_T$, after which all equilibrium gradients are removed in both DIII-D and ITER configurations. GTC simulations use gyrokinetic thermal and fast ions described with a $\delta f$ method, and massless fluid electrons evolved with an electron continuity equation \cite{Dong2017}. The electron contribution to zonal density is artificially removed in nonlinear simulations, based on their adiabatic response and to avoid numerical instabilities. This aspect will be detailed further in section 5.1. M3D-C1 also uses a $\delta f$ method with both thermal ion and fast ion kinetic effects. XTOR-K only describes kinetically the fast ions, with a full-f approach and a 6D full-orbit method resolving the EP gyroradius. Only the n=0 and n=1 modes are retained in GTC and M3D-C1 simulations, as to specifically study the interplay existing between n=1 fishbone modes and $n=m=0$ zonal flows. XTOR-K simulations are however restricted to the n=1 modes only, for two reasons. First because realistic simulations of zonal flows need to take into account their collisionless damping, due to both thermal and fast ions kinetic effects. Neglecting thermal ion kinetic effects would then underestimate the zonal flows residual levels \cite{Rosenbluth1998}. Moreover, since XTOR-K employs a full-f method, the code does not split the $n=0$ mode into equilibrium and perturbed parts to only evolve the perturbed components, as performed with codes using a $\delta f$ method. When considering an isotropic MHD equilibrium with an anisotropic EP distribution, this leads to an important evolution of the $n=0$ equilibrium that significantly perturbs the growth of the n=1 fishbone modes. For both these reasons, $n=0$ modes are currently filtered out in XTOR-K simulations. Convergence studies have been carried out over spatial grid size, time step, number of particles per cells and radial boundary treatment in all codes. Specifically, the radial, poloidal and parallel/toroidal grid resolution used for GTC and XTOR-K were respectively $N_{\psi}=100, 200$; $N_{\theta}=250 $ at $r=0.5a$ ($r\Delta\theta$ is constant on each flux surface), 64; $N_{\zeta |\varphi}=24,12$. M3D-C1 uses a 3D mesh with 8 poloidal planes at different toroidal angles, each poloidal plane containing 5629 triangle elements. In GTC we imposed a Gaussian radial boundary decay from $\rho_T\sim0.8$ to $\rho_T=1$, while XTOR-K and M3D-C1 use respectively at the edge a free-slip and a no-slip boundary condition.
\section{Nonlinear validation against DIII-D experiment}
\subsection{Linear stability of the $n=1$ fishbone}
\begin{figure}[h!]
\begin{subfigure}{.5\textwidth} 
   \centering
   \includegraphics[scale=0.4]{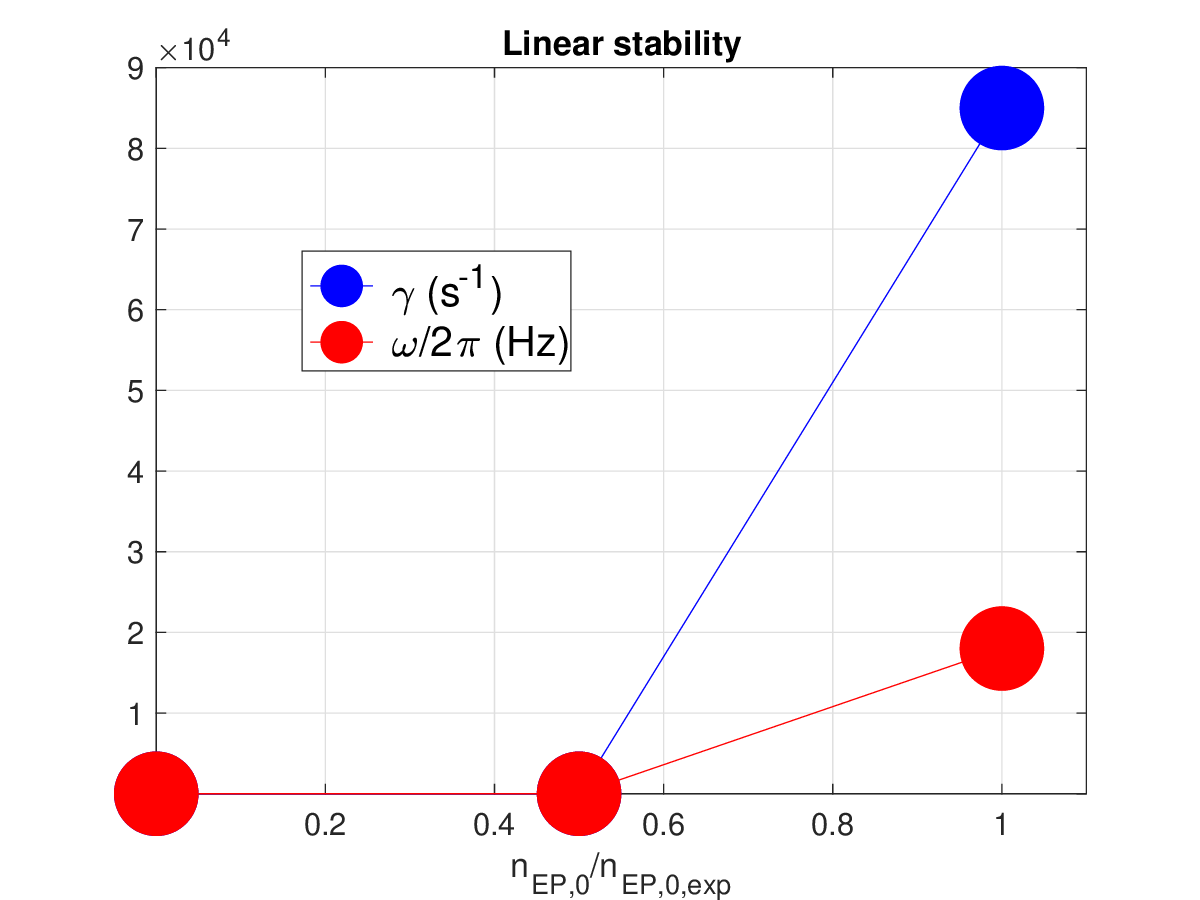}
   \caption{}
\end{subfigure}     
\begin{subfigure}{.5\textwidth} 
   \centering
   \includegraphics[scale=0.22]{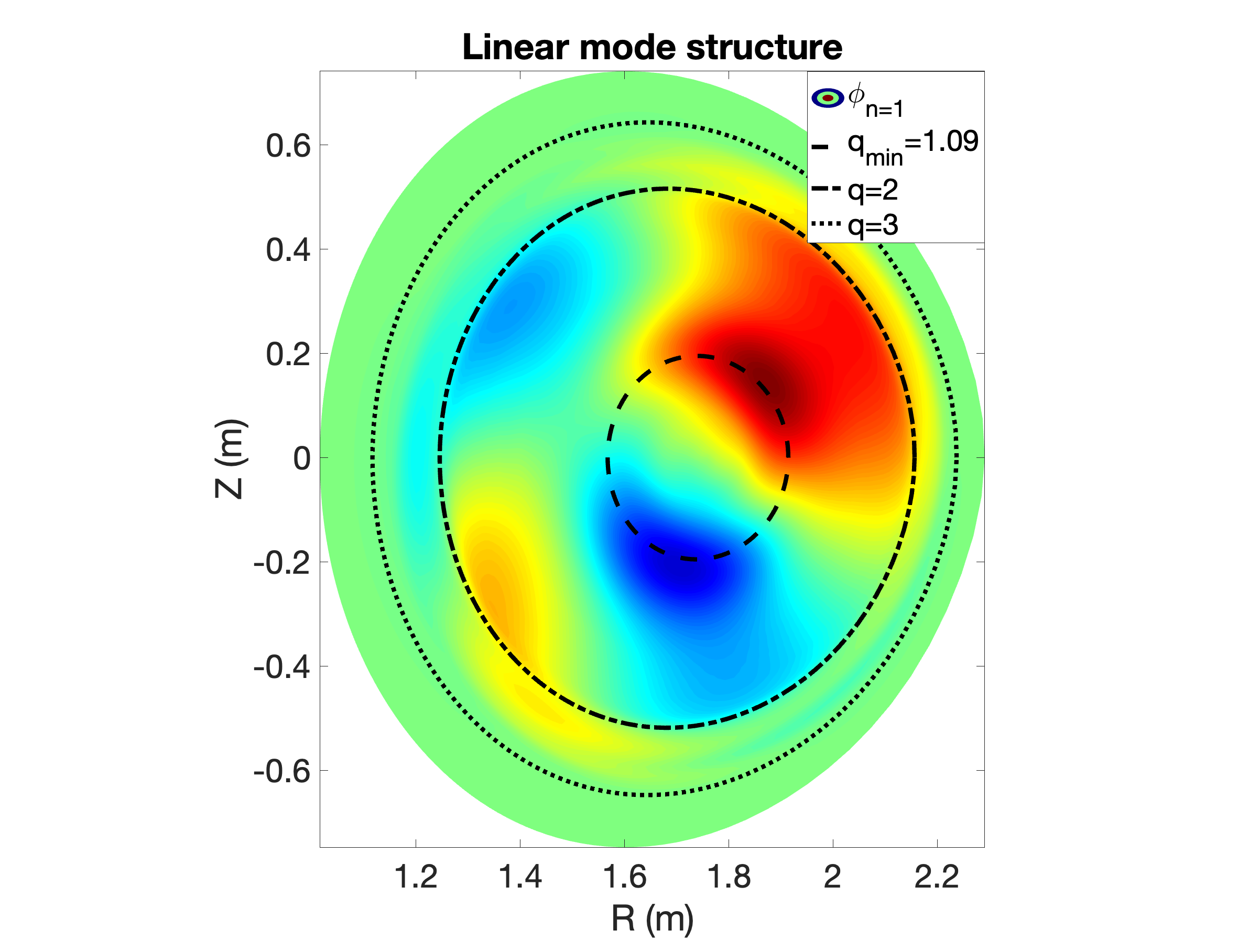}
   \caption{}
\end{subfigure}
\begin{subfigure}{.5\textwidth} 
   \centering
      \includegraphics[scale=0.4]{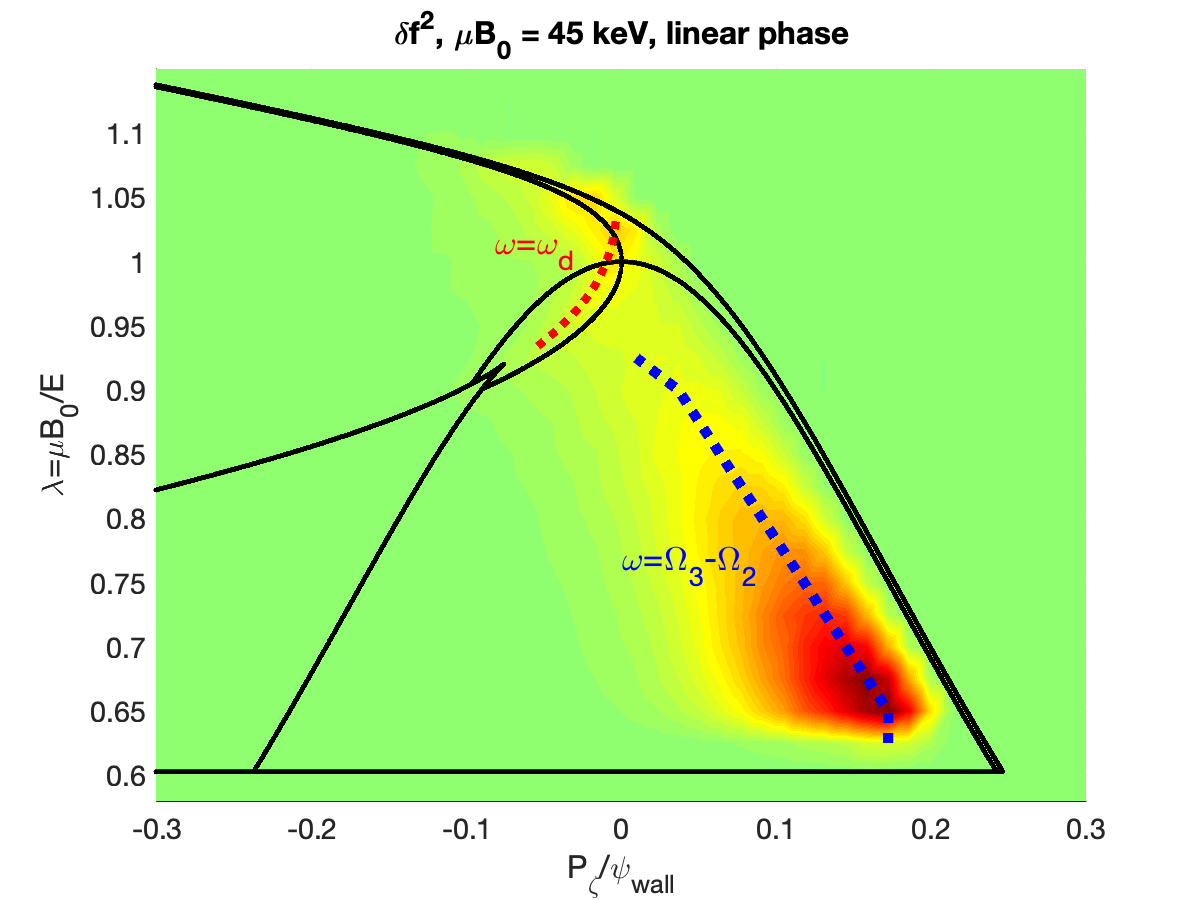}
   \caption{}
\end{subfigure}  
\begin{subfigure}{.5\textwidth} 
   \centering
      \includegraphics[scale=0.4]{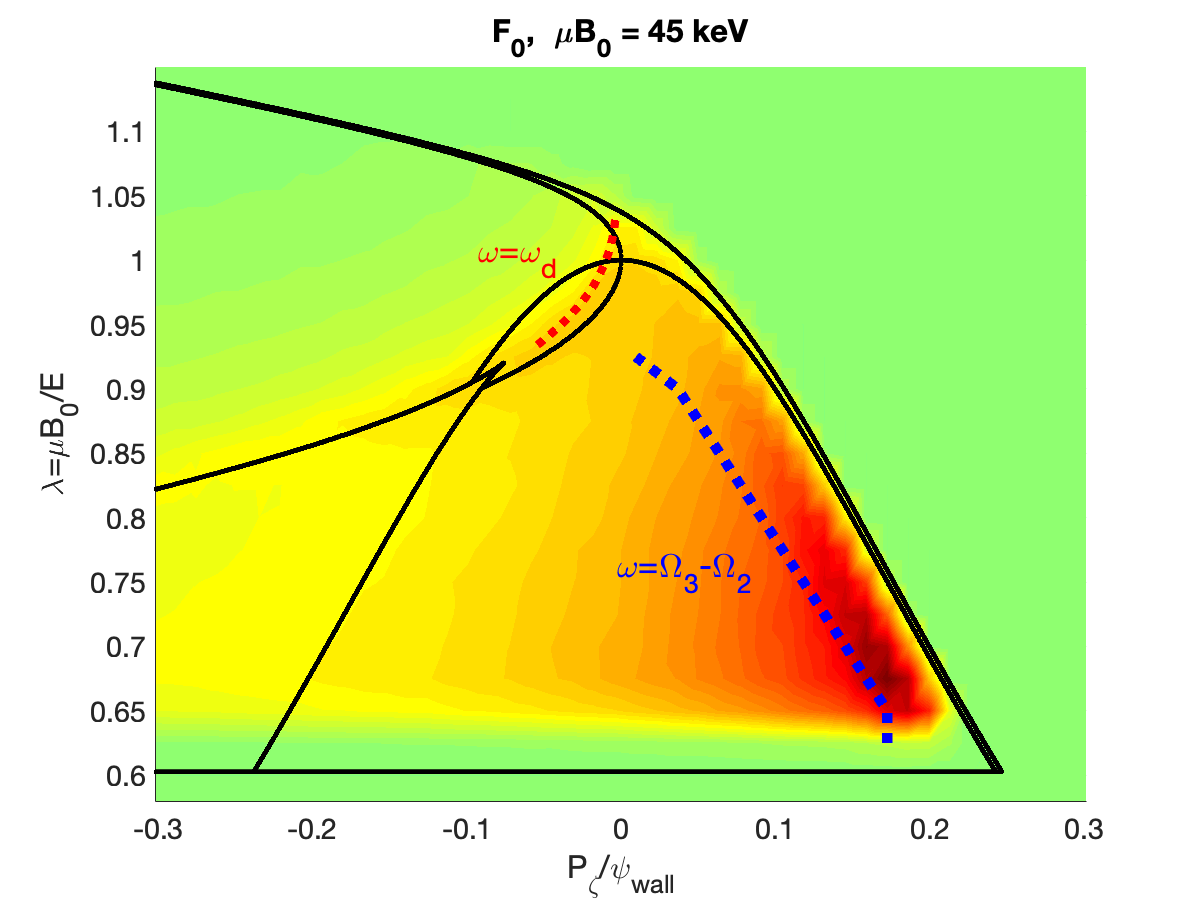}
   \caption{}
\end{subfigure}  
\caption{Kinetic-MHD stability of the DIII-D discharge from GTC gyrokinetic simulations. (a) Linear stability of the $n=1$ fishbone mode against on-axis EP density. (b) n=1 mode structure of the electrostatic potential $\phi_{n=1}$. (c) $\delta f^2$ and (d) $F_0$ EP histograms in the $(P_{\zeta},\lambda)$ phase space diagram at $\mu B_0=45keV$.}
\label{DIIID_linear}
\end{figure}
The linear stability of the DIII-D configuration to $n=1$ fishbone modes is described on Fig. \ref{DIIID_linear} using GTC linear simulations. In the absence of EPs, the n=1 internal kink modes is found stable for this configuration. Fishbone modes are destabilized only when the realistic beam distribution described by Eq. (\ref{SDani}) is used in GTC, past a EP pressure threshold at $p_{f,thres}\sim0.8p_f$. When the equivalent Maxwellian distribution is used instead of the anisotropic slowing-down distribution, the fishbone mode is fully stabilized. Slowing-down distributions therefore strongly enhances the drive of fishbones modes compared to maxwellian distributions, which stresses the fact that global EP-driven modes can indeed be very sensitive to the considered EP distributions. A scan of the fishbone linear stability against the EP on-axis density is displayed in Fig \ref{DIIID_linear} (a). At the experimental on-axis density $n_{f,exp}$, the fishbone mode has a growth rate of $\gamma=8.5\times10^4$s$^{-1}$, and a mode frequency of $\omega/2\pi=17kHz$. In this paper, the positive sign convention for frequencies corresponds to the ion diamagnetic rotation direction. The fishbone mode structure can be observed in the poloidal plane in Fig \ref{DIIID_linear} (b). The mode has a dominant m=1 harmonic that peaks at $q_{min}=1.09$, and a subdominant yet significant $m=2$ harmonic that vanishes past the $q=2$ surface. \\Fishbone modes are driven unstable by wave-particles resonances between the fishbone frequency and the EPs poloidal and toroidal frequencies, characterised by the following resonance condition in angle-action coordinates \cite{Brochard2020b}
\begin{equation}\label{res_cond}
\omega=l\Omega_2(P_{\zeta},\lambda,\mu)+n\Omega_3(P_{\zeta},\lambda,\mu)
\end{equation}
where $l$ is a relative integer,  $\Omega_2$ is the poloidal bounce/transit frequency, $\Omega_3$ is the toroidal precessional/transit frequency, and $\textbf{J}=(P_{\zeta},\lambda,\mu)$ the actions i.e. the conserved quantities of the tokamak configuration, with $P_{\zeta}$ the toroidal canonical momentum, and $\mu$ the magnetic moment. The relevant resonances lines can be identified by projecting the linearly perturbed EP distribution squared $\delta f^2$ in the $(P_{\zeta},\lambda)$  phase space at a given $\mu$, the resonances being positioned at the locations in phase space where $\delta f^2$ peaks. Two distinct resonances have been found to drive the mode. Since both resonances co-exist at $\mu B_0=45$keV, this $\mu$ value has been chosen to highlight the nature of the resonances, and more generally to charactarize the nonlinear phase space dynamics for this DIII-D configuration in the rest of the paper. The $\delta f^2$ and the $F_0$ histograms are respectively displayed in the $(P_{\zeta},\lambda)$ diagram at $\mu B_0=45$keV in Fig. \ref{DIIID_linear} c-d.
\begin{figure}[h!]
\begin{subfigure}{.5\textwidth} 
   \centering
   \includegraphics[scale=0.185]{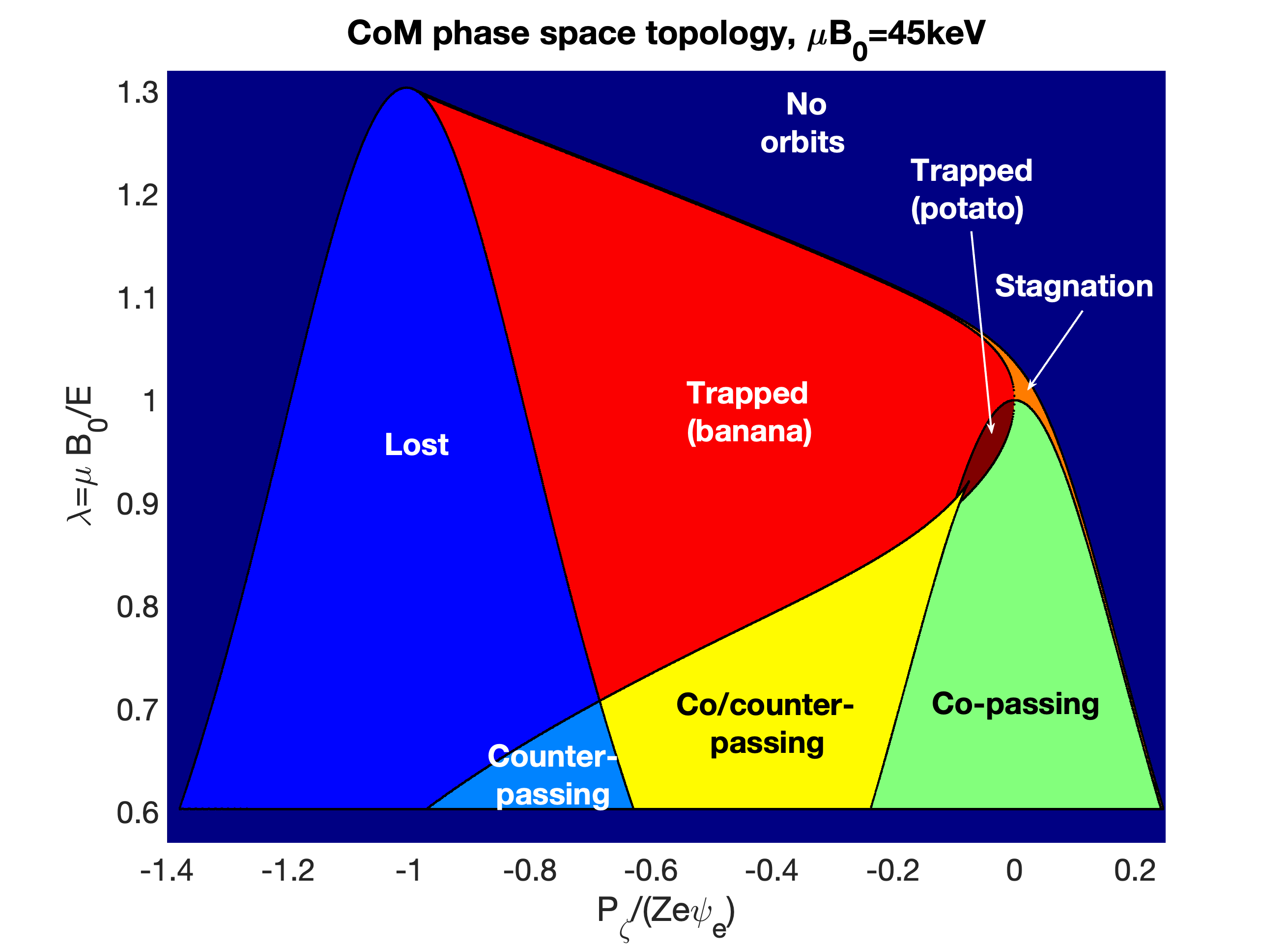}
   \caption{}
\end{subfigure}     
\begin{subfigure}{.5\textwidth} 
   \centering
   \includegraphics[scale=0.185]{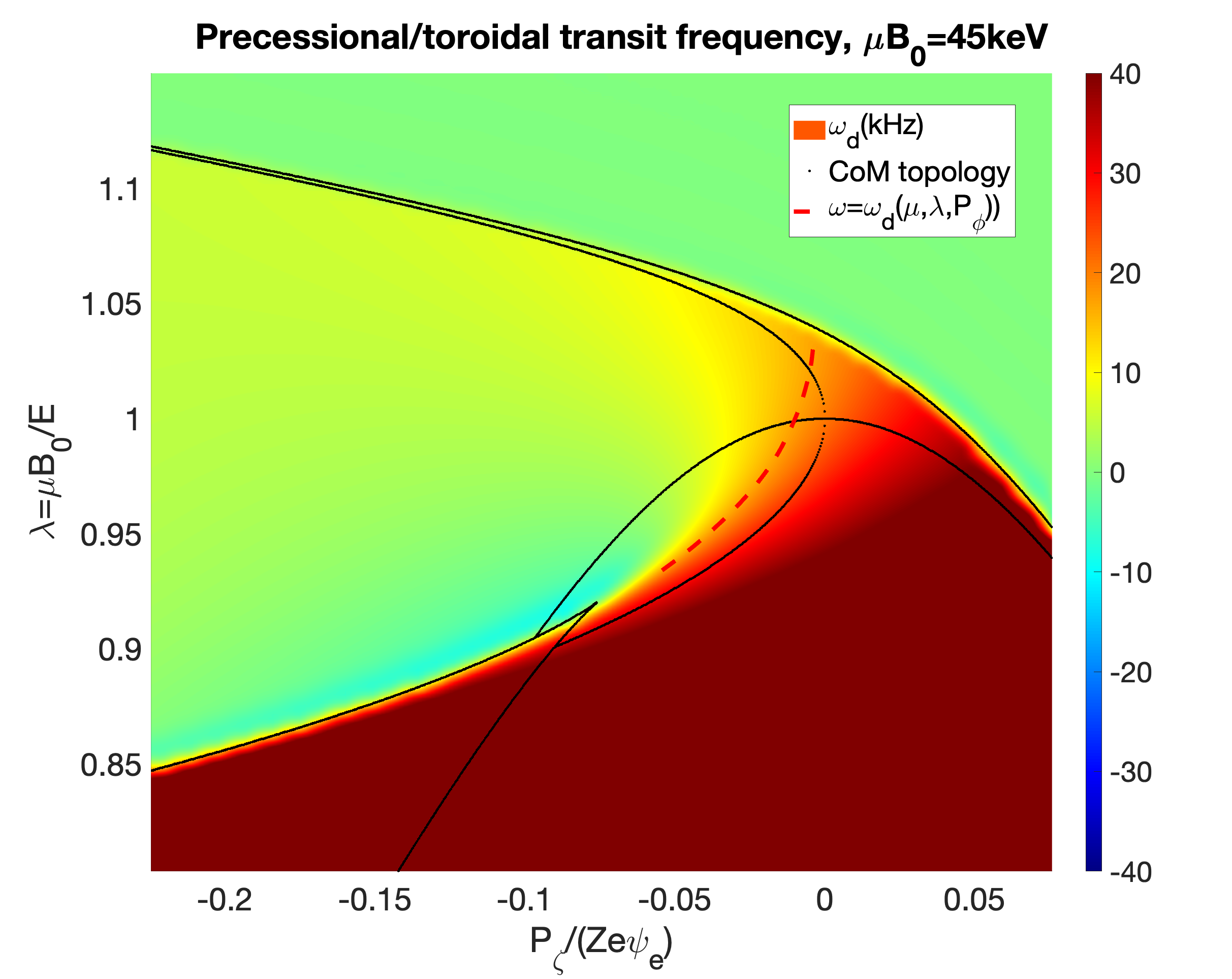}
   \caption{}
\end{subfigure}
\caption{(a) Topology of the $(P_{\zeta},\lambda,\mu B_0$=45kev) CoM phase space. (b) Precessional/toroidal frequency in the $(P_{\zeta},\lambda)$ diagram at $\mu B_0=45keV$.}
\label{DIIID_CoM}
\end{figure}
Two resonances can be identified in Fig. \ref{DIIID_linear} c, which belong to two distinct zones of the constants of motion (CoM) phase space topology, zones which are delimited by black lines in Fig. \ref{DIIID_linear} c-d. These lines have been obtained by initializing EPs on a fine cartesian grid (2000$\times$2000) in the CoM space, and evolving them onto the equilibrium magnetic field to recover their orbit in the poloidal plane. The CoM space topology, displayed on Fig. \ref{DIIID_CoM}a, can then be obtained as a function of the particles $(R,v_{\parallel})$ values when their orbits cross the midplane on both the low field and high field sides. The topology in Fig. \ref{DIIID_CoM}a is similar to what is described in \cite{White2006}, Fig 3.3. This classification is performed here using a code currently being developed at the ITER organisation, based on XTOR-K's particle pusher, which converts numerical/experimental distributions into smooth CoM distribution inputs for linear and first-principle codes. As mentioned in section 3, this work will be discussed in an upcoming publication. According to Fig. \ref{DIIID_linear}c and Fig. \ref{DIIID_CoM}a, one resonance is purely due to co-passing particles, while the other is mostly caused by trapped particles (both banana and potato orbits) and partly by stagnation orbits. \\ The resonances identification is performed by following the time evolution of the particles poloidal and toroidal angles over their orbits in order to explicitly compute $\Omega_2$ and $\Omega_3$ in CoM space.  $\Omega_3$ is displayed in Fig. \ref{DIIID_CoM}b. As can be observed in this figure, the EP precessional frequency mostly goes in the ion diamagnetic direction, to the exception of a small zone near the trapped-passing boundary where the precession reverses. This zone is known to destabilize electronic fishbones \cite{Vlad2016}. The red line in Fig. \ref{DIIID_CoM}b highlights the location of the $l=0$ precessional resonance $\omega=\Omega_3=\omega_d$, crossing CoM zones belonging to potato, banana and stagnation orbits. The same line is also plotted on Fig. \ref{DIIID_linear} c-d, identifying the resonance in the trapped-stagnation region as the precessional resonance, as it coincides with the $\delta f^2$ structure located in this region. The co-passing resonance is somewhat harder to identify, as the $\Omega_2,\Omega_3$ frequencies have similar values in this CoM zone, with $\Omega_2/2\pi\sim\Omega_3/2\pi\sim1.5\times10^5$kHz. The fishbone frequency being about a tenth of these frequencies, the resonance in the co-passing part of phase space in Fig. \ref{DIIID_linear} c is most likely the $l=-1$ drift-transit resonance $\omega=\Omega_3-\Omega_2$. The $\Omega_2$ frequency cannot however yet be computed with a large enough precision to draw a resonance line in CoM space. Instead, the blue line in Fig. \ref{DIIID_linear} c-d and Fig. \ref{DIIID_CoM} is obtained from the position of the maximal $\delta f^2$ value at each $\lambda$, and therefore represents the $l=-1$ drift-transit resonance. Lastly, it can be noted in Fig. \ref{DIIID_linear} d that at each resonance location $\partial F_0/\partial P_{\zeta}<0$, which is a necessary condition to drive kinetic-MHD modes in analytical theory \cite{Porcelli1994}, further illustrating that $l=0$ precessional and $l=-1$ drift-transit resonances are driving the fishbone mode in this DIII-D configuration.
\subsection{Fishbone saturation dominated by self-generated zonal flows}
As mentioned in section 3, the saturation of $n=1$ fishbone modes is analysed here with nonlinear GTC simulations only keeping $n=1$ modes, with and without $n=m=0$ zonal flows. This setting is meant to specifically study the two-way nonlinear interplay existing between fishbone modes and zonal flows. The excitation of zonal flows by fishbone bursts is a well-known aspect of this interplay, which was reported both experimentally in CHS \cite{Ohshima2007} and HL-2A \cite{Chen2016a} plasmas, and numerically in low-n global kinetic-MHD simulations \cite{Shen2015}\cite{Brochard2020b}\cite{Ge2022}. The nonlinear impact of zonal flows on the $n=1$ fishbone saturation, if any, is however less clear, the fishbone saturation mechanism being mostly attributed to the flattening of the EP distribution around linear resonances \cite{Fu2006}\cite{Brochard2020b}. Previous numerical works showed that the inclusion of MHD nonlinearities reduces the saturation of $n=1$ fishbone instability, because of the additional dissipation brought by the n=0-4 side-bands in kinetic-MHD simulations \cite{Ge2022}\cite{Fu2006}. Nonetheless, the role played specifically by zonal flows in the fishbone saturation was not identified, and needs to be further studied. Such a study requires the inclusion of kinetic thermal ions effects to estimate realistically the zonal flows levels \cite{Rosenbluth1998}, study which can be self-consistently performed with GTC gyrokinetic simulations.
\begin{figure}[h!]  
\begin{subfigure}{.5\textwidth} 
   \centering
   \includegraphics[scale=0.35]{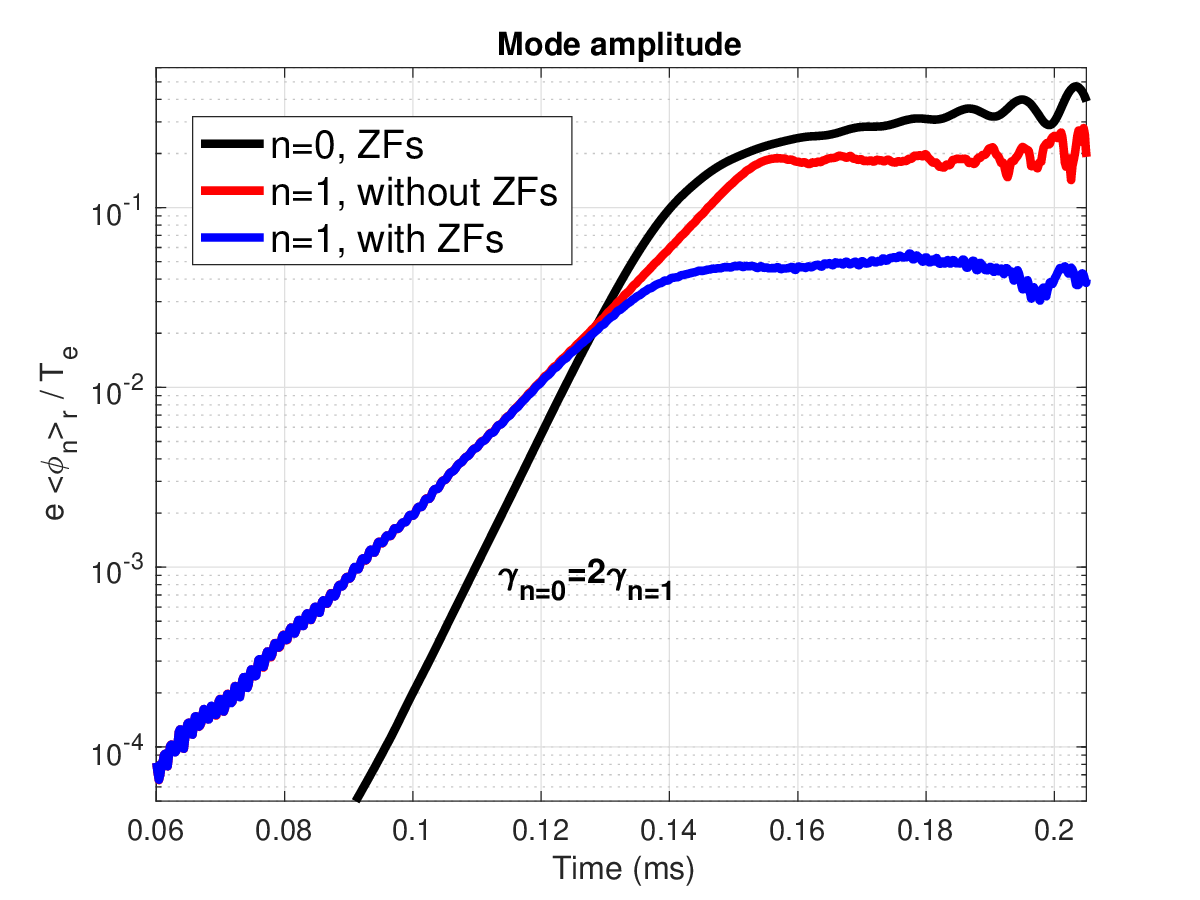}
   \caption{}
\end{subfigure}     
\begin{subfigure}{.5\textwidth} 
   \centering
      \includegraphics[scale=0.17]{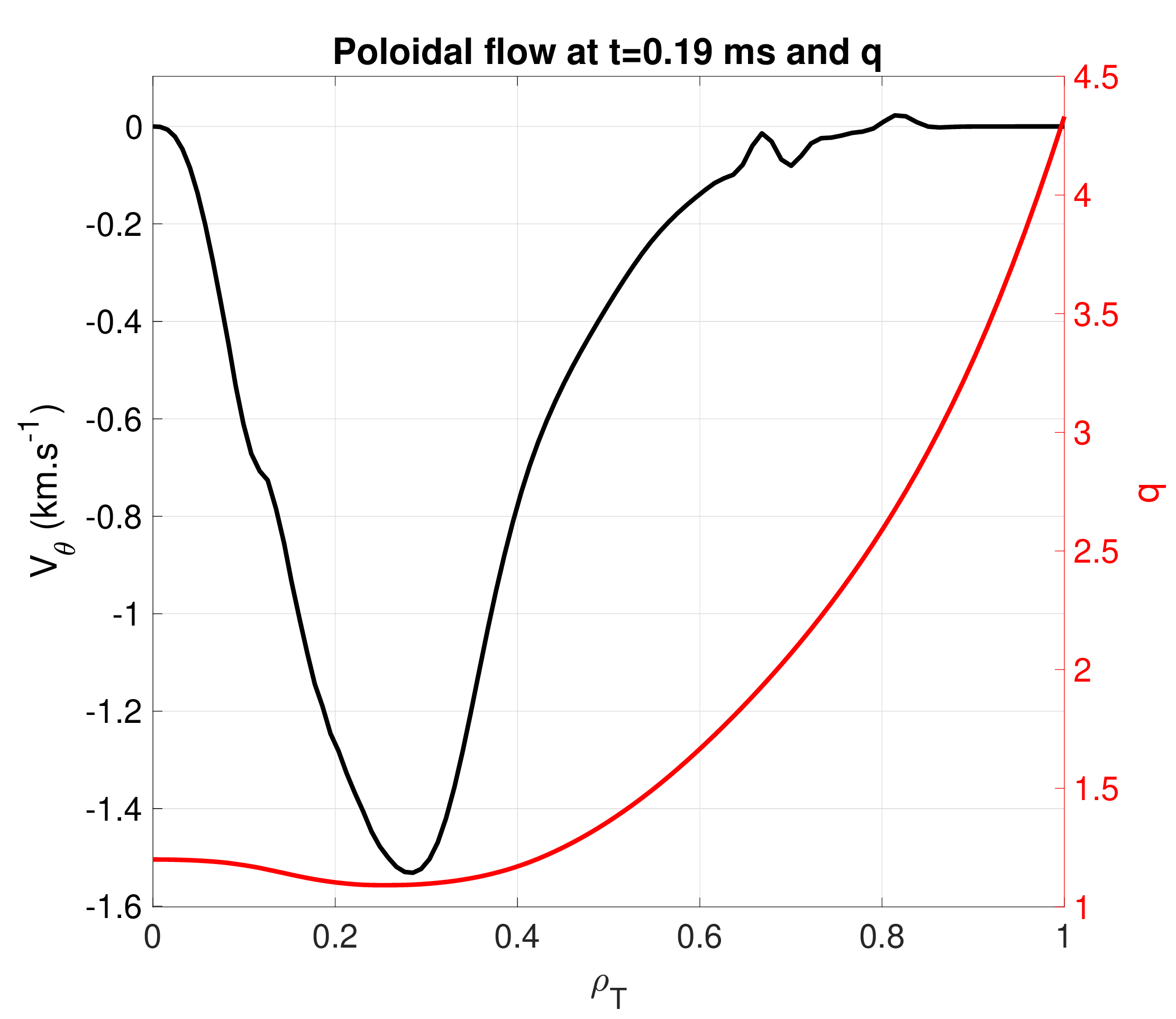}
   \caption{}
\end{subfigure}  
\caption{(a) Time evolution of volume-averaged perturbed electrostatic potential $e\langle \phi\rangle/T_e$ (n=0,1).(b) Zonal poloidal flow $V_{\theta}$ (km.s$^{-1}$) after saturation at t=0.19ms in GTC simulations. Figure (a) is reproduced from \cite{Brochard2024}}
\label{DIIID_time}
\end{figure}
\\
The time evolution of the volume-averaged electrostatic potential $e\langle \phi\rangle/T_e$ in GTC simulations is displayed in Fig. \ref{DIIID_time} a. The blue and red lines stand respectively for the $n=1$ mode with and without  the inclusion of zonals flows, while the $n=m=0$ zonal flows themselves are represented by the black line. From these results, it appears that the zonal flows inclusion dominates the fishbone saturation, which indicates that the underlying fishbone saturation mechanism is possibly more complex than the sole wave-particle resonant interaction. In both simulations, the $n=1$ mode saturates around $t=0.15$ms with an amplitude of $\delta B/B_0\sim2\times10^{-3}$ at $q_{min}$ with zonal flows, and $\delta B/B_0\sim8\times10^{-3}$ without. The zonal flows are found to be force-driven by the $n=1$ fishbone modes through $n=\pm1$ coupling, their linear growth rate being twice that of the fishbone. Zonal flows also experience a spontaneous growth after the saturation of the primary wave, which is a common feature in gyrokinetic simulations \cite{Chen2016}\cite{Liu2022a}. This zonal flows generation is reminiscent of that of TAEs (Toroidal Alfv\'en eigenmodes), which can likewise generate zonal flows through force-driven processes. It should however be noted that TAEs can also destabilise zonal flows through modulational instability \cite{Chen2012}, similarly to microscopic drift-waves such as ITG modes (Ion Temperature Gradients) and TEM (Trapped Electron modes) \cite{Chen2000}. Such results stress that zonal flows can be driven by instabilities occurring at every spatial scales in tokamak plasmas, and play a regulatory role on modes arising at these diffferent scales, as microturbulence and AEs saturation levels are also impacted by zonal flows \cite{Lin1998}\cite{Todo2012}. It is therefore expected that the overall bulk and EP transport in burning plasmas will depend on complex cross-scale interactions, as recently shown in DIII-D plasmas \cite{Liu2022a}, and discussed theoretically in \cite{Zonca2014}. It also implies that the inclusion microturbulence and meso-scale AEs can impact the fishbone saturation by affecting the overall zonal flow state. Cross-scale fishbone simulations will be conducted in a future study to quantify these effects. The mode structure of the $n=m=0$ can be observed on Fig. \ref{DIIID_time} b, where the radial profile of the zonal poloidal flow is displayed. In GTC the zonal flow $\textbf{V}_{00}$ is defined as the $E\times B$ flow resulting from the zonal electrostatic potential  $\textbf{V}_{00}=\textbf{b}_0\times\nabla\delta\phi_{00}/B_0$. The zonal poloidal flow have a macroscopic extent, which differs from the usual microscopic and mesoscopic structures recovered with microturbulence and AEs \cite{Liu2024}. This difference is due to the spatial scale of the primary wave driving the zonal flows. The zonal flows peak near the $q_{min}$ location with $V_{\theta}\sim1.5$km.s$^{-1}$. They are strongly sheared within $\rho\in[0,0.6]$, which can affect microturbulent transport. This aspect will be discussed in greater length in section 7.\\\\
The simulated fishbone saturation amplitudes from GTC, M3D-C1 and XTOR-K simulations have been successfully compared to DIII-D experimental measurements, as discussed in \cite{Brochard2024}, supporting the novel saturation mechanism of fishbone modes by self-generated zonal flows. In the limit without zonal flows, the radial envelope of $\delta T_e$ agrees well between XTOR-K and GTC simulations. However their saturation amplitude  $\delta T_e\sim 500-600$eV are still larger by about a factor three compared to the experimental level, $\delta T_e\sim200$eV. With zonal flows in M3D-C1 and GTC simulations, both codes recover a quantitative agreement with the experimental $\delta T_e$. Successful comparisons have also been obtained between the experimental neutron drop and the simulated one in GTC simulations. Without zonal flows, the simulated drop is $\delta\Gamma_N\geq6\%$, while with zonal flows $\delta\Gamma_N\sim 1.1\%$, which lies within experimental levels $\delta\Gamma_{N,exp}=0.9\%\pm0.3\%$. The zonal flows inclusion therefore strongly reduces the EP transport after saturation, $15\%$ of core EPs being redistributed outside of the $q_{min}$ surface without zonal flows, while only $3\%$ are redistributed with zonal flows.\\
The fishbone instability being very sensitive to the q profile, whose EFIT reconstruction has a certain degree of uncertainty, a nonlinear scan of the fishbone saturation amplitude has been conducted by varying the $q_{min}$ value, as displayed in Fig. \ref{saturation}(a). Such a scan is necessary to confirm that the fishbone-induced zonal flows dominate the saturation of the $n=1$ fishbone modes. In order to vary the q profile while computing self-consistently the related MHD equilibrium, the EFIT equilibrium has been reprocessed with the equilibrium solver code CHEASE \cite{Luetjens1996}. The CHEASE q profile have been fitted analytically on the EFIT one to have the same $q_{min}$ and $\rho_{q_{min}}$ values. Small variations can be observed between the two profiles, that mostly affect the magnetic shear profile. Based on the CHEASE reconstruction, two additional CHEASE equilibria have been computed, with q profiles that have been shifted as a whole by a constant factor $\pm0.05$ to study the impact of the $q_{min}$ value on the fishbone saturation. In all CHEASE reconstructions, the MHD pressure is kept constant to preserve the drive of the fishbone instability. GTC nonlinear simulations with and without zonal flows have been carried out with these new MHD equilibria. The related fishbone saturation amplitudes are displayed on Fig. \ref{saturation} (b). Overall, the inclusion of zonal flows in GTC simulations systematically reduces the fishbone amplitude by a factor 2-3, confirming their dominant role in the fishbone saturation. The fishbone amplitude also mostly decreases with increasing $q_{min}$ values, which is consistent with the stabilisation of fishbone modes with $q_{min}$ values far from unity in analytical theory. Moreover, the small differences in magnetic shear between the CHEASE and EFIT equilibria lead to a 50\% increase of the saturation amplitude with zonal flows using the CHEASE equilibrium, highlighting the strong sensitivity of the fishbone dynamics to the $q$ profile.
\begin{figure}[h!]
\begin{subfigure}{.5\textwidth} 
   \centering
      \includegraphics[scale=0.21]{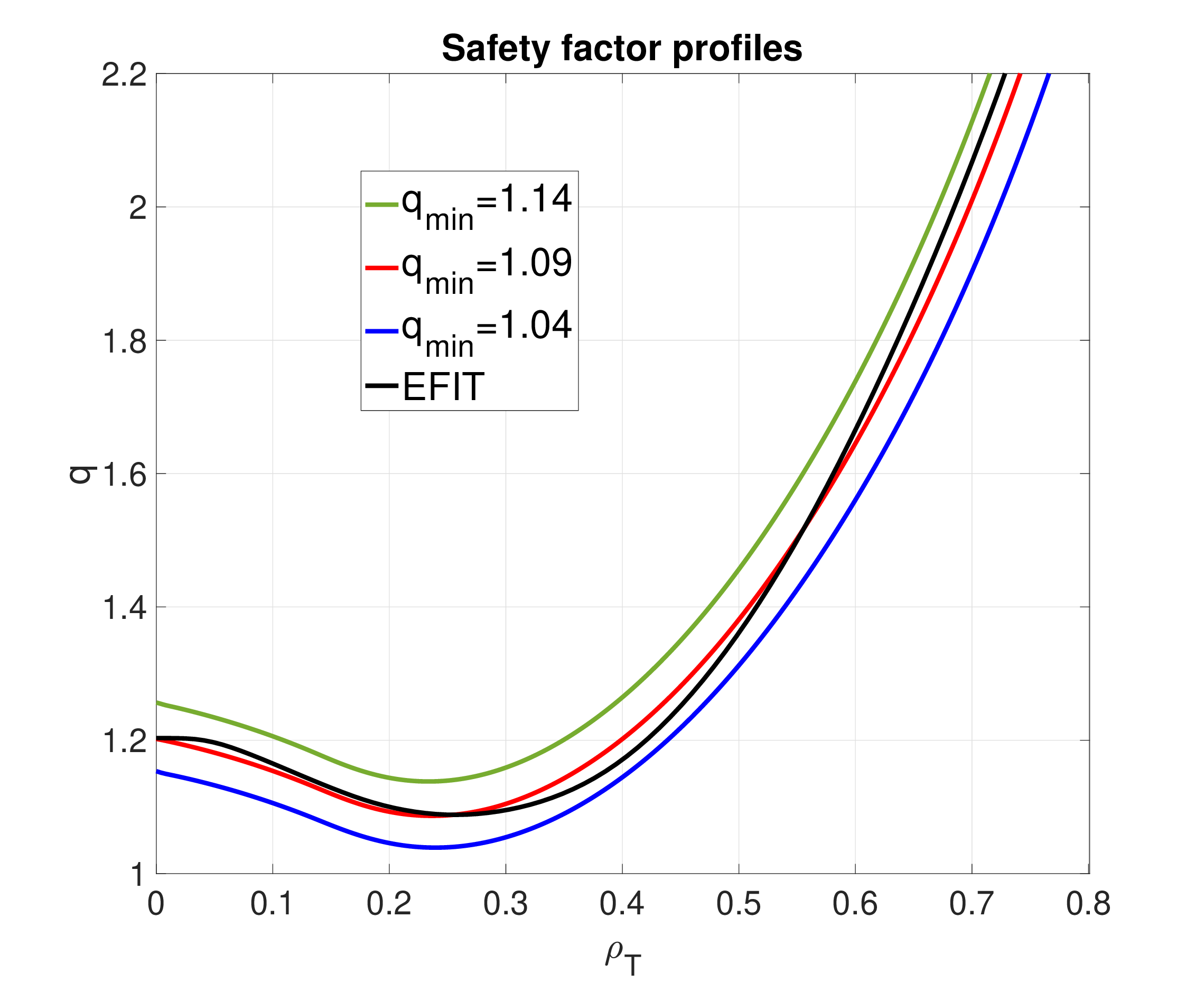}
   \caption{}
\end{subfigure} 
\begin{subfigure}{.5\textwidth} 
   \centering
      \includegraphics[scale=0.42]{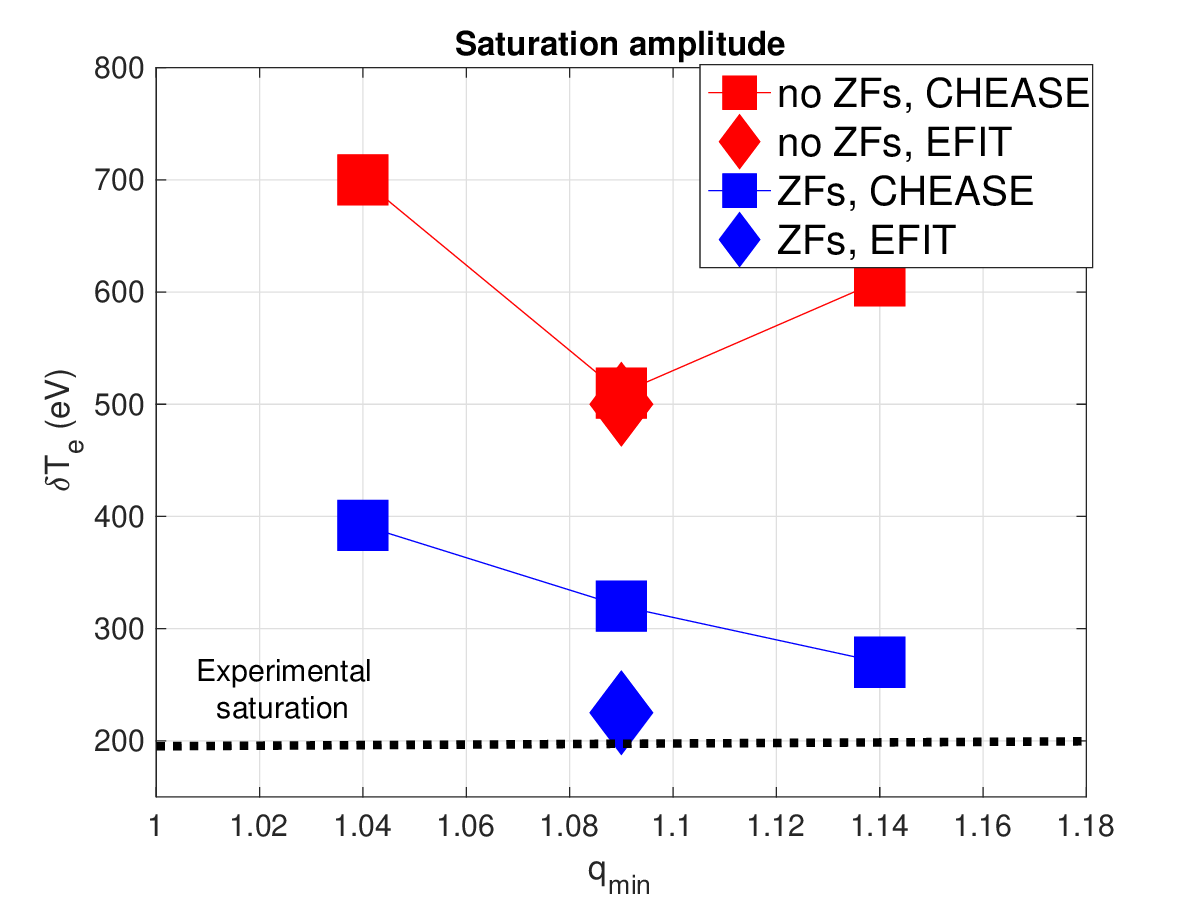}
   \caption{}
\end{subfigure}   
\caption{(a) Safety factor profiles used for sensitivity scan (b) Scan of saturation amplitude against $q_{min}$ in GTC simulations.}
\label{saturation}
\end{figure}
\section{Underlying mechanisms of the two-way fishbone-zonal flows interplay}
While we showed self-consistently in section 4 that fishbone modes can generate zonal flows that in turn dominate the fishbone nonlinear saturation, the underlying mechanisms leading to this two-way interplay between zonal flows and fishbone have however not been unveiled. In this section, both the zonal flows generation and fishbone saturation mechanisms are described in details.
\subsection{Generation of zonal flows by fishbone-induced EP transport}
Fishbone modes have been suspected for a long time \cite{Guenter2001}\cite{Pinches2001}\cite{Liu2023} to generate zonal flows through the resonant redistribution of EPs. From the MHD perspective, this redistribution leads to a radial current that can indeed drive a $\textbf{J}\times\textbf{B}$ torque, leading to the generation of poloidal flows. This can be equivalently formulated from the gyrokinetic perspective, by saying that the resonant EP transport is suspected to create a gyrocenter charge separation, creating a radial electric field that generates poloidal rotation through $E\times B$ flow. This assumption have been investigated with reduced models using an imposed fishbone mode evolution \cite{Pinches2001} or a predator-prey model \cite{Liu2023}, coupled with fluid equations to link the fishbone-induced radial current to the poloidal rotation \cite{Rosenbluth1996}\cite{Peeters1998}. To confirm that the EP resonant redistribution is indeed the underlying mechanism for the fishbone-induced generation of poloidal flows, a gyrokinetic formalism is needed to evolved self-consistently both the fishbone instability and the zonal electric field resulting from the Poisson equation. The GTC code is therefore well suited to confirm this mechanism.\\\\
Two important approximations in these GTC simulations first need to be reported. As mentioned in section 3, a fluid description is adopted for the electronic population, and the contribution of fluid electrons to zonal density is artificially removed. This first assumption is based on the adiabatic response of electrons, the electrons remaining at lowest order confined to their flux surface. The electron zonal density is removed to prevent the onset of numerical instabilities. These numerical issues are most likely due to the second assumption used in GTC simulations. In the code formulation used \cite{Xiao2015}, the zonal part of the Poisson equation is computed separately for accuracy purposes. When solving the zonal components of the Poisson equation, the zonal perturbed densities are computed with a flux-surface average considering the equilibrium flux surfaces. At the mode saturation however, the flux surfaces within the fishbone mode structure depart non-negligibly from the equilibrium ones, with $\delta\psi_{n=1,max}\sim0.3\psi_0$ at $q_{min}$. For this reason, short nonlinear simulations are conducted with GTC to limit the impact of this approximation. Longer simulations, necessary for cross-scale analysis, will require to use a different formulation that does not split the zonal from the non-zonal response, as employed in \cite{Fang2019} for the simulation of the cross-scale interaction between microturbulence and magnetic islands. The effects of the electron dynamics on the fishbone-induced zonal flows generation will also be investigated in a future study, using this formulation.\\\\
\begin{figure}[h!]
\begin{center}
\begin{subfigure}{\textwidth} 
   \centering
   \includegraphics[scale=0.25]{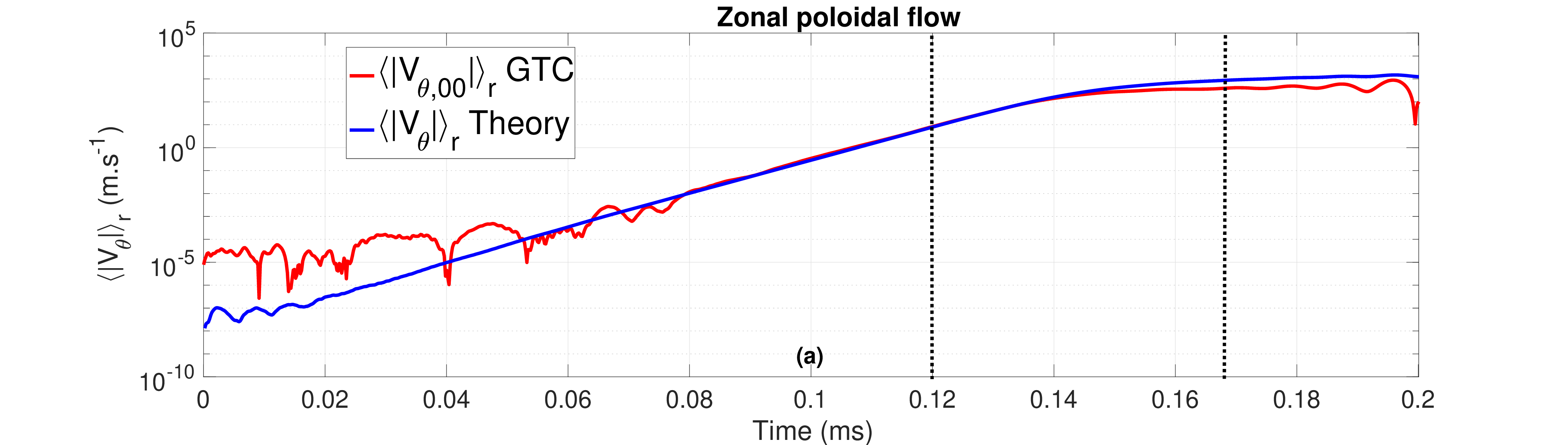}
\end{subfigure}
\begin{subfigure}{.32\textwidth} 
   \centering
   \includegraphics[scale=0.14]{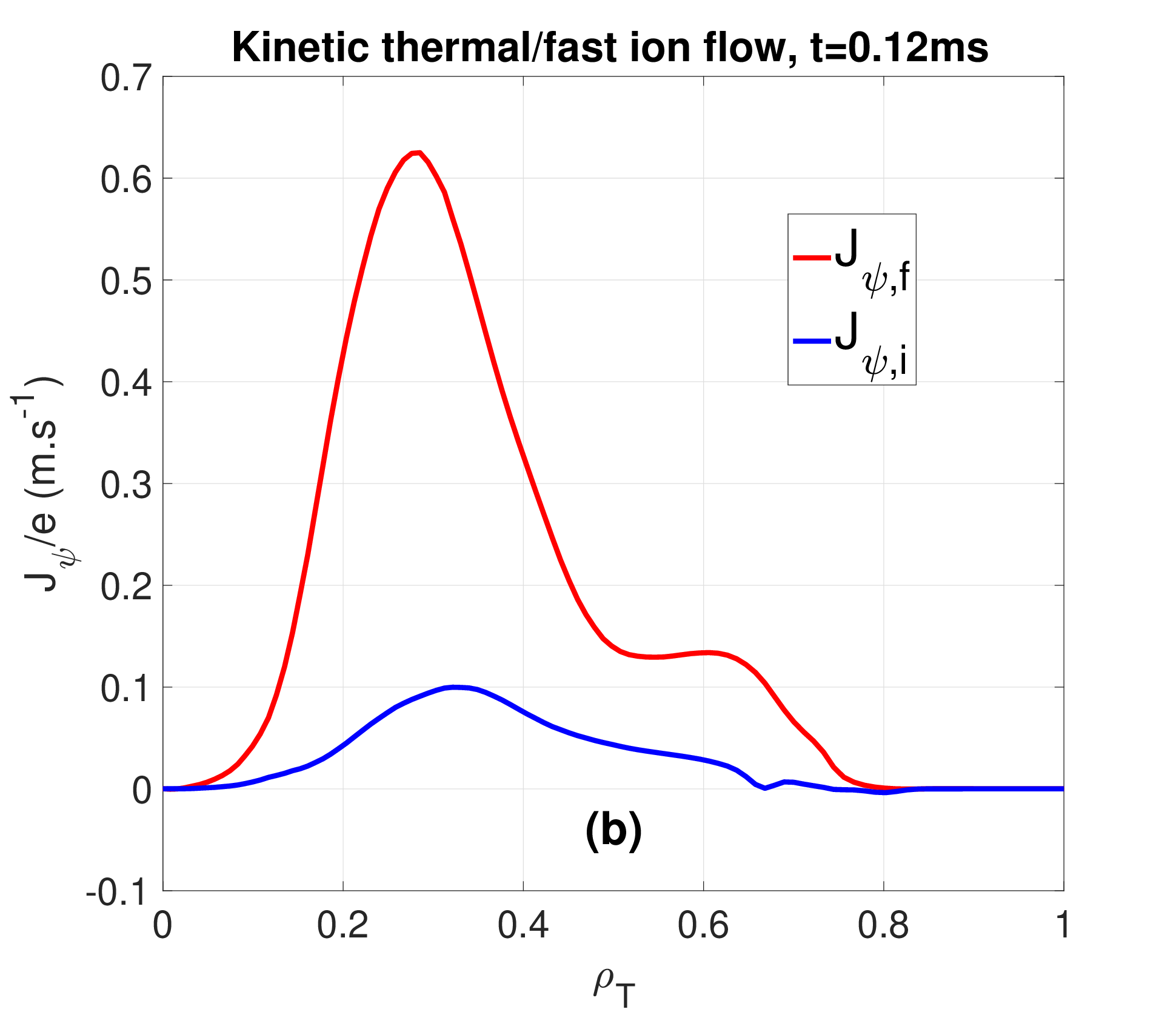}
\end{subfigure}     
\begin{subfigure}{.32\textwidth} 
   \centering
   \includegraphics[scale=0.14]{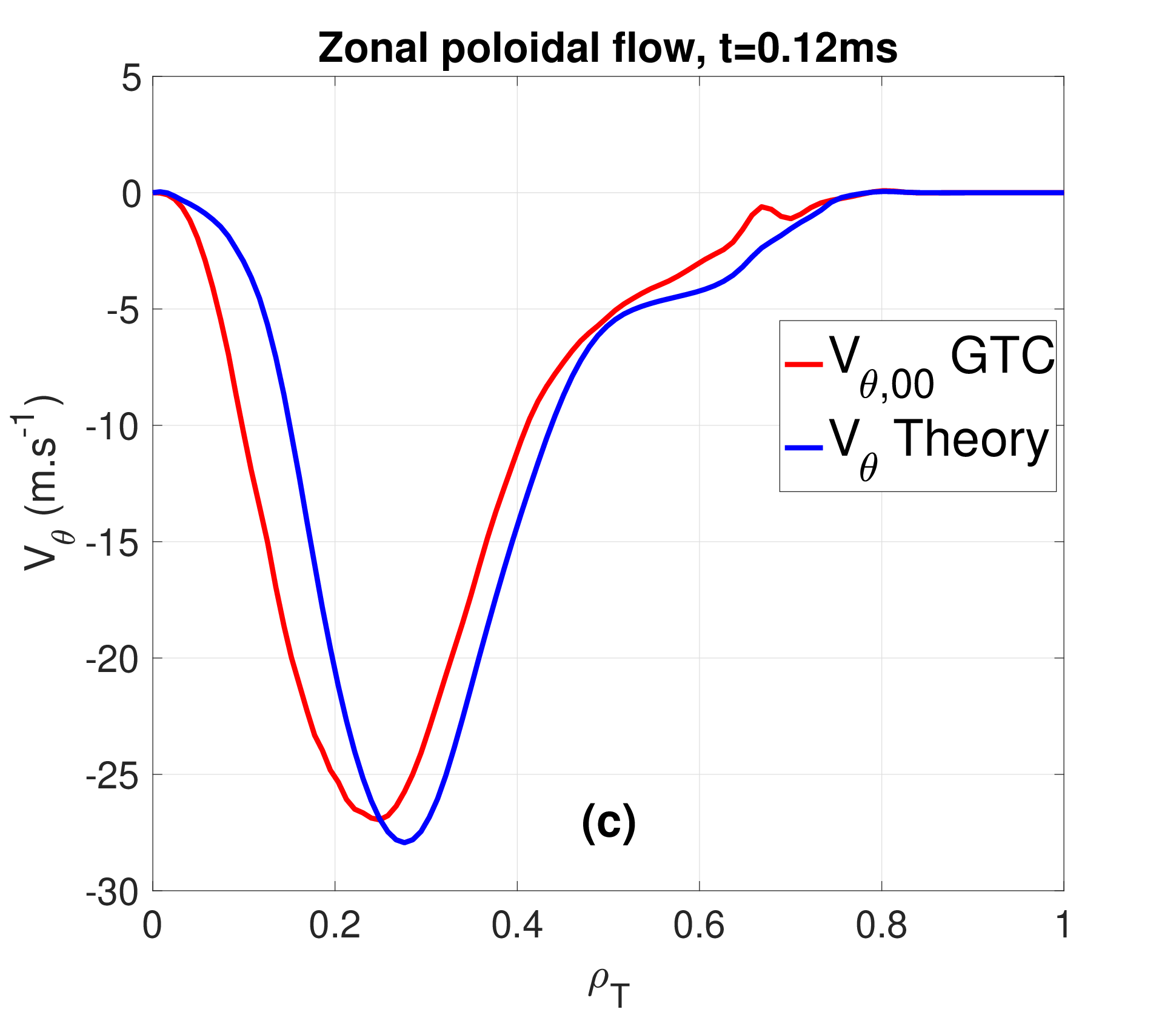}
\end{subfigure}
\begin{subfigure}{.32\textwidth} 
   \centering
      \includegraphics[scale=0.14]{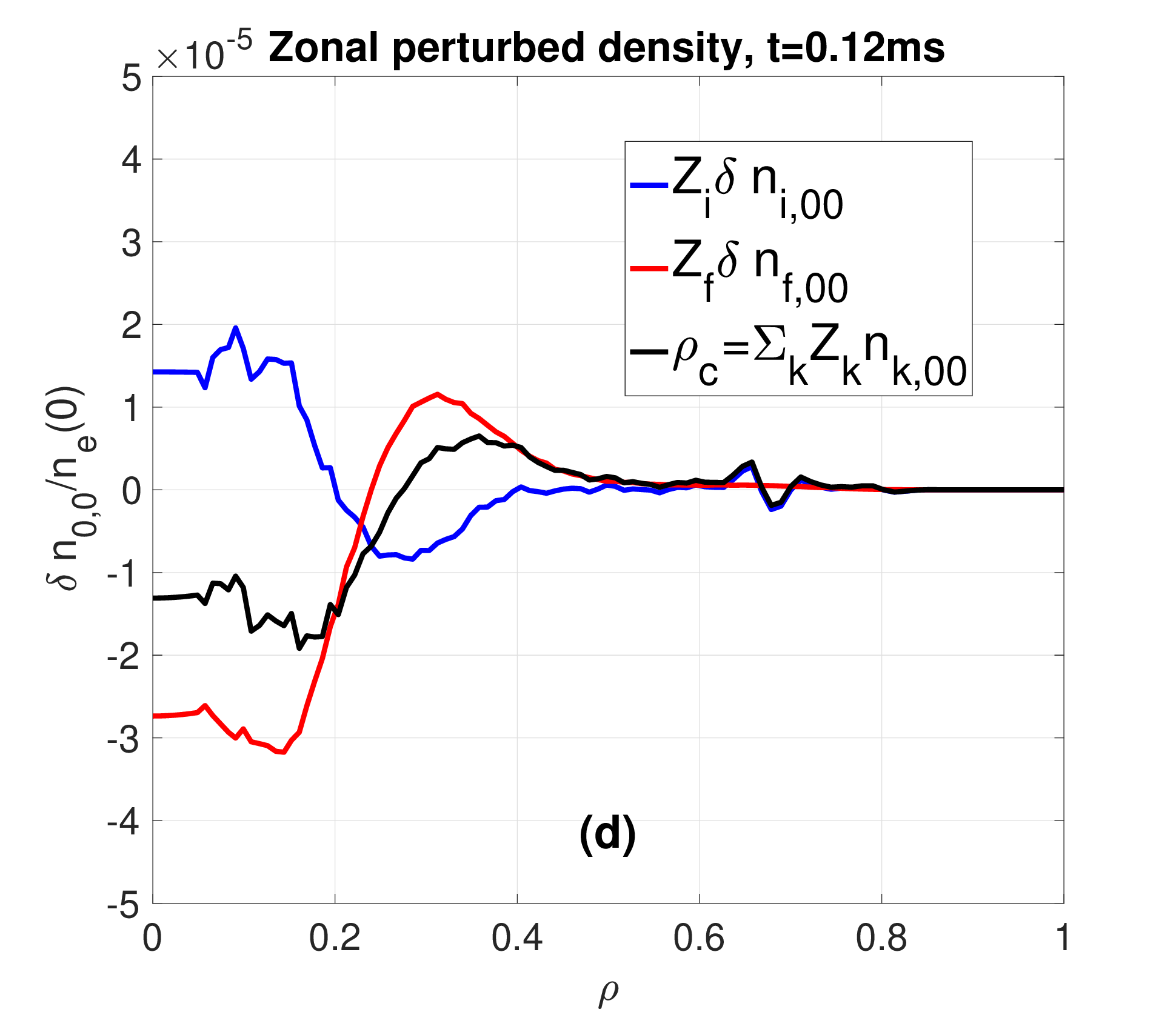}
\end{subfigure}       
\begin{subfigure}{.32\textwidth} 
   \centering
   \includegraphics[scale=0.14]{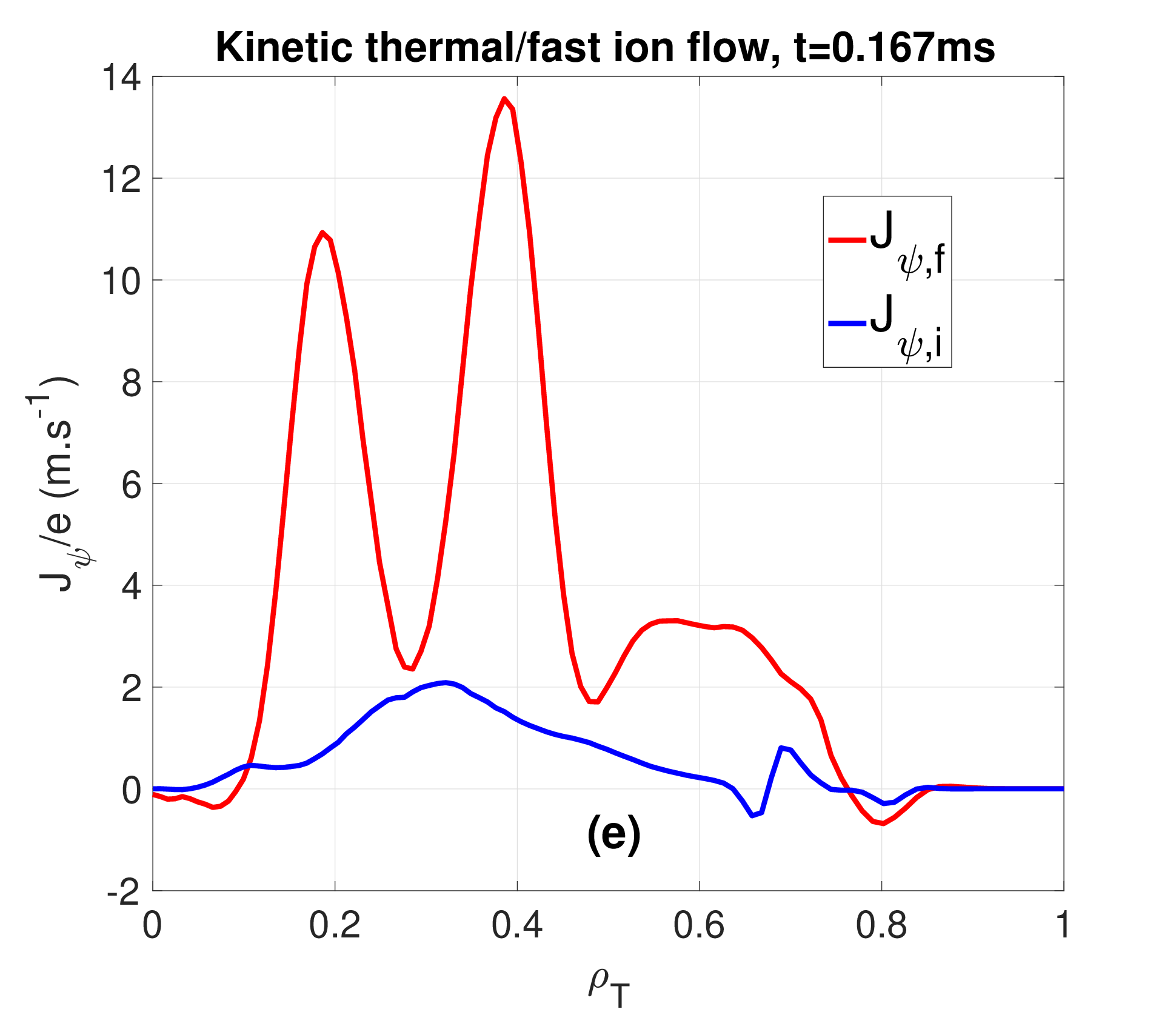}
\end{subfigure}     
\begin{subfigure}{.32\textwidth} 
   \centering
   \includegraphics[scale=0.14]{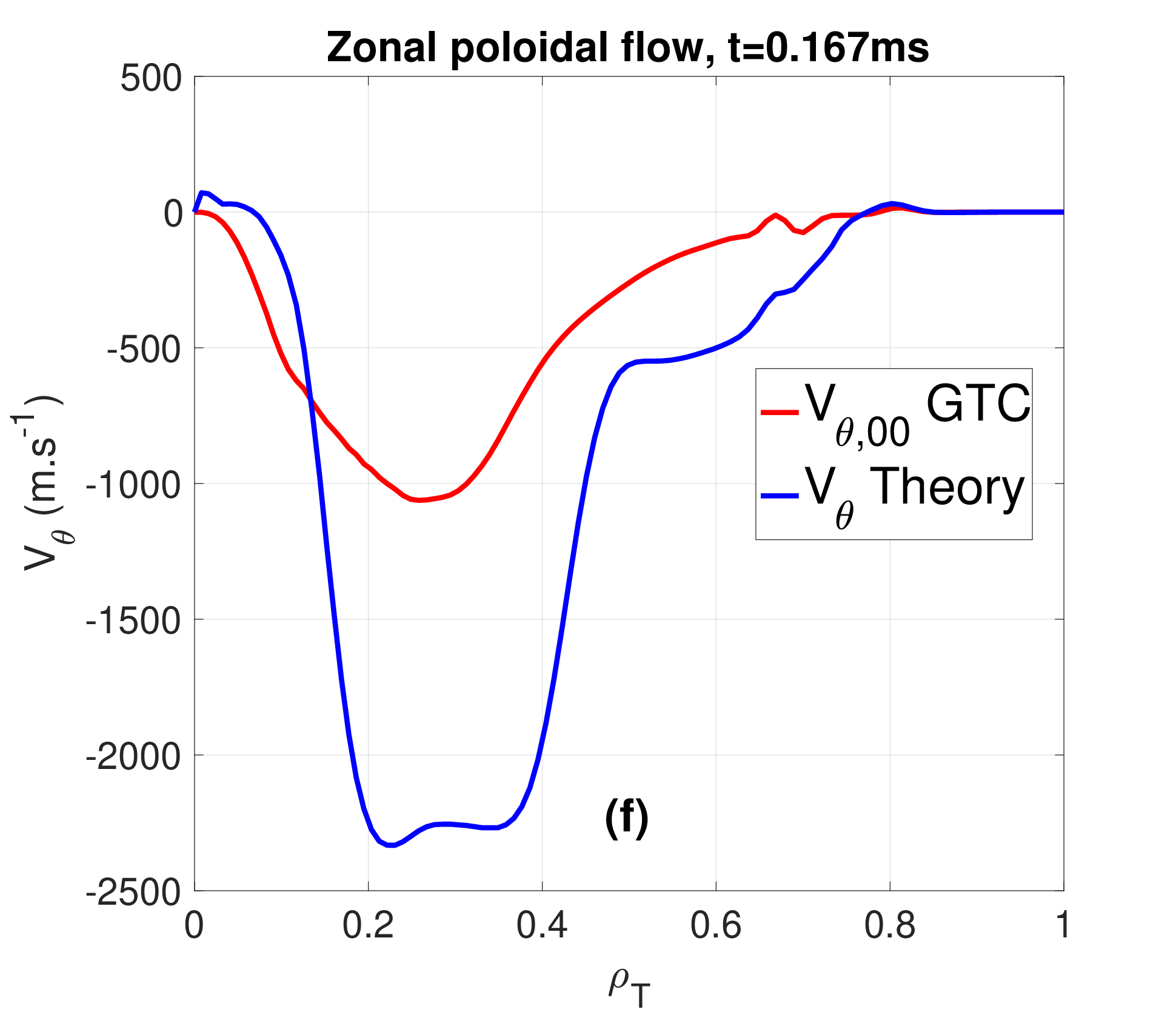}
\end{subfigure}
\begin{subfigure}{.32\textwidth} 
   \centering
      \includegraphics[scale=0.14]{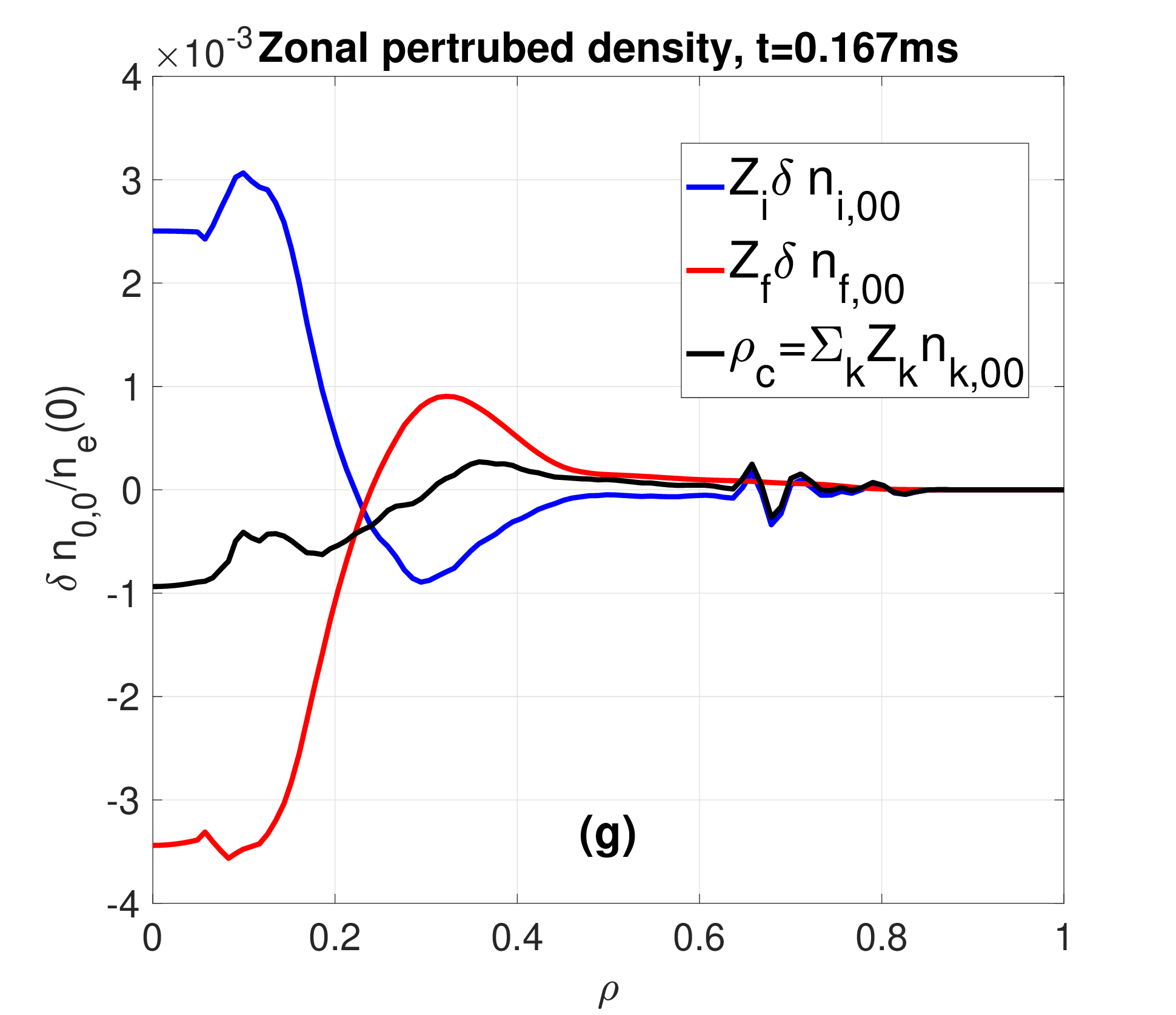}
\end{subfigure}     
\end{center}
\caption{(a) Time evolution of the volume-averaged poloidal flow in GTC simulations and analytical theory. (b,e) Thermal and fast ions current profiles in GTC simulations, respectively in the linear and nonlinear phase. (c,f) Poloidal flow profiles in GTC simulations and analytical theory, in linear and nonlinear phase (d,g) Zonal perturbed density profiles in GTC simulations, in linear and nonlinear phase.}
\label{ZFs_generation}
\end{figure}
To isolate the contribution of the EP redistribution to the zonal flows generation in GTC simulations, the perturbed flux-averaged radial currents of thermal and fast ions $\langle J_i^{\psi}\rangle$ and $\langle J_f^{\psi}\rangle$ are computed at each time step. The poloidal flow generation solely due to this radial ion flow can be explicitly computed using the electron and ion momentum equations, including the perturbed thermal and fast ion current as external currents \cite{Rosenbluth1996}\cite{Peeters1998}. Retaining the neoclassical contribution of $\langle \textbf{B}\cdot\nabla\cdot\Pi\rangle$, where $\Pi$ is the anisotropic pressure tensor, the poloidal flow equation can be cast in the following form, using a large aspect ratio circular approximation for the flux-averaged quantities of the MHD equilibrium quantities \cite{Liu2023}
\begin{equation}\label{flow_p}
\fl
\frac{\partial V_{\theta}}{\partial t} = \frac{1}{\epsilon^2q^{-2}(1+2q^2+1.63q^2\epsilon^{-1/2})}\bigg[-1.1\nu_{ii}\epsilon^{1/2}V_{\theta} - \frac{\epsilon}{n_im_iR_0q}\bigg(1-\frac{\epsilon^2}{q^2}[1+2q^2]\bigg)\langle J^{\psi}_{f,i}\rangle \bigg]
\end{equation}
where $\epsilon$ is the inverse aspect ratio, $\nu_{ii}\sim3.3\times 10^{1}$ s$^{-1}$ the ion-ion collision frequency and $J^{\psi}_{f,i}=J^{\psi}_f + J^{\psi}_i$ the contravariant radial current. It should be noted here that the inclusion of $\nu_{ii}$ has a negligible impact of the flow evolution over the considered GTC simulation time ($\sim0.2$ms). Given the marker gyrocenter velocity equation in GTC, its perturbed contravariant radial component reads
\begin{equation}
v^{\psi}=\dot{\textbf{R}}\cdot\nabla\psi= (\textbf{v}_E+\textbf{v}_{\delta A_{\parallel}} + \textbf{v}_{\delta B_{\parallel}} )\cdot\nabla\psi
\end{equation}
with $\textbf{v}_E$, $\textbf{v}_{\delta A_{\parallel}} $ and $\textbf{v}_{\delta B_{\parallel}}$ respectively the $E\times B$ and the magnetic flutter perpendicular and parallel velocities, each associated to the perturbed scalar potentials $\delta\phi$, $\delta A_{\parallel}$ and $\delta B_{\parallel}$. The $J^{\psi}_{f,i}$ profile is simply obtained by projecting radially the thermal and fast ion marker $v^{\psi}$ contributions. \\
Comparisons between the self-consistently evolved poloidal flow in GTC and the theoretical one taking only into account ion redistributions are displayed on Fig. \ref{ZFs_generation}. The time evolution of the volume-averaged poloidal flows is shown on Fig. \ref{ZFs_generation} (a). As discussed the theoretical evolution is obtained from the integration of Eq. (\ref{flow_p}), taking into account the ion radial current profiles displayed respectively in the fishbone linear and nonlinear phases on Fig. \ref{ZFs_generation} (b) and (e). The EP radial current dominates the total ion response, the resonant interaction of EPs with the fishbone mode being stronger due to their larger energy. Among the three different radial drifts, the EP radial current is almost entirely due to the $E\times B$ velocity, while the thermal ion radial current has both $E\times B$ and parallel magnetic flutter contributions, even though the $E\times B$ velocity also dominates the thermal ions response. Such an observation is consistent with the fact that zonal flows $\delta\phi_{00}$ are larger amplitude than the zonal fields $\delta A_{\parallel,00}, \delta B_{\parallel,00}$ in this DIII-D fishbone simulation.\\  As can be seen on Fig. \ref{ZFs_generation} (a), in the linear phase the theoretical time evolution matches quantitatively with the GTC one, confirming that the EP redistribution is the underlying mechanism for the fishbone-induced zonal flows generation. This is further shown in Fig. \ref{ZFs_generation} (c), displaying the poloidal flow profiles in the linear phase, where it can be observed that these profiles are also in quantitative agreement. Small radial differences subsist, which can be either attributed to the geometrical approximations used in deriving Eq. (\ref{flow_p}), or sub-dominant contributions from other physical mechanisms. The zonal ion density profiles in the linear phase are displayed on Fig. \ref{ZFs_generation} (d). Following \cite{Falessi2023}, the time evolution of the zonal density for a given plasma specie $s$ must satisfy the following continuity equation 
\begin{equation}
\frac{\partial\delta n_{s,00}}{\partial t} = \frac{\partial\delta n_{s,pol,00}}{\partial t} - \nabla\cdot\Gamma_{NL}
\end{equation}
where $\delta n_{s,pol,00}$ is the zonal polarisation density and $\Gamma_{NL}$ the nonlinear particle flux which corresponds to $J^{\psi}_s$. For EPs, the perturbed polarization density is negligible which leads to an outward EPs transport, as $\partial_{\psi}J^{\psi}>0$ inside the $q_{min}$ surface and $\partial_{\psi}J^{\psi}<0$ outside of it. The thermal ion polarization density is however significant and corresponds to the return current counter-balancing the resonant radial current $J^{\psi}_{f,i}$. As $J^{\psi}_{f,i}>J^{\psi}_i$, the thermal ions undergo an inward pinch that is opposite to the EPs dynamics. Overall, as the amplitude of the EP zonal density response is larger than the thermal ions response, it leads to a gyrocenter charge separation which is consistent with the zonal electric field in \cite{Brochard2024} Fig. 1d, or equivalently with the poloidal flow in Fig. \ref{DIIID_time} (b). In the nonlinear phase it can be noticed on Fig. \ref{ZFs_generation} (a) and (f) that Eq. (3) over-estimates the poloidal flow by about a factor of 2. This may be explained by the fact that the zonal flows computation tends to become inaccurate in GTC simulations during the nonlinear phase, due to the distorsion of the flux surfaces induced by the fishbone. On Fig. \ref{ZFs_generation} (a), the theoretical and numerical results start indeed to depart from one another when the fishbone mode reaches saturation at $t\sim0.14$ms, which could be suggestive of such an effect.\\\\
These results from self-consistent gyrokinetic simulations therefore confirm that the EP redistribution is the main mechanism for the destabilization of zonal flows during fishbone bursts. They indicate that wave-particle nonlinearity dominates the zonal flow generation in this DIII-D experiment with relatively weak fishbones. Wave-wave nonlinearity may become more important far away from marginality. For example, zonal flow generation is dominated by thermal plasma radial current driven by a strong RSAE in another DIII-D experiment \cite{Liu2022a}. Kinetic electrons effects and flux surface distortion \cite{Fang2019} will also need to be taken into account in a future work, to estimate quantitatively the saturated zonal flows levels in the fishbone nonlinear phase.
\subsection{Saturation of fishbone modes through zonal flows-induced Doppler shift} 
\begin{figure}[h!]
\begin{subfigure}{\textwidth}
   \centering
   \includegraphics[scale=0.34]{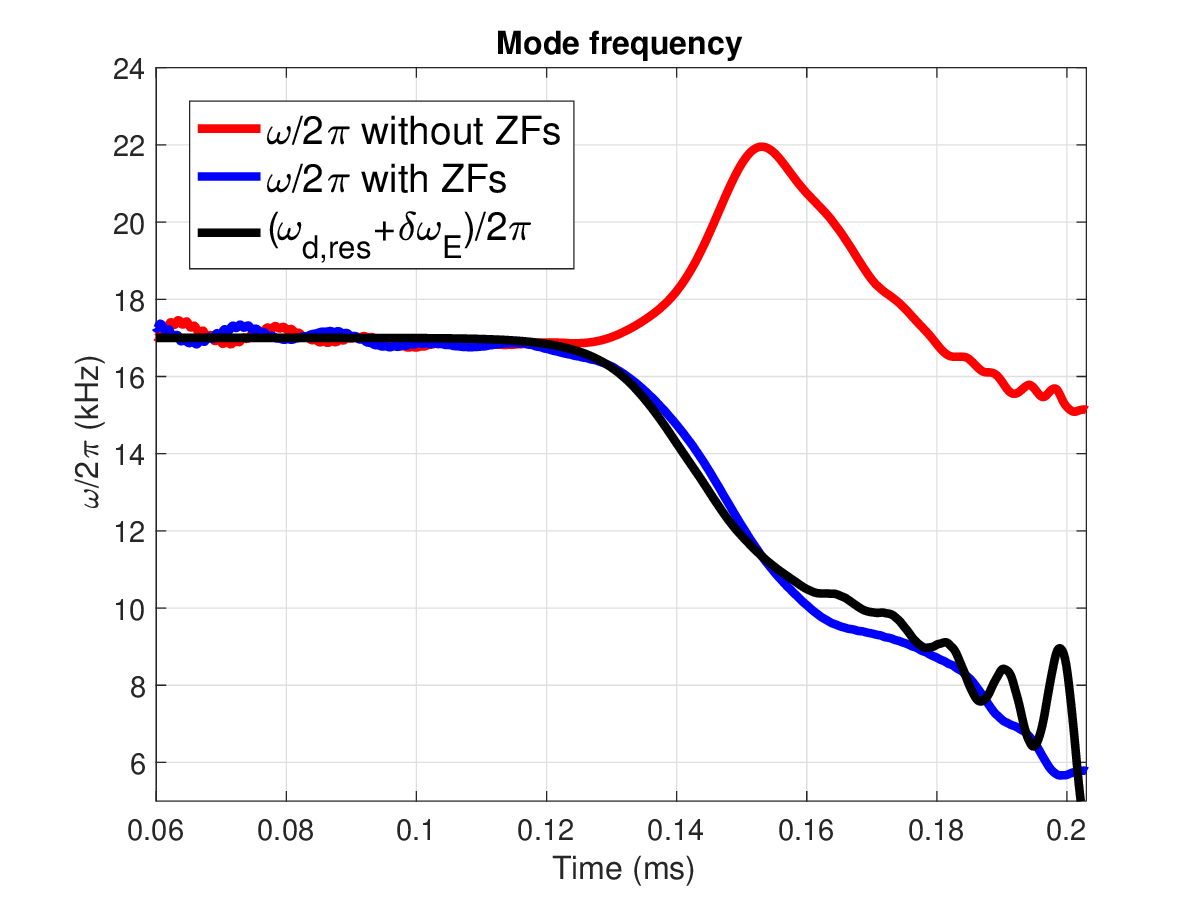}
   \caption{}
\end{subfigure}
\begin{subfigure}{.5\textwidth} 
   \centering
   \includegraphics[scale=0.38]{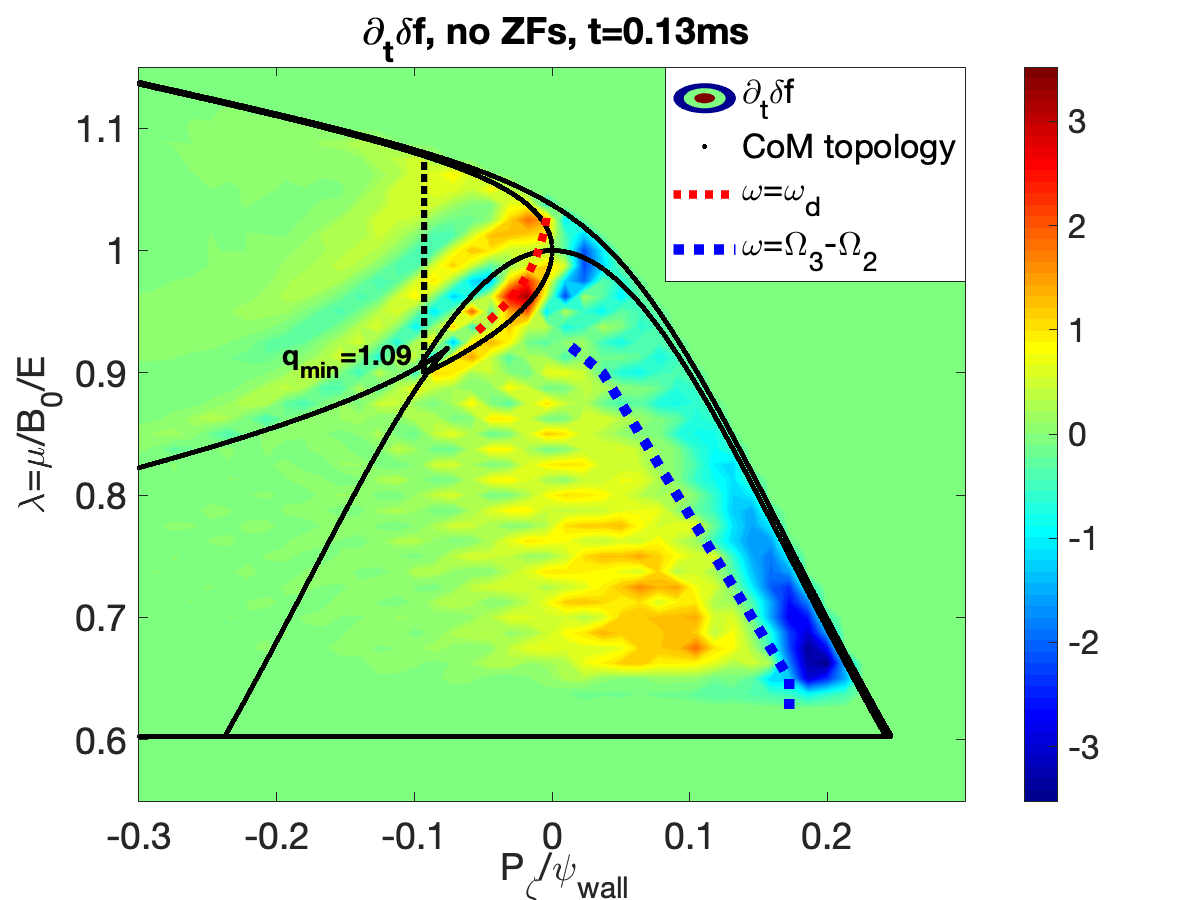}
   \caption{}
\end{subfigure}     
\begin{subfigure}{.5\textwidth} 
   \centering
   \includegraphics[scale=0.38]{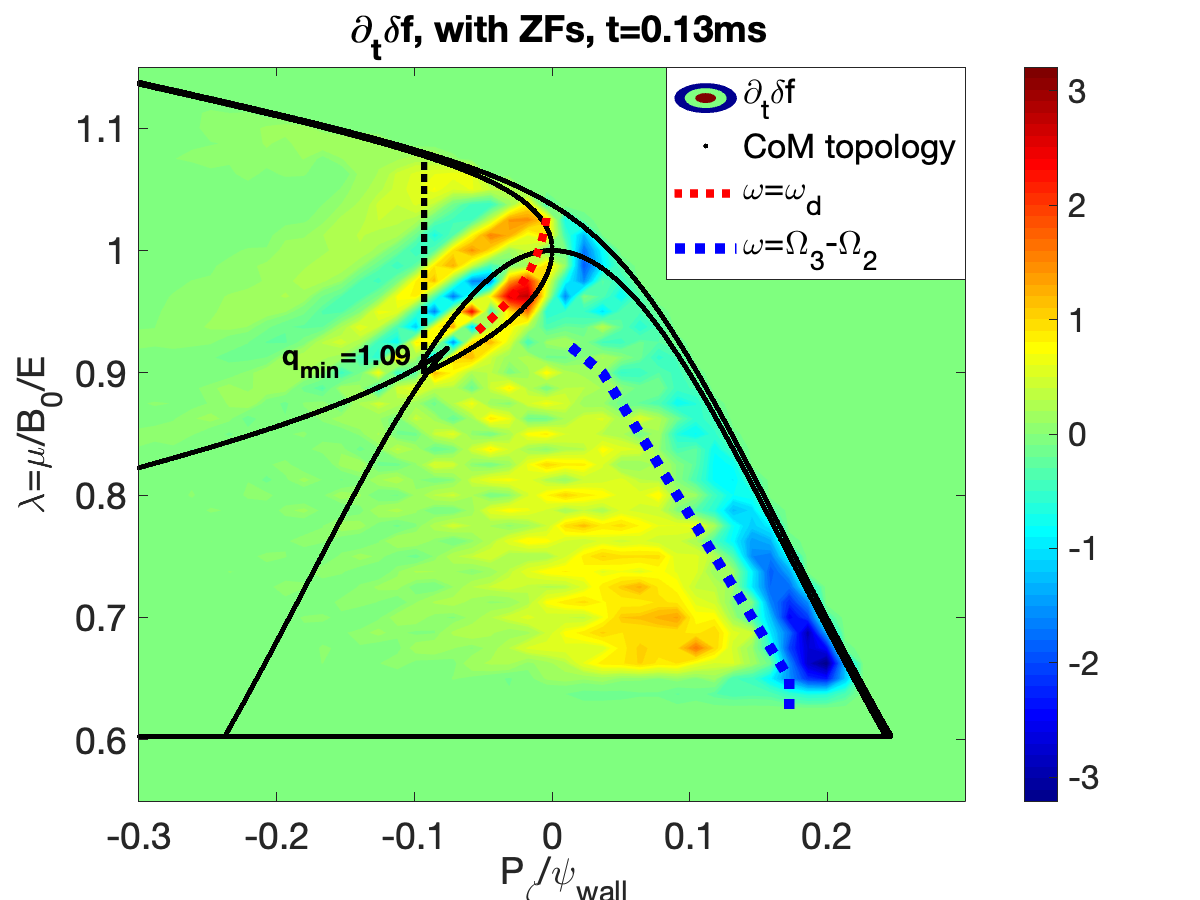}
   \caption{}
\end{subfigure}
\begin{subfigure}{.5\textwidth} 
   \centering
      \includegraphics[scale=0.38]{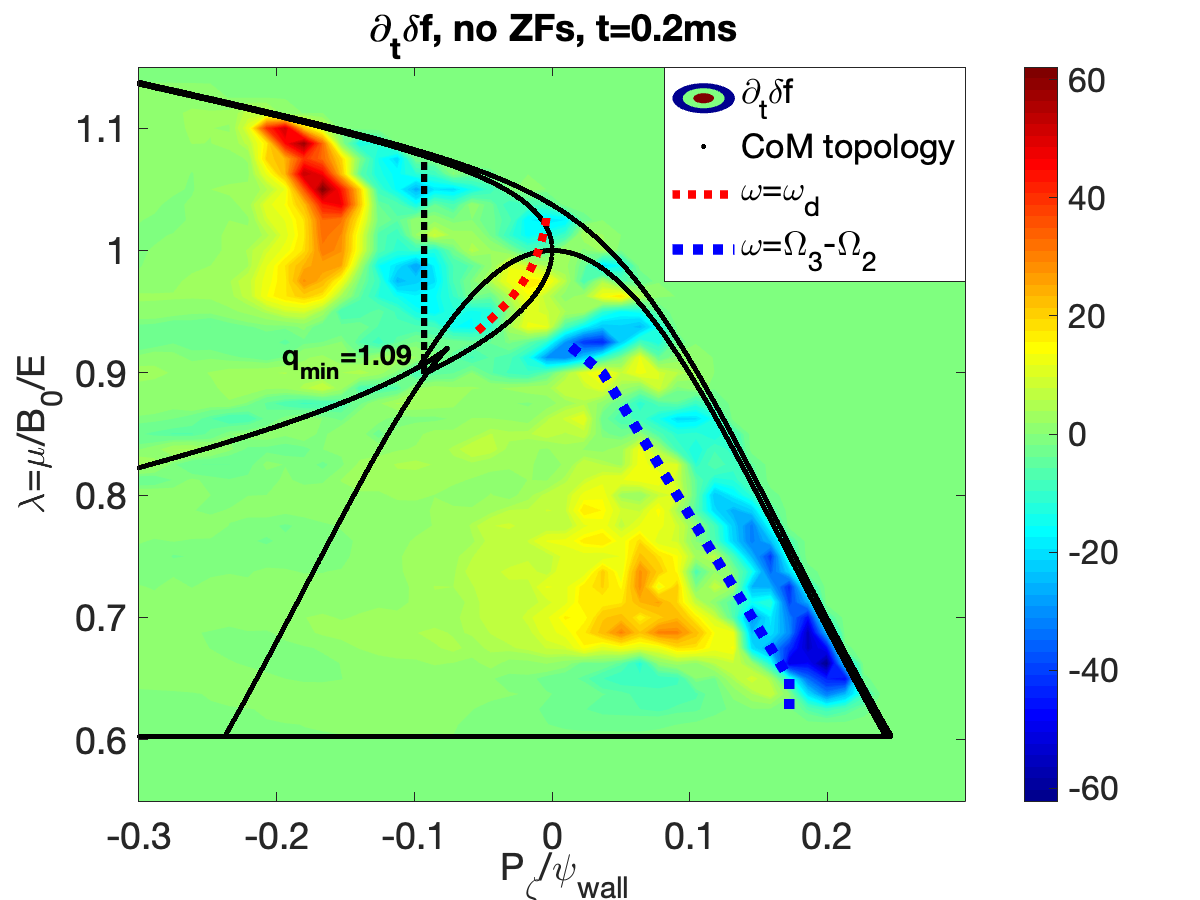}
   \caption{}
\end{subfigure}  
\begin{subfigure}{.5\textwidth} 
   \centering
      \includegraphics[scale=0.38]{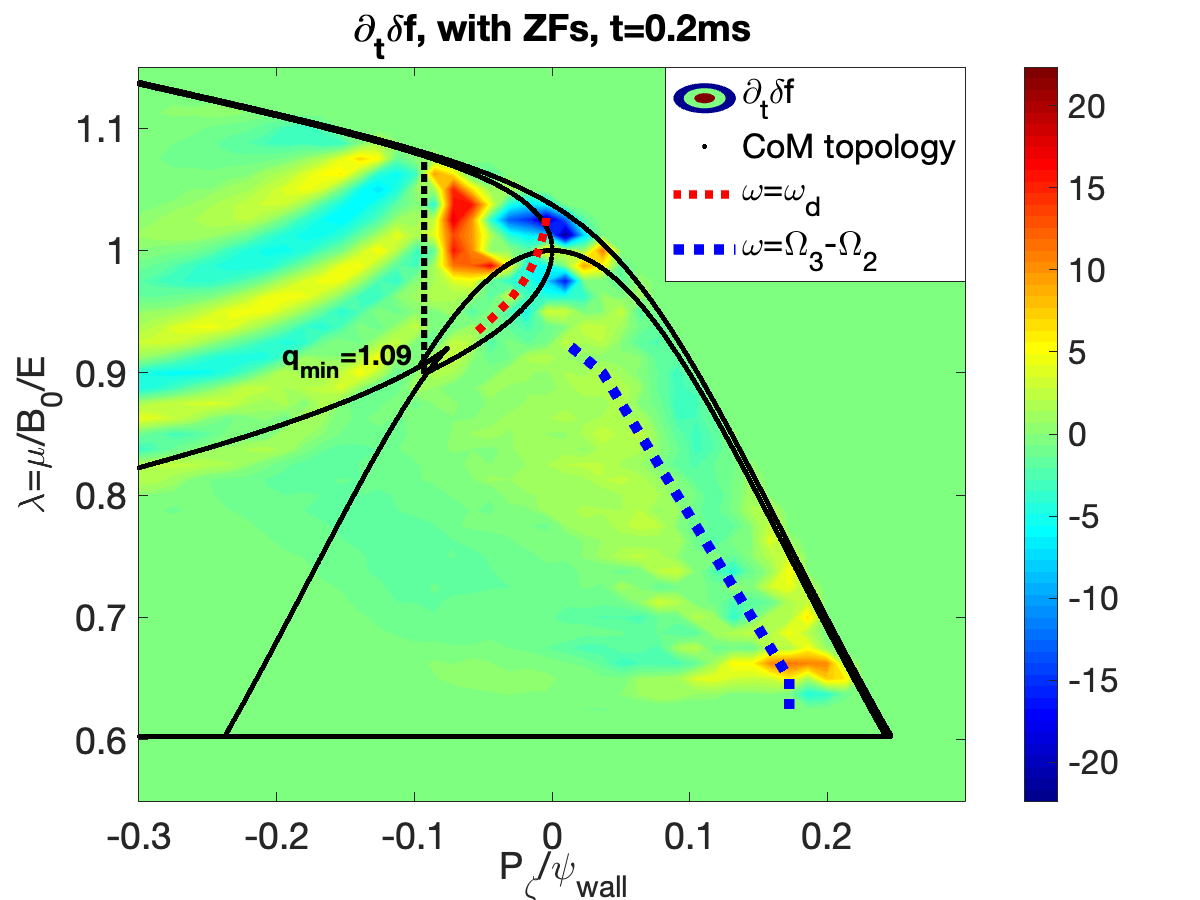}
   \caption{}
\end{subfigure}  
\caption{(a) Time evolution of n=1 mode frequency $\omega_{n=1}$ and linearly resonant precessional frequency $\omega_{d,res}$ plus zonal $E\times B$ frequency $\omega_E$ at $q_{min}$ in GTC simulations, reproduced from \cite{Brochard2024}. (b-e) Instantaneous EP distribution $\partial_t\delta f$ in linear (top) and nonlinear (bottom) phases, without (left) and with (right) zonal flows in the $(P_{\zeta},\lambda)$ CoM diagram at $\mu B_0 = 45$ in GTC simulations.}
\label{PSZS}
\end{figure}
The impact of zonal flows on the fishbone saturation can be characterized by looking at the time evolution of both the fishbone mode frequency and the phase space zonal structures (PSZS) \cite{Falessi2019}\cite{Falessi2023} in CoM space, displayed in Fig. \ref{PSZS}. As can be seen on Fig. \ref{PSZS} (a), at the nonlinear fishbone saturation near $t\sim0.15$ms, the mode frequency chirps down by about 10 kHz with and without zonal flows, which is typical of EP-driven instabilities in tokamak plasmas \cite{Chen2016}. Just before saturation without zonal flows, the mode frequency experiences a brief up-chirping phase that may be attributed to ideal MHD nonlinear effects \cite{Cowley1996}, related to the large mode amplitude near saturation. The dominant fishbone down-chirping has been theoretically predicted \cite{Zonca2015}\cite{Chen2016} and observed in kinetic-MHD simulations \cite{Wang2016}\cite{Vlad2016}\cite{Brochard2020b}\cite{Wang2023} to be related to a synchronisation between the fishbone mode frequency and the EP resonant frequencies. This synchronisation occurs to maximize the wave-particle power exchange by preserving the resonance conditions, which leads to a convective EP transport through a process referred to as an EPM (Energetic Particle Mode) avalanche. As a result, the resonance positions moves radially, generally outward in tokamak plasmas due to the negative equilibrium gradients, to include more EPs that were linearly unable to resonate with the mode. \\ In addition to the $n=1$ mode frequency down-chirping, a Doppler-shift induced by zonal flows, defined as $\omega_E=\textbf{V}_{00}\cdot(mq\nabla\theta-n\nabla\zeta)$ in GTC simulations, can be observed in Fig. \ref{PSZS} (a). The zonal Doppler shift leads to the modification of the resonance conditions as discussed in \cite{Chen2016} (Eq. 4.182) and \cite{Brochard2020b}, the precessional frequency yielding in particular $\omega=\omega_d+\omega_E$. The black line in Fig. \ref{PSZS} (a) corresponds to the time evolution of the zonal Doppler shift plus the linear resonant precessional frequency $\omega_{d,res}=17$kHz. Its time evolution is almost exactly in phase with that of the $n=1$ mode frequency $\omega$ in the simulations with zonal flows, which implies that the linear position in CoM space of the precessional resonance is almost preserved in the nonlinear phase, despite the mode-down chirping. PSZS linked to the precessional resonance should therefore remain static during the fishbone nonlinear saturation, instead of drifting in CoM space to reduce $\omega_d$ in order to preserve the mode resonance during down-chirping.\\\\
This result is confirmed by the time evolution of the instantaneous EP transport $\partial_t\delta f$ in the $(P_{\zeta},\lambda)$ diagram at $\mu B_0=45$keV, displayed on Fig. \ref{PSZS} (b-e). This quantity is used instead of the usual perturbed EP distribution $\delta f$ \cite{Brochard2020b}, in order to precisely capture the evolution of the resonance positions under mode chirping and zonal flows-induced Doppler shift. During the late linear phase at $t=0.13$ms, described  by both Fig. \ref{PSZS} (b) and (c) without and with zonal flows, a hole and clump structure \cite{Berk1999} develop around both resonance positions described in section 4.1. These phase space zonal structures are characteristic of a resonant outward EP redistribution, the holes being located at larger $P_{\zeta}$ values than the clumps, with $P_{\zeta}\propto -\psi$. \\During the nonlinear phase at $t=0.2$ms, the PSZS experience different dynamics with and without zonal flows. As predicted above for the precessional resonance, without zonal flows, the associated hole and clump moves to lower $P_{\zeta}$ to stay in resonance during the mode down-chirping, as $\omega_d\propto1\sqrt{\psi}$. With zonal flows, the hole and clump stays indeed locked-in around the linear resonance position. Zonal flows are therefore able to significantly reduce the EP resonant drive by preventing the precessional resonance from exploring parts of the distributions function that were linearly non-resonant, thus limiting the extent of the EPM avalanche. This reduction in resonant drive is illustrated by the weaker amplitude of the hole and clump structure with and without zonal flows. This trapping of PSZS structures by zonal flows is reminiscent of the trapping of turbulence eddies by zonal flows in microturbulence \cite{Diamond2005}\cite{Guo2009}.\\ Regarding the drift-transit resonance, the associated hole and clump structure persists around its linear resonance position without zonal flows, potentially because the reduction in mode frequency is much lower than both the poloidal and transit frequencies, and because the weak magnetic shear broadens the resonance width in $P_{\zeta}$. However with zonal flows, the structure vanishes, which is typical of a resonance detuning. Two mechanisms could account for this resonance detuning. The zonal Doppler shift being a function of $\psi$, the drift-transit resonance is locally affected in the $(P_{\zeta},\lambda)$ diagram, and $\omega=\Omega_3-\Omega_2+\omega_E$ may not have a solution due to the cancellation between $\Omega_3$ and $\Omega_2$ in the linear phase. The $E\times B$ flow shear could also affect the EP poloidal transit frequency $\Omega_2$, modifying as well the drift-transit resonance. \\
In conclusion, the zonal flows are able to strongly affect the dynamics of PSZS, by preventing them to either persist or drift in the CoM space, which reduces the fishbone resonant drive and dominates the fishbone saturation by limiting the EPM avalanche process. While the fishbone saturation mechanism remains the flattening of the EP distribution in the phase space resonance region, zonal flows can affect the locations in phase space where the wave-particle interactions are able to flatten the EP distribution.
\section{Chirping rates comparison during mode locking}
The non-adiabatic frequency chirping of waves in plasma physics is not limited to tokamak plasmas, but also extends to astrophysical plasmas \cite{Teng2023}. Chorus whistler waves in the Earth's magnetosphere are driven by electronic wave-particle interactions, that also couple electron transport to wave chirping \cite{Tao2021}\cite{Zonca2022}. The chirping rate of such waves is determined by the phase locking occurring between the electronic population and the waves, as demonstrated in \cite{Zonca2022} through quantitative agreements for the chirping rate between analytical estimations and nonlinear PIC simulations. The mode-locking being also suspected to be key in the chirping dynamics of EPMs (as noted in \cite{Chen1994}) and AEs in fusion plasmas \cite{Zonca2000}\cite{Zhang2012}\cite{Wang2016}\cite{Vlad2016}\cite{Brochard2020b}\cite{Yu2022}\cite{Wang2023}, a universal mechanism is possibly at play for the non-perturbative chirping of waves in plasmas physics \cite{Zonca2023}. The identification of such a mechanism is important in both astrophysical and fusion plasmas, as it can improve our understanding of magnetospheres on Earth and other planets, as well as help predicting the EPs transport in burning plasmas. \\
For these reasons, analytical comparisons based on mode-locking are here conducted for the fishbone chirping rate in GTC simulation, a mode-locking occurring in GTC simulations at $t\in[0.14, 0.16]$ms between the fishbone frequency and the precessional EP frequency under the influence of zonal flows. Following \cite{Zonca2015}\cite{Chen2016}, the mode-locking condition can be expressed as $\dot{\omega}-\dot{\omega}_{res}=0$, with $\omega_{res}=\omega_d+\omega_E$ the precessional resonance. The precessional frequency and the zonal Doppler-shift being 3D functions of the constant of motions $(E,P_{\zeta},\mu)$, with $\dot{\mu}=0$ as the magnetic moment is a nonlinear invariant of motion, the mode-locking condition can be cast as \cite{Zonca2023}
\begin{equation}\label{odot}
\dot{\omega} = \dot{P}_{\zeta}\frac{\partial}{\partial P_{\zeta}}(\omega_d+\omega_E)+\dot{E}\frac{\partial}{\partial E}(\omega_d+\omega_E)
\end{equation}
Note that this mode-locking condition is referred to a single particle, and therefore yields a different chirping rates at different locations in CoM space along the considered resonant structure. The physical chirping rate is obtained as proper average among all particles that participate in the wave-particle power exchange \cite{Zonca2015}\cite{Chen2016}. Neglecting the time evolution of the phase space island width associated with the precessional resonance \cite{Brochard2020b}, the time evolution of the particles toroidal canonical momentum and kinetic energy can be linked as $\dot{P}_{\varphi}=n\dot{E}/\omega$. Then using the concept of nonlinear equilibrium \cite{Falessi2019}\cite{Falessi2023}, the wave-particle power exchange reads 
\begin{equation}\label{Edot}
\dot{E}(E,\mu,P_{\zeta})=-\bigg\langle e\textbf{v}_d\cdot\nabla\delta\phi_n\bigg\rangle_{\alpha_2}
\end{equation}
where $\langle\cdot\cdot\cdot\rangle_{\alpha_2}$ is the time averaging operator over the particles' orbit, $\alpha_2$ being the second angle of the angle-action formalism in tokamaks \cite{Brochard2020b}, and $\textbf{v}_d$ the magnetic drift velocity that can be cast as \cite{Dong2017}
\begin{equation}
\textbf{v}_d=\frac{mv_{\parallel}^2}{ZB_0^*}\nabla\times\textbf{b}_0 + \frac{\mu}{ZB_0^*}\textbf{b}_0\times\nabla B_0
\end{equation}
Note that Eqs. (\ref{odot},\ref{Edot}) are consistent with the "mode-particle pumping" mechanism, originally conjectured to account for the EP ejection rate proportional to the fishbone amplitude \cite{Chen1984}\cite{White1983} and to explain the experimental evidence of EP losses \cite{McGuire1983}. Combining Eqs. (\ref{odot},\ref{Edot}), the chirping rate associated with the precessional resonance for $n=1$ fishbones due to mode-locking reads
\begin{equation}\label{chirp_eq}
\dot{\omega}=-\frac{1}{2}\bigg|\frac{e\textbf{v}_d\cdot\nabla\delta\phi_{n=1}}{\omega}\bigg|_{\alpha_2}\bigg(\frac{\partial\omega_d}{\partial P_{\zeta}} + \frac{\partial\omega_E}{\partial P_{\zeta}} + \omega\bigg[\frac{\partial\omega_d}{\partial E}+\frac{\partial\omega_E}{\partial E}\bigg]\bigg)
\end{equation}
Where $|\cdot\cdot\cdot|_{\alpha_2}$ refers to the maximum of $\left\langle\cdot\cdot\cdot\right\rangle_{\alpha_2}$; i.e. the peak value that is independent of the wave-particle phase. This condition is what maximises wave-particle power exchange \cite{Zonca2022}\cite{Falessi2023}. In GTC, the average over a marker orbit, orbit that is fully determined by a given triplet of invariants $(E_i,\mu_j,P_{\zeta,k})$, is performed by summing over the contribution of each markers belonging to the same $(i,j,k)$ CoM volume (through tri-linear interpolation), divided by the number of contributions. This approach is exactly equivalent to a time average over one bounce/transit time, as markers in the same CoM volume are non-uniformly distributed along the corresponding orbit, with a weight characterised by the Hamiltonian equation of motions (\cite{White2006} Eqs 3.28-3.31). The wave-particle power exchange $W_{n=1}=-e\textbf{v}_d\cdot\nabla\delta\phi_{n=1}$ is implemented in Boozer coordinates, in which $W$ can be explicitly computed as
\begin{eqnarray}
\fl
W_{n=1} = \frac{2E}{Z(gq+I)}\bigg[\bigg(1-\frac{\lambda}{H}\bigg)\bigg(\frac{\partial I}{\partial\psi}\frac{\partial\delta\phi_{n=1}}{\partial\zeta}-\frac{\partial g}{\partial\psi}\frac{\partial\delta\phi_{n=1}}{\partial\theta}\bigg) + \frac{1}{B_0}\bigg(1-\frac{\lambda}{2H}\bigg)\times \nonumber\\\bigg(g\bigg[\frac{\partial B_0}{\partial\psi}\frac{\partial\delta\phi_{n=1}}{\partial\theta}-\frac{\partial B_0}{\partial\theta}\frac{\partial\delta\phi_{n=1}}{\partial\psi}\bigg] + I \bigg[\frac{\partial B_0}{\partial\zeta}\frac{\partial\delta\phi_{n=1}}{\partial\psi}-\frac{\partial B_0}{\partial\psi}\frac{\partial\delta\phi_{n=1}}{\partial\zeta}\bigg]\bigg)\bigg]
\end{eqnarray}
with $\textbf{B}_0=g\nabla\zeta+I\nabla\theta$ the magnetic field, and $H=B_{axis}/B_0$.
\begin{figure}[h!]
\begin{center}
\begin{subfigure}{.49\textwidth} 
   \centering
   \includegraphics[scale=0.2]{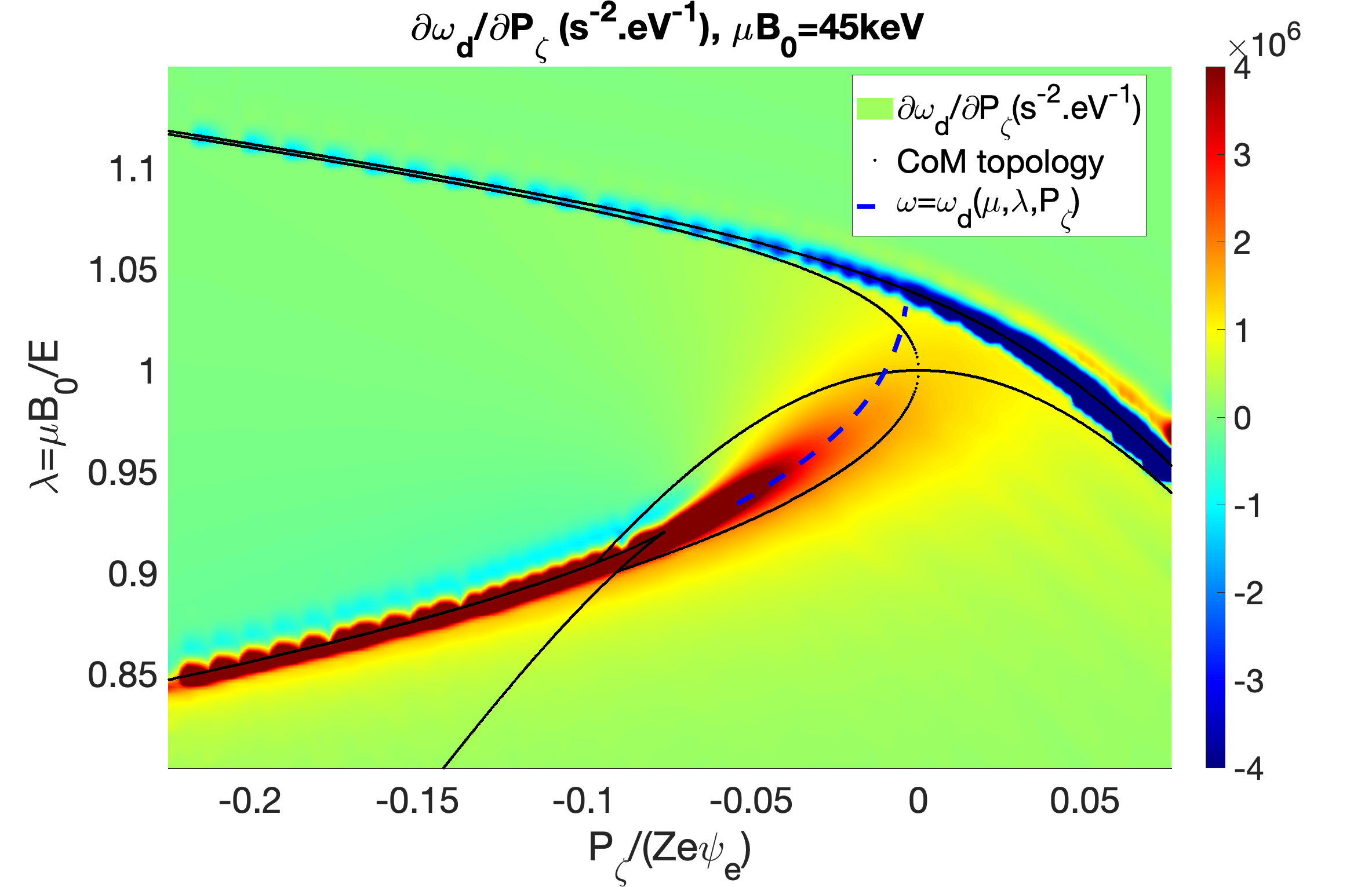}
   \caption{}
\end{subfigure}     
\begin{subfigure}{.49\textwidth} 
   \centering
   \includegraphics[scale=0.2]{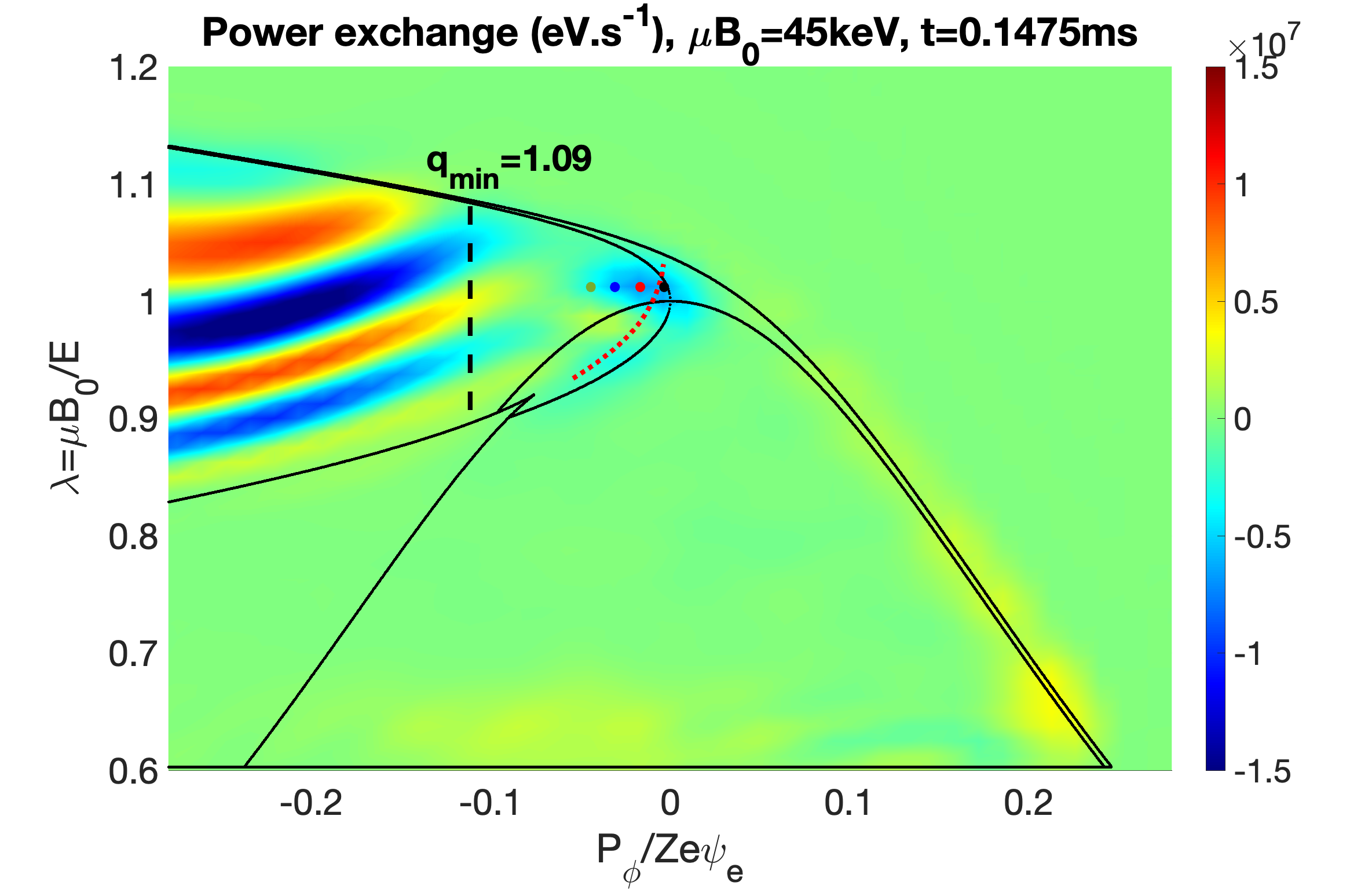}
   \caption{}
\end{subfigure}
\begin{subfigure}{.49\textwidth} 
   \centering
      \includegraphics[scale=0.2]{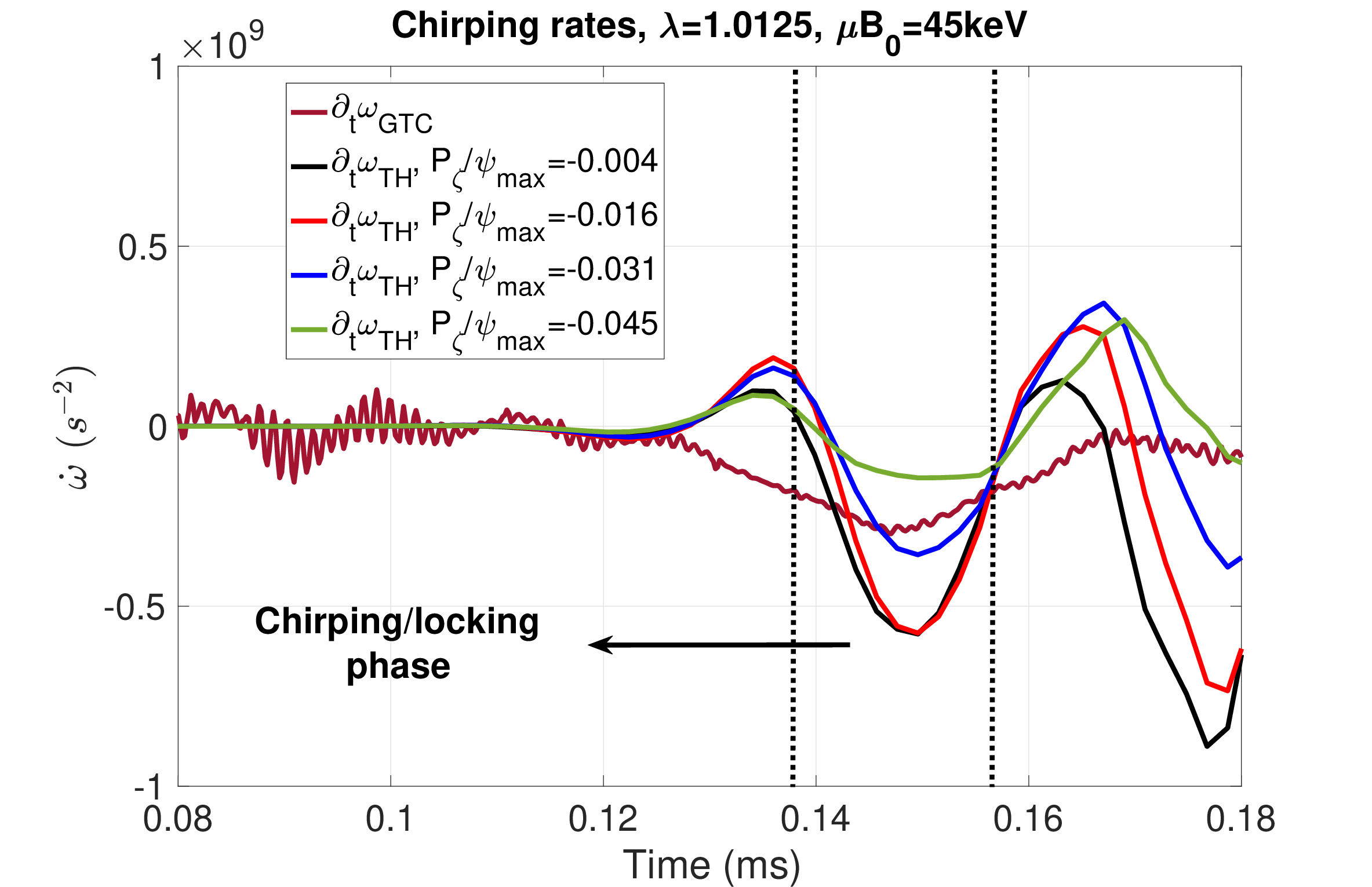}
   \caption{}
\end{subfigure}  
\begin{subfigure}{.49\textwidth} 
   \centering
      \includegraphics[scale=0.2]{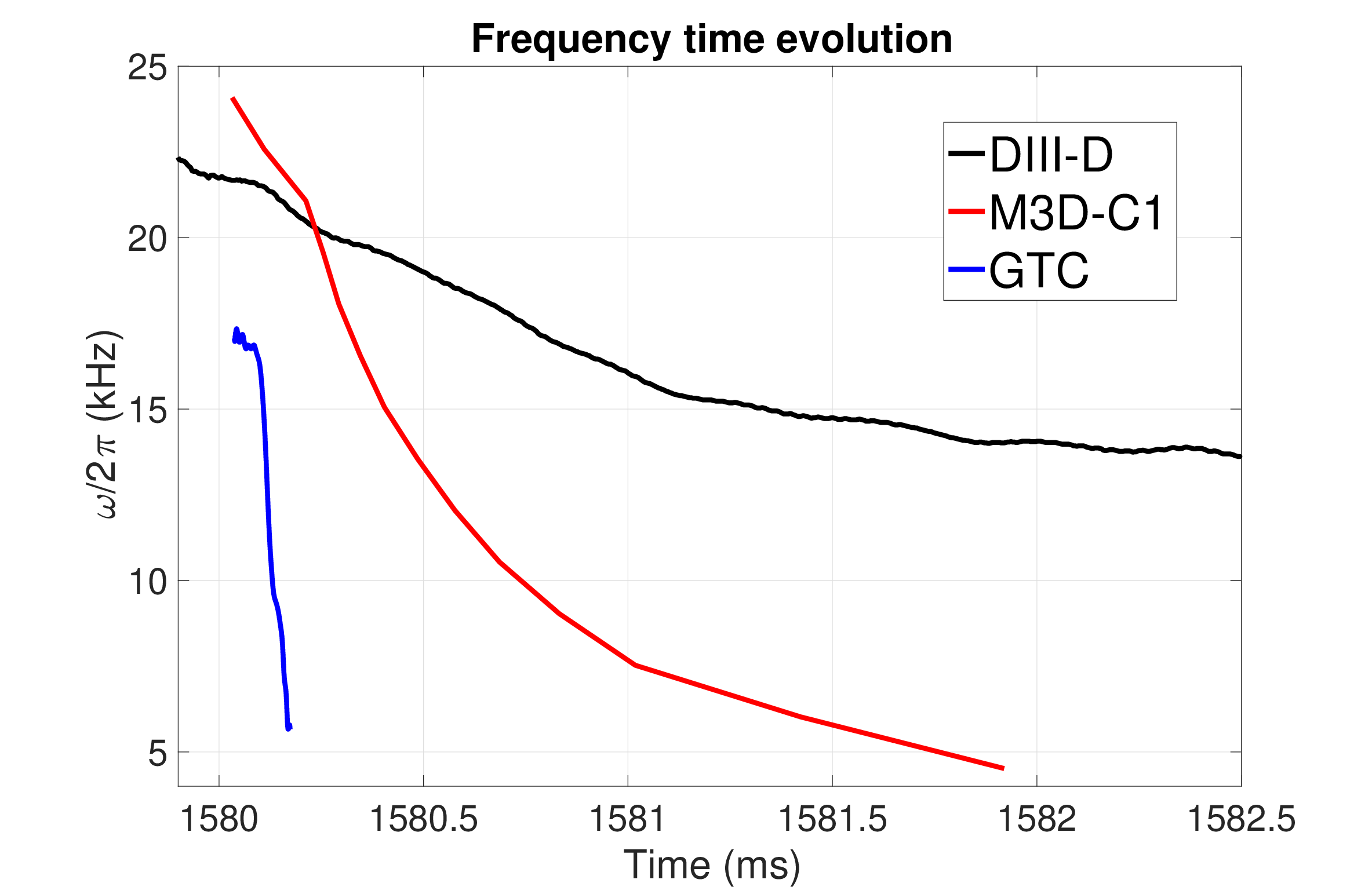}
   \caption{}
\end{subfigure}  
\end{center}
\caption{(a) $\partial\omega_d/\partial P_{\zeta}$ derivative and (b)  the perpendicular energy exchange $W_{n=1}/2$ in CoM phase space. (c) Comparison between the measured chirping rate in GTC simulation and the theoretical predictions over a $P_{\zeta}$ range. (d) Comparison of the frequency time evolution between experimental measurements, M3D-C1 and GTC}
 \label{chirping}
\end{figure}
\\\\The comparison between the analytical, numerical and experimental chirping rates are reported on Fig. \ref{chirping}. The two terms associated respectively with the first order CoM derivatives of $\omega_{res}$ and the wave-particle power exchange in Eq. \ref{chirp_eq} are displayed on Fig. \ref{chirping} (a)(b). The $\partial \omega_d/\partial P_{\zeta}$ derivative is the dominant one in Eq. \ref{chirp_eq} at the CoM space postion of interest, where the precessional PSZS is located on Fig. \ref{PSZS}(d) $(P_{\zeta}/e\psi_{max}\sim-0.03, \lambda\sim1.01)$. In this zone indeed, $\partial \omega_d/\partial P_{\zeta}\sim1.5\times10^6$s$^{-2}$.eV$^{-1}$, while $\partial \omega_E/\partial P_{\zeta}\sim5\times10^4$s$^{-2}$.eV$^{-1}$, $\omega\partial\omega_d/\partial E\sim4\times10^3$s$^{-2}$.eV$^{-1}$, $\omega\partial\omega_E/\partial E\sim1\times10^3$s$^{-2}$.eV$^{-1}$. At this CoM position, a negative structure can be observed for $W_{n=1}$ during the chirping phase at $t=0.1475$ms with $W_{n=1}\sim-6\times10^6$ eV.s$^{-1}$, which corresponds to resonant EPs giving out energy to the $n=1$ fishbone mode, therefore experiencing outward radial transport following $\dot{P}_{\zeta}=n\dot{E}/\omega$. Large amplitude structures can also be observed on Fig. \ref{chirping} (b) at lower $P_{\zeta}$ values in the trapped domain outside of the $q_{min}$ volume, i.e. mostly outside the fishbone mode structure. They correspond to a phase mixing process characteristic of EP Landau damping, as they oscillate in CoM space as a function of $v_{\parallel}$, $\mu$ being fixed in Fig. \ref{chirping} (b). A comparison between analytical and numerical $\dot{\omega}$ is displayed on Fig. \ref{chirping} (c), where the time evolution of the GTC chirping rate and the analytical chirping rates at different locations in CoM space over the precessional PSZS are plotted. The positions in CoM where the different analytical chirping rates are computed are represented by color dots in Fig. \ref{chirping} (b), using the same color code as in Fig. \ref{chirping} (c). During the chirping phase, i.e. when Eq. (\ref{chirp_eq}) is valid, a quite good comparison is recovered between the numerical and analytical approaches at $t=0.1475$ms, with $\dot{\omega}_{GTC}\sim-3\times10^8$ s$^{-2}$, and $\dot{\omega}_{TH}\in[-5.7,-1.4]\times10^8$ s$^{-2}$. A quantitative agreement is obtained with the weighted average around the $P_{\zeta}/e\psi_e=-0.031$ location with $\dot{\omega}_{TH}=-3,4\times10^8$ s$^{-2}$. These results confirm that mode-locking is the underlying mechanism leading to fishbone down-chirping in this GTC simulation, which implies it is indeed possibly an universal mechanism for the non-adiabatic chirping of waves in plasmas physics. Similar comparisons are currently being conducted with other EPMs \cite{Wang2023} and EP-driven geodesic acoustic modes (EGAMs) \cite{Biancalani2018} in tokamaks, to investigate the universal aspect of this mechanism.\\\\
To conclude the chirping rates comparison in this DIII-D plasma, results from GTC and M3D-C1 simulations are compared with the time evolution of the experimental mode frequency in Fig. \ref{chirping} (d). It can observed that both M3D-C1 and the experimental chirping rates are much lower, with respectively $\dot{\omega}_{M3D-C1}\sim 1.8\times10^7$ s$^{-2}$ and $\dot{\omega}_{exp}\sim 7.5\times10^6$ s$^{-2}$. The qualitative difference between simulated and experimental chirping rates may be explained by the lower dissipation existing in these $n=0,1$ simulations, a cross-scale analysis being required to incorporate contribution from a wide spectrum of toroidal modes. The absence of particle source and collisions may also impact the dynamics, by competing with the readjustment of the resonance to maximize the wave-particle power exchange, thus affecting the chirping rate. Additionally, the differences between GTC and M3D-C1 could be attributed to the absence of resistivity in GTC simulations, which introduces a larger dissipation.
\section{Ion-ITB formation during fishbone bursts in DIII-D discharge}
\begin{figure}[h!]
   \centering
      \includegraphics[scale=0.25]{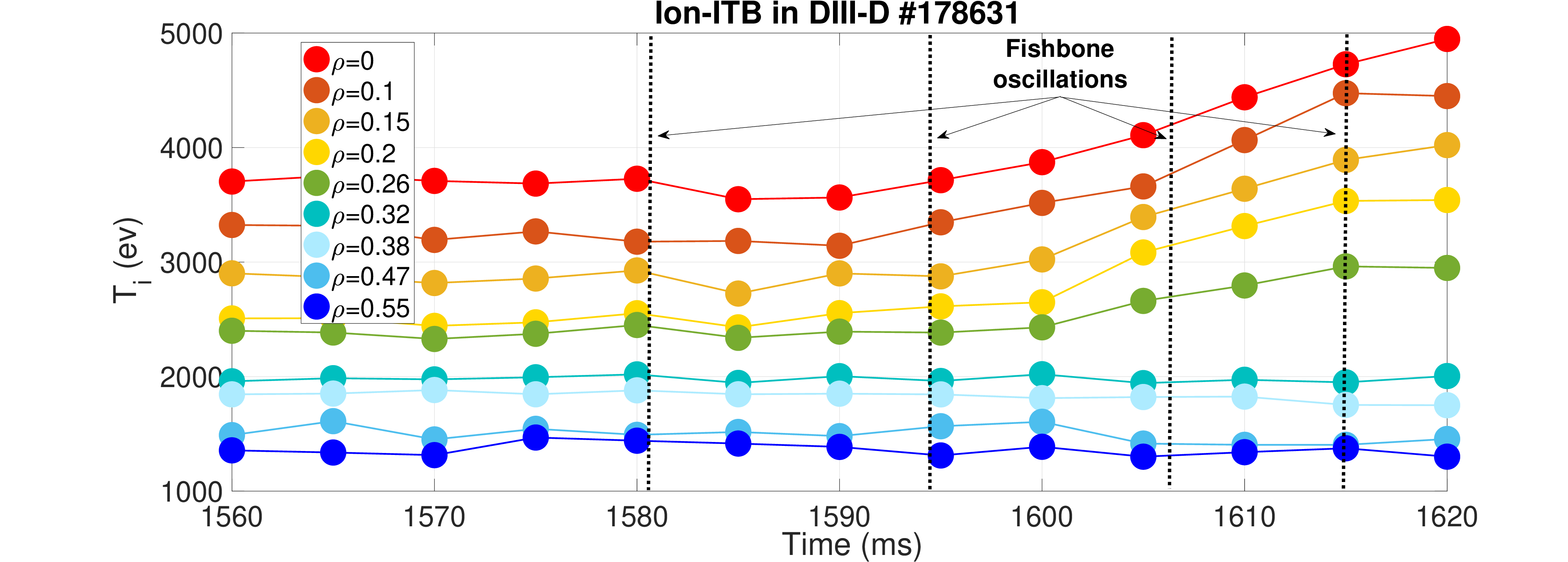}
\caption{Time evolution of $T_i$ channels during fishbone bursts from CXRS measurements in DIII-D shot \#178631.}
\label{ITB}
\end{figure}
As mentioned in section 2, the onset of fishbone bursts in this DIII-D discharge leads to an increase of the core $T_i$ temperature. Such an increase cannot be explained by additional power brought by the beams, as they were at constant power for $\sim1300$ms before the sharp $T_i$ increase at $t\sim 1600$ms in Fig. \ref{DIIID_signals}. A causality between the fishbone bursts starting at $t\sim1580$ ms and the increase in thermal ion temperature is therefore plausible, as the $n=1$ fishbones are the dominant instabilities in this DIII-D plasma over $t\in[1580,1700]$ ms. \\To investigate further the link between fishbone modes and increased $T_i$ confinement, the time evolution of different $T_i$ channels on the low field slide, obtained from the charge exchange recombination spectroscopy (CXRS) diagnostic, are displayed in Fig. \ref{ITB} over $t\in[1560,1620]$ ms. Each channel corresponds to a given radial position, and the four fishbone bursts occurring over this time interval at t=1581, 1594, 1607 and 1615 ms are marked by dashed lines. An ion-ITB starting at $t\sim1595$ ms, i.e. 14 ms after the first fishbone burst, can clearly be observed  in Fig. \ref{ITB}, as only core channels within $\rho\in[0,0.26]$ measure an increase in ion temperature. The maximum amplitude of the fishbone mode being located around $q_{min}$ at $\rho\sim0.25$, the foot of the ion-ITB seems related to the fishbone instability, which reinforces the possible causality between fishbone bursts and ion-ITB formation. These experimental results were reproduced in four other DIII-D discharges (\#178632, \#178640, \#178641, \#178642) using similar heating power, density, current and $q_{min}$ parameters compared to \#178631. Ion-ITBs were also observed in these plasmas after fishbone bursts, the ITB formation usually taking place $\sim10-20$ ms after the first fishbone burst. \\
Since fishbone modes are found in GTC simulations to destabilise zonal flows in the DIII-D discharge \#178631, ion-ITB formations in DIII-D plasmas could be explained by microturbulence suppression caused by a large fishbone-induced zonal flows shearing rates $\omega_{E\times B}$, if $\omega_{E\times B}$ exceeds the growth rate of the most unstable drift-wave \cite{Hahm1995} for these configurations. Evidences supporting this ITB formation mechanism were recently reported in \cite{Ge2022}, where fishbones were observed in kinetic-MHD simulations to have large enough shearing rates to suppress ITG turbulent transport in EAST plasmas \cite{Yang2017}, featuring ITB formation after the onset of fishbones. To confirm whether a similar mechanism could also explain the ITB formation in these DIII-D plasmas, high-n electrostatic GTC simulations with kinetic trapped electrons are performed to identify the most unstable drift-wave mode. In these simulations the radial and poloidal grid size spacings are respectively $\Delta r=0.35\rho_i$ and $r\Delta\theta=0.7\rho_i$, with $\rho_i=4\times10^{-3}$m the thermal ion Larmor radius, and 32 grid points are used in the parallel direction, and the toroidal mode domain considered is $n\in[30, 50]$. The most unstable drift-wave is a collisionless trapped electron mode (CTEM) \cite{Adam1976} localized at $\rho=0.41$, with a growth rate of $\gamma_{TEM}=1.38\times 10^5$ s$^{-1}$ a wavelength of $k_{\theta}\rho_i\sim 0.5$ and a $n\sim40$ dominant toroidal mode number. The fishbone-induced shearing rate at saturation in the GTC electromagnetic simulation is larger than the TEM growth rate over $\rho\in[0.2,0.55]$, as reported in \cite{Brochard2024} Fig. 4b. The fishbone-induced shearing rate indeed peaks at $\rho=0.32$ with $\omega_{E\times B}=8.3\times10^5$ s$^{-1}$, and at the TEM location $\omega_{E\times B}/\gamma_{TEM}\sim3$. The ratio of TEM radial to poloidal wavelength is also much larger than one as can be observed in \cite{Brochard2024} Fig. 4a, which implies that the effective shearing rate \cite{Hahm1995} of the fishbone-induced zonal flows is large enough to  suppress the TEM turbulence and explain the ion-ITB formation in DIII-D plasmas. \\ 
Cross-scale GTC simulations involving simultaneously fishbones and TEM turbulence will however be necessary, in order to demonstrate that microturbulence can be suppressed by fishbone modes by simulating self-consistently a transport barrier. Additional DIII-D experiments are also proposed to be conducted, in order to disentangle the different mechanisms that could lead to ITB formation. Weakly reversed magnetic shear configurations and equilibrium flows, present in these DIII-D plasmas, are also known to lead to ITB formation \cite{Wolf1999}. Reproducing the DIII-D discharge \#178631 with monotonic q profiles would enable to isolate the impact of fishbones on the ITB formation, ITB formation preceded by fishbone modes having also been observed in EAST plasmas with monotonic q profiles \cite{Zhang2023}.
\section{Prediction of fishbone dynamics in ITER prefusion baseline scenario}
\begin{figure}[h!]
\begin{subfigure}{.5\textwidth} 
   \centering
   \includegraphics[scale=0.4]{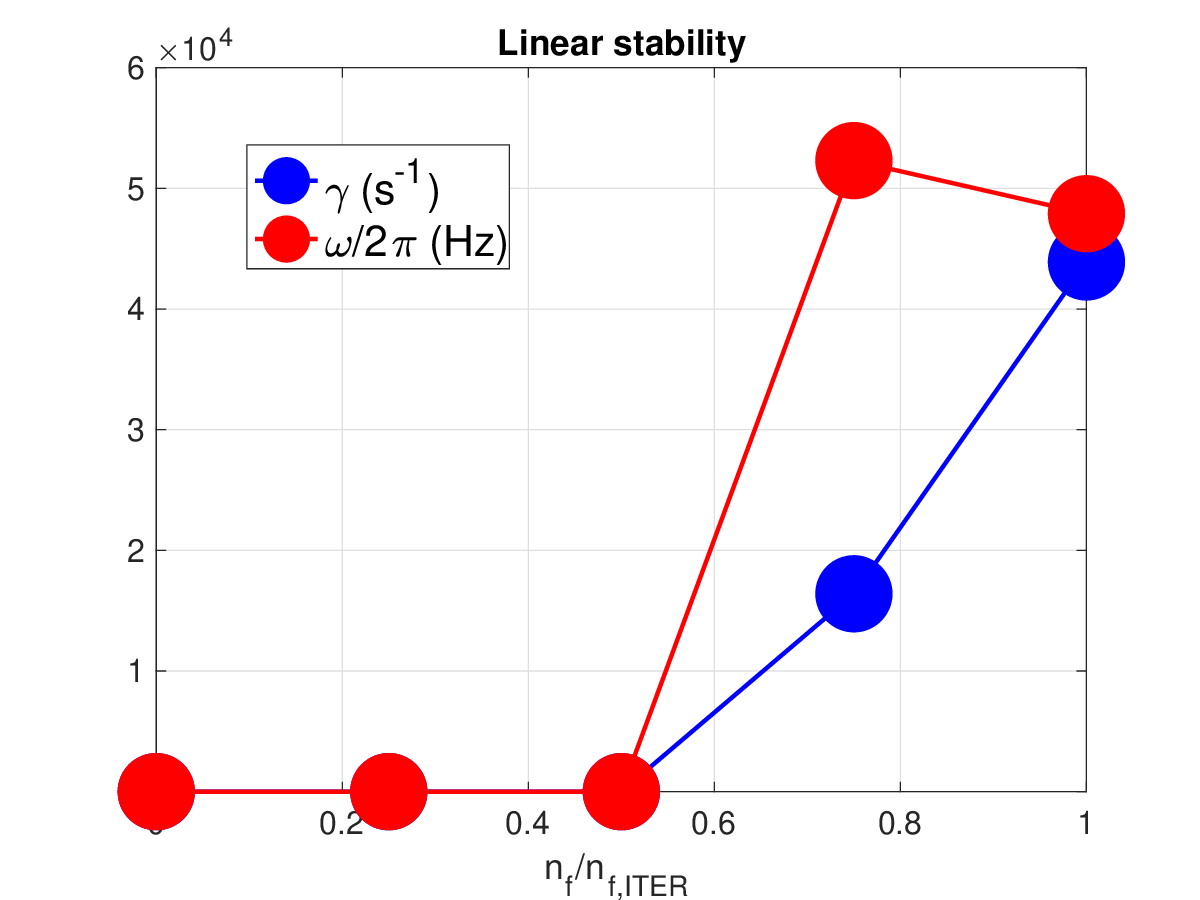}
   \caption{}
\end{subfigure}     
\begin{subfigure}{.5\textwidth} 
   \centering
   \includegraphics[scale=0.22]{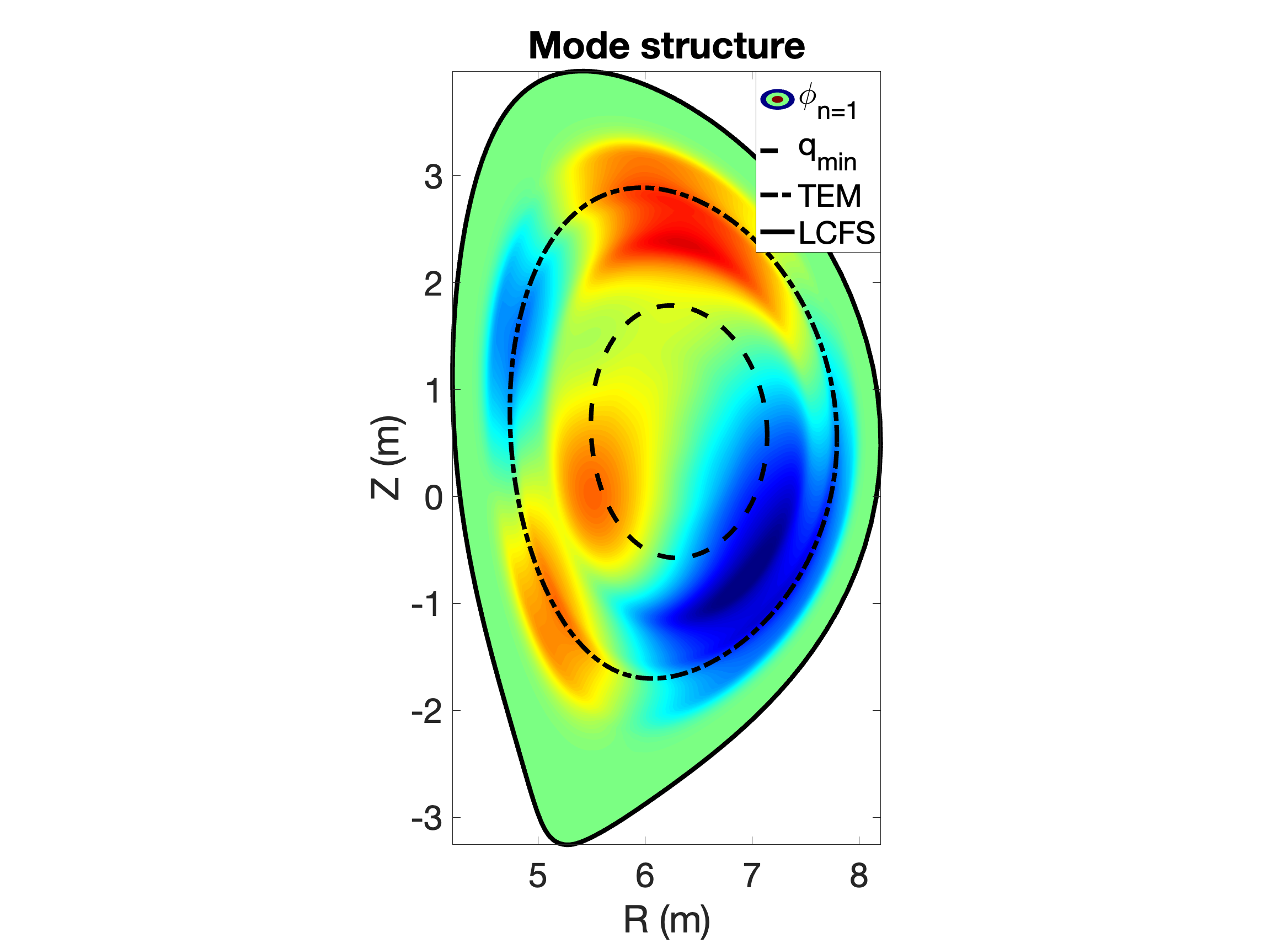}
   \caption{}
\end{subfigure}
\begin{subfigure}{.5\textwidth} 
   \centering
      \includegraphics[scale=0.4]{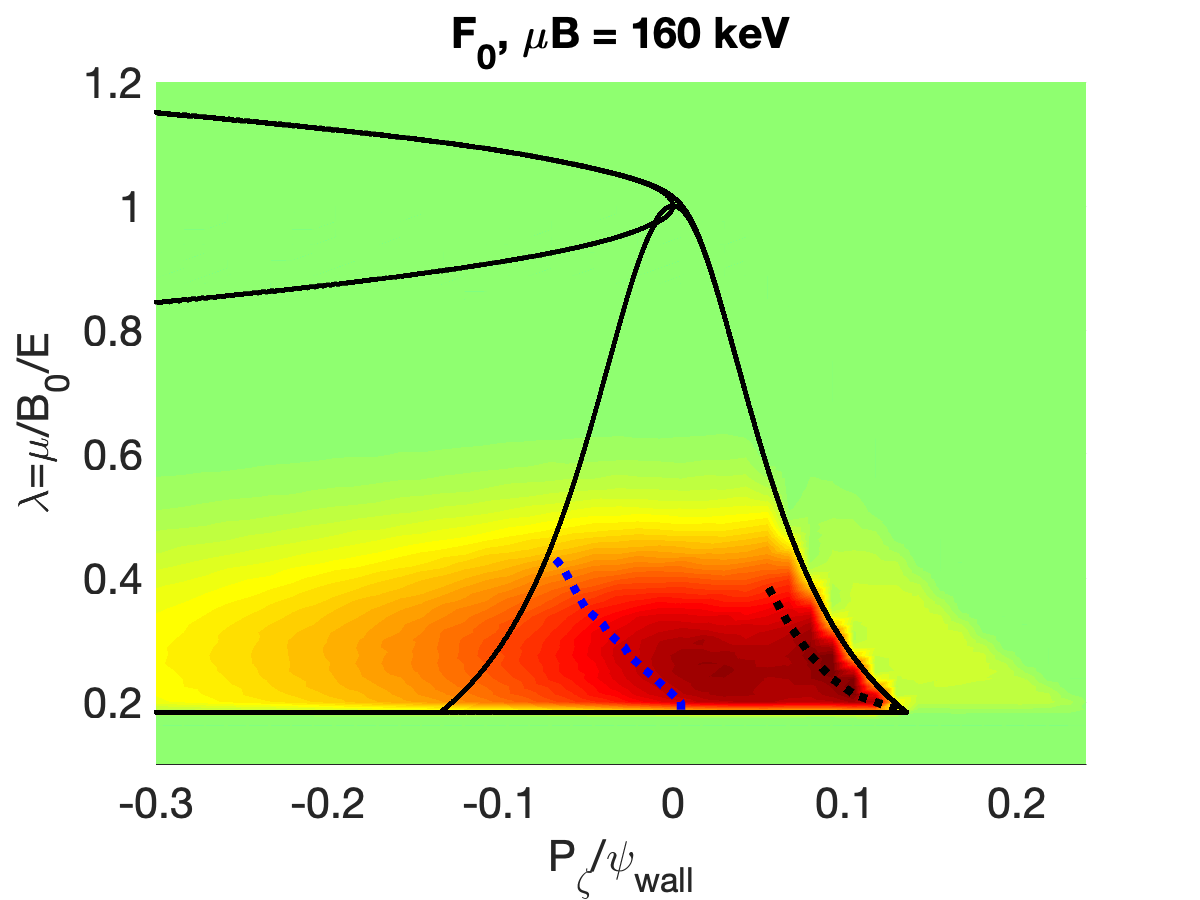}
   \caption{}
\end{subfigure}  
\begin{subfigure}{.5\textwidth} 
   \centering
      \includegraphics[scale=0.4]{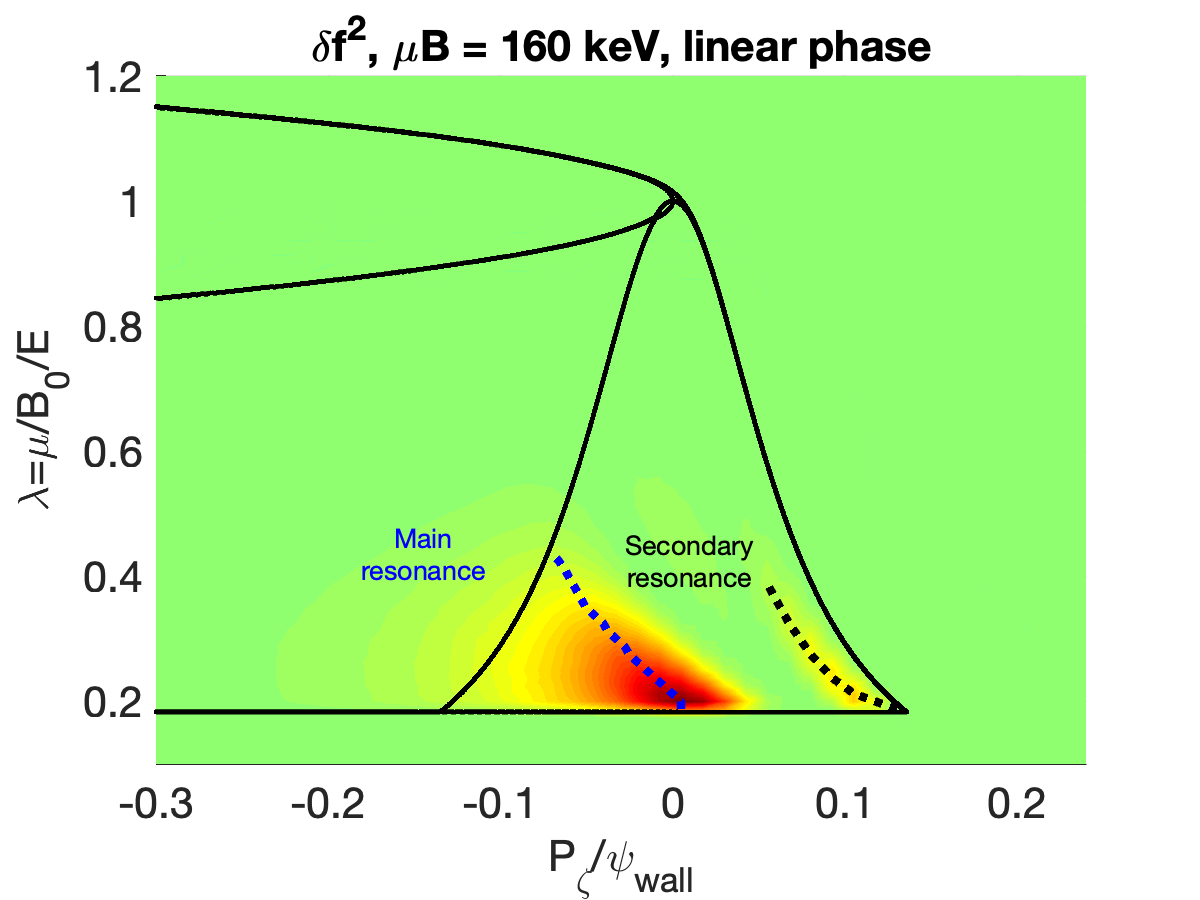}
   \caption{}
\end{subfigure}  
\caption{Kinetic-MHD stability of the ITER scenario from GTC gyrokinetic simulations. (a) Linear stability of the $n=1$ fishbone mode against on-axis EP density. (b) n=1 mode structure of the electrostatic potential $\phi_{n=1}$. (c) $F_0$ and (d) $\delta f^2$ EP histograms in the $(P_{\zeta},\lambda)$ phase space diagram at $\mu B_0=160keV$.}
\label{ITER_linear}
\end{figure}
With fishbone simulations having been validated against the DIII-D experiment, GTC can now be applied to predict realistically the fishbone-induced EP dynamics in the selected ITER prefusion scenario. The linear stability of the configuration described in Fig. \ref{ITER_eq} is examined in Fig. \ref{ITER_linear}. Similarly to the DIII-D case, when an equivalent maxwellian distribution is used instead of the realistic beam in Fig. \ref{ITER_eq} (c), n=1 modes are stable in this ITER configuration. With the realistic beam, an EP density scan is performed for the $n=1$ mode growth rate and frequency, displayed in Fig. \ref{ITER_linear} (a), A $n=1$ fishbone is destabilized past a EP beta threshold $\beta_{EP}\sim 0.75\beta_{EP,exp}$ with $\gamma=4.4\times10^{-4}$ s$^{-1}$ and $\omega/2\pi=48$kHz at nominal EP density. The fishbone mode structure is shown in Fig. \ref{ITER_linear} (b). Again similarly to the DIII-D case, the mode has a dominant $m=1$ harmonic that peaks at $q_{min}=1.05$, and a subdominant $m=2$ sideband centered around $q=2$. The resonance driving the fishbone mode is however different for this configuration, as the tangential beams inject mostly co-passing particles. The integrated $\delta f^2$ signal being largest for $\mu B_0\sim 160$ keV, this value is used to identify the resonance in CoM phase space. The $F_0$ and $\delta f^2$ histograms are displayed respectively in Fig. \ref{ITER_linear} (c-d). As shown in Fig. \ref{ITER_linear} (c), the EP distribution is indeed purely co-passing, the trapped CoM space domain being empty. Two resonant structures can be observed in Fig. \ref{ITER_linear} (d), which are most likely belonging to the same $l=-1$ drift-transit resonance $\omega=\Omega_3-\Omega_2$. Indeed, in this CoM space zone, $\Omega_3/2\pi\sim\Omega_2/2\pi\sim3\times10^5$ Hz. Since the fishbone frequency is again only a tenth of the
particle orbital frequencies, resonance lines cannot be drawn precisely for $\omega=\Omega_3-\Omega_2$ due the current accuracy in computing $\Omega_2$. It is however clear that the $l=-1$ resonance is the only one which can resonate with the fishbone mode, $\Omega_2$ and $\Omega_3$ having similar amplitudes. Both resonance locations are driving the fishbone mode as $\partial F_0/\partial P_{\zeta}>0$ in their vicinity, as shown in Fig. \ref{ITER_linear} (c).
\begin{figure}[h!]
\begin{subfigure}{.4\textwidth} 
   \centering
   \includegraphics[scale=0.16]{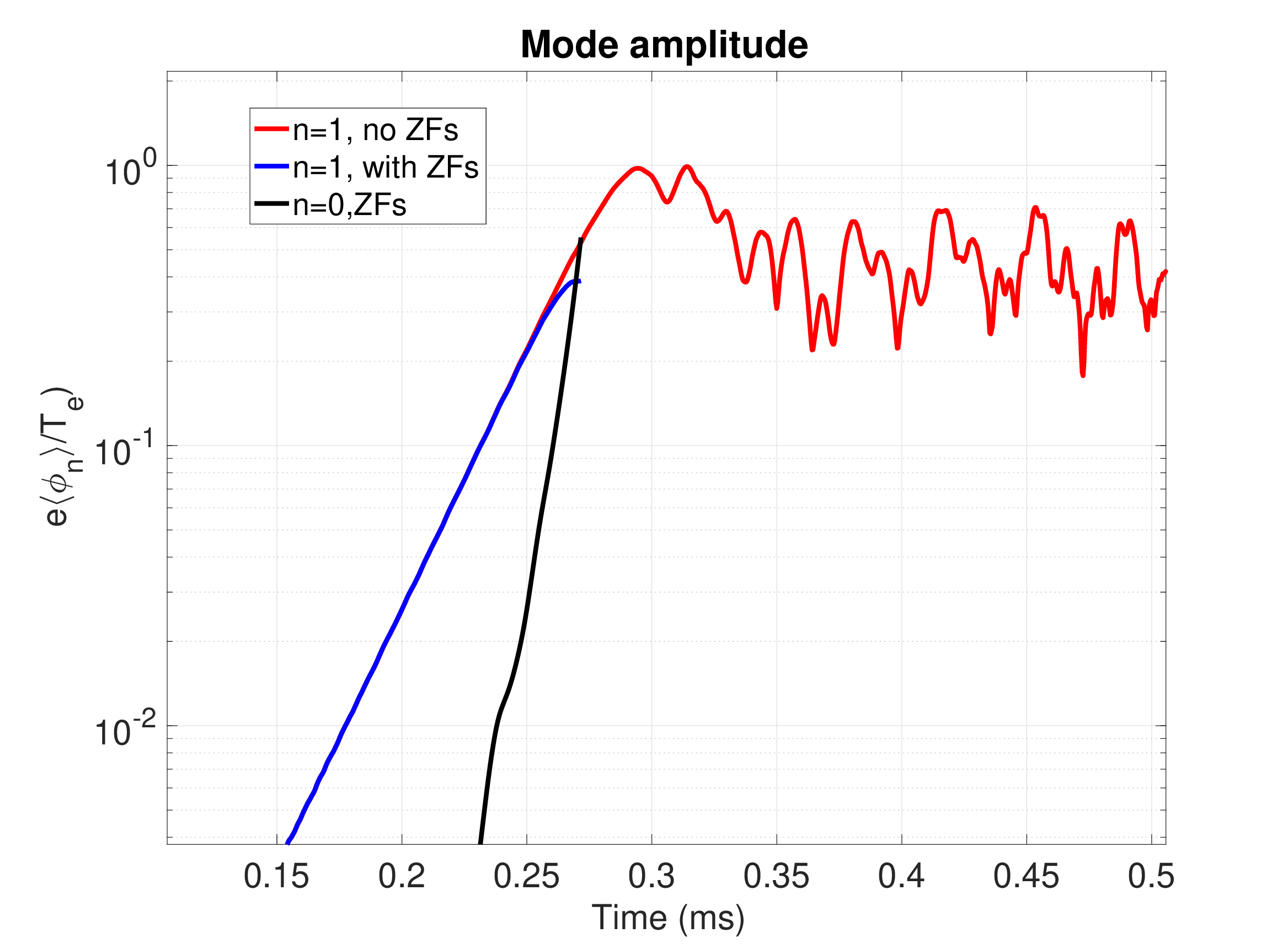}    \caption{}
\end{subfigure}     
\begin{subfigure}{.2\textwidth} 
   \centering
   \includegraphics[scale=0.18]{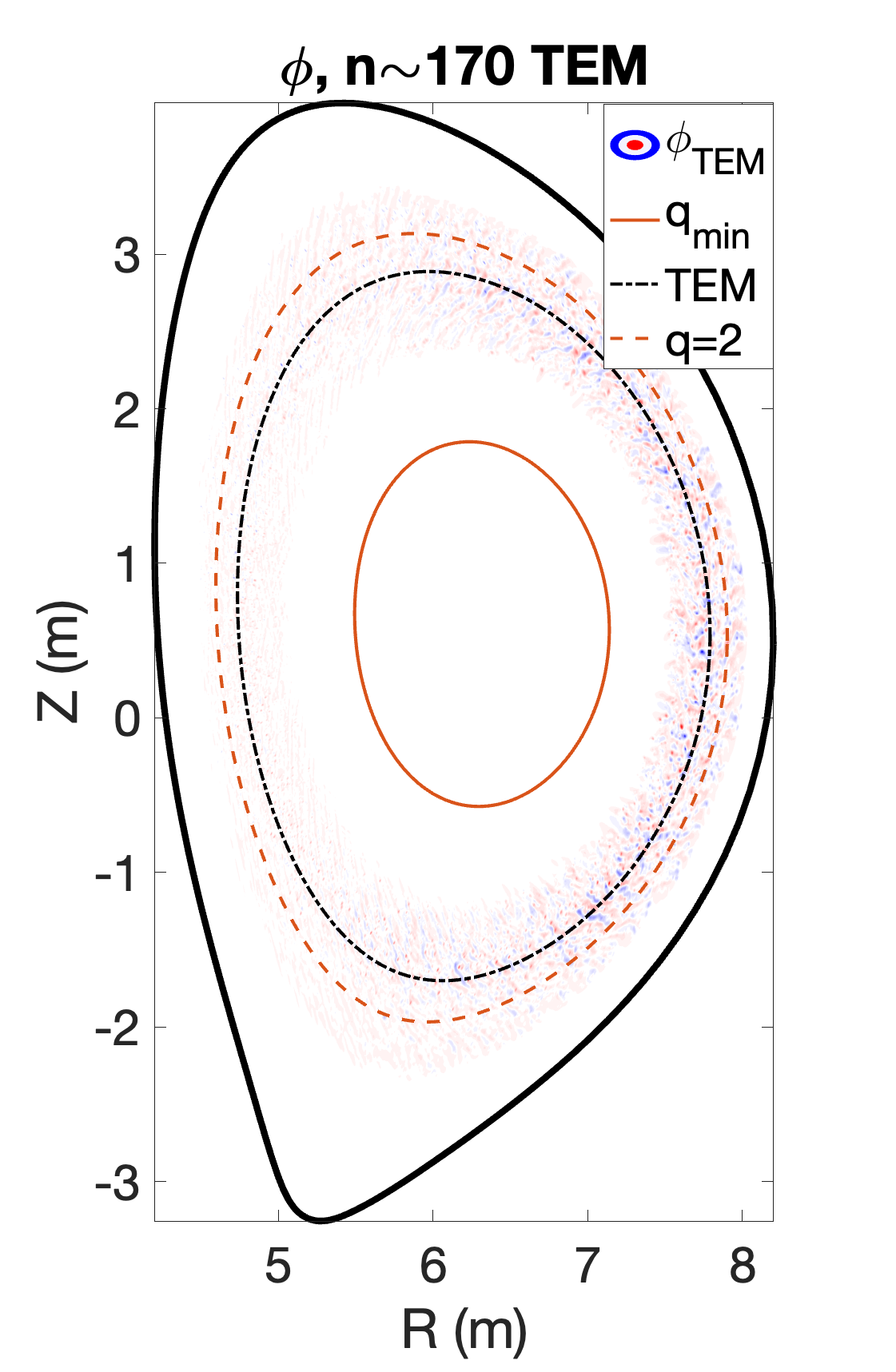}
   \caption{}
\end{subfigure}
\begin{subfigure}{.4\textwidth} 
   \centering
      \includegraphics[scale=0.16]{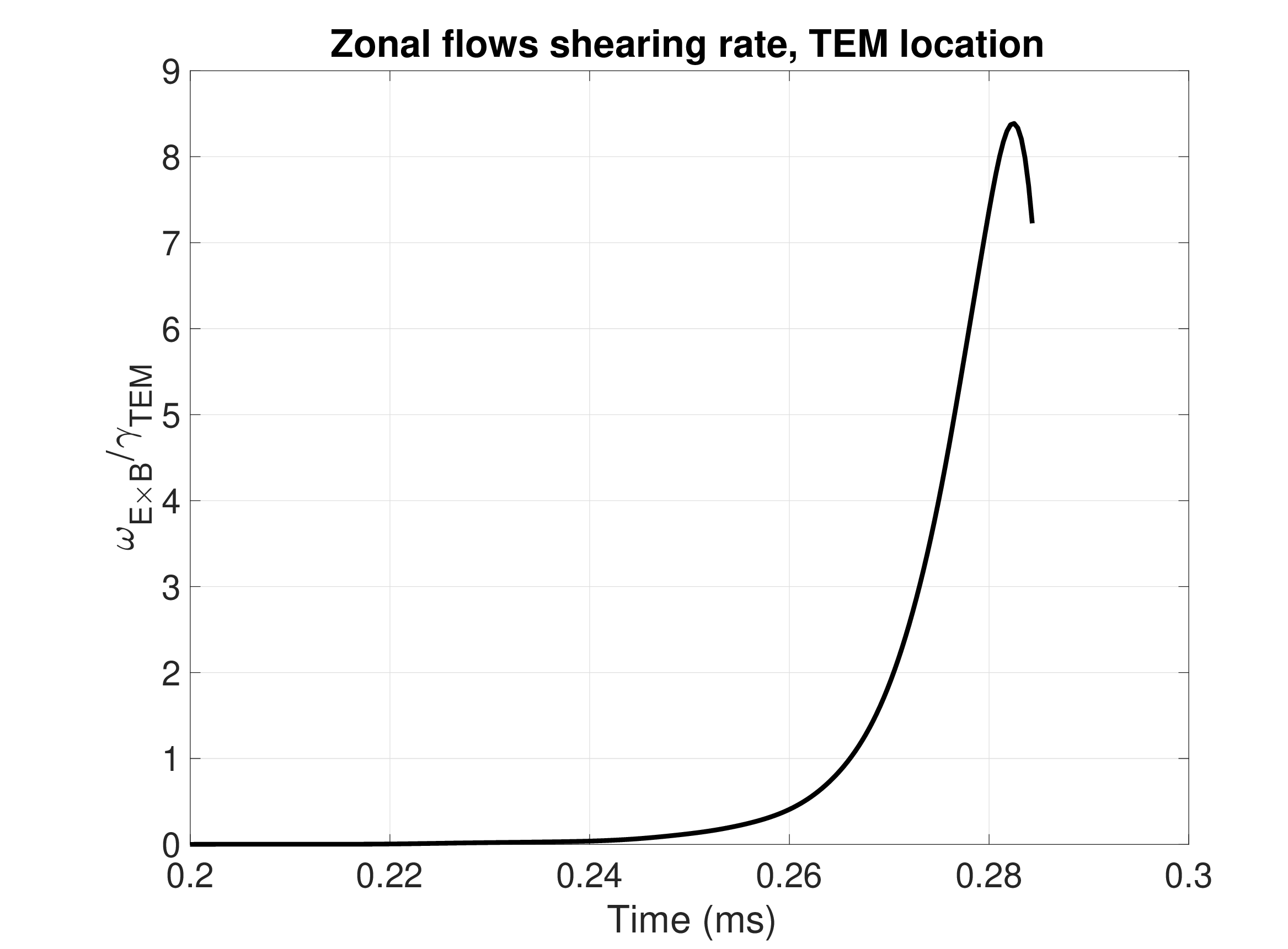}
   \caption{}
\end{subfigure}  
\caption{(a) Time evolution of the volume-averaged perturbed electrostatic potential $e\langle\phi\rangle_r/T_e$ (n=0,1) from GTC simulation of fishbone in ITER. (b) GTC simulation of TEM microturbulence (c) Fishbone-induced shearing rate at the TEM location}
\label{ITER_NL}
\end{figure}
\\ Nonlinear $n=1$ GTC simulations are conducted on this ITER scenario, with and without zonal flows. The time evolution of the volume-averaged electrostatic potential is shown in Fig. \ref{ITER_NL} (a). Zonal flows are again found to be forced-driven, with a growth rate twice that of the $n=1$ fishbone mode. The inclusion of zonal flows also leads to an earlier saturation of the fishbone mode towards $t\sim0.27$ ms, with $\delta B/B_0\sim1\times10^{-4}$ with zonal flows and $\delta B/B_0\sim4\times10^{-4}$ without zonal flows. The simulations with zonal flows cannot however be pushed further in the nonlinear phase, due to the onset of numerical instabilities. This issue is most likely due to the GTC code formulation used \cite{Xiao2015}, which computes zonal densities on equilibrium flux surfaces, while at $t\sim0.27$ ms the flux surfaces are significantly impacted at the core plasma with $\delta\psi_{n=1}\sim0.4\psi_0$ at $\rho=0.25$. As mentioned in section 5.1, the code formulation employed in \cite{Fang2019} will be utilized in upcoming cross-scale GTC simulations involving both global kinetic-MHD and microturbulence. Nonetheless, it is still relevant to compare the fishbone-induced zonal flows shearing rate at the fishbone saturation with the linear growth rate of the most unstable drift-wave mode for this configuration, to see whether fishbone modes could also impact the turbulent transport in this ITER plasma. GTC electrostatic simulations with kinetic trapped electrons are therefore conducted, with a grid resolution of $N_{\psi}=500$, $N_{\theta}=3600$ at $r=0.5a$ ($r\Delta\theta$ is constant on each flux surface) and $N_{\parallel}=32$. The toroidal mode domain retained in these simulations is $n\in[100, 250]$. The most unstable drift-wave is a TEM located at $\rho=0.71$, within the fishbone mode structure as shown in Fig. \ref{ITER_linear} (b), with $n\sim170$, and a growth rate of $\gamma_{TEM}=3\times10^4$. The microturbulence associated with the TEM in its nonlinear phase is displayed in Fig. \ref{ITER_NL} (b). Since the TEM and the fishbone modes overlap in configuration space, zonal flows produce by the fishbone mode may suppress the turbulent transport induced by the TEM. To quantify this aspect, the time evolution of the fishbone-induced shearing rate $\omega_{E\times B}$ at the TEM location is shown in Fig. \ref{ITER_NL} (c). At the fishbone saturation, $\omega_{E\times B}/\gamma_{TEM}\sim 7$, which suggests that the formation of a fishbone-induced ITB in this ITER prefusion plasma near $\rho=0.71$ is possible.
\begin{figure}[h!]
\begin{subfigure}{.5\textwidth} 
   \centering
   \includegraphics[scale=0.4]{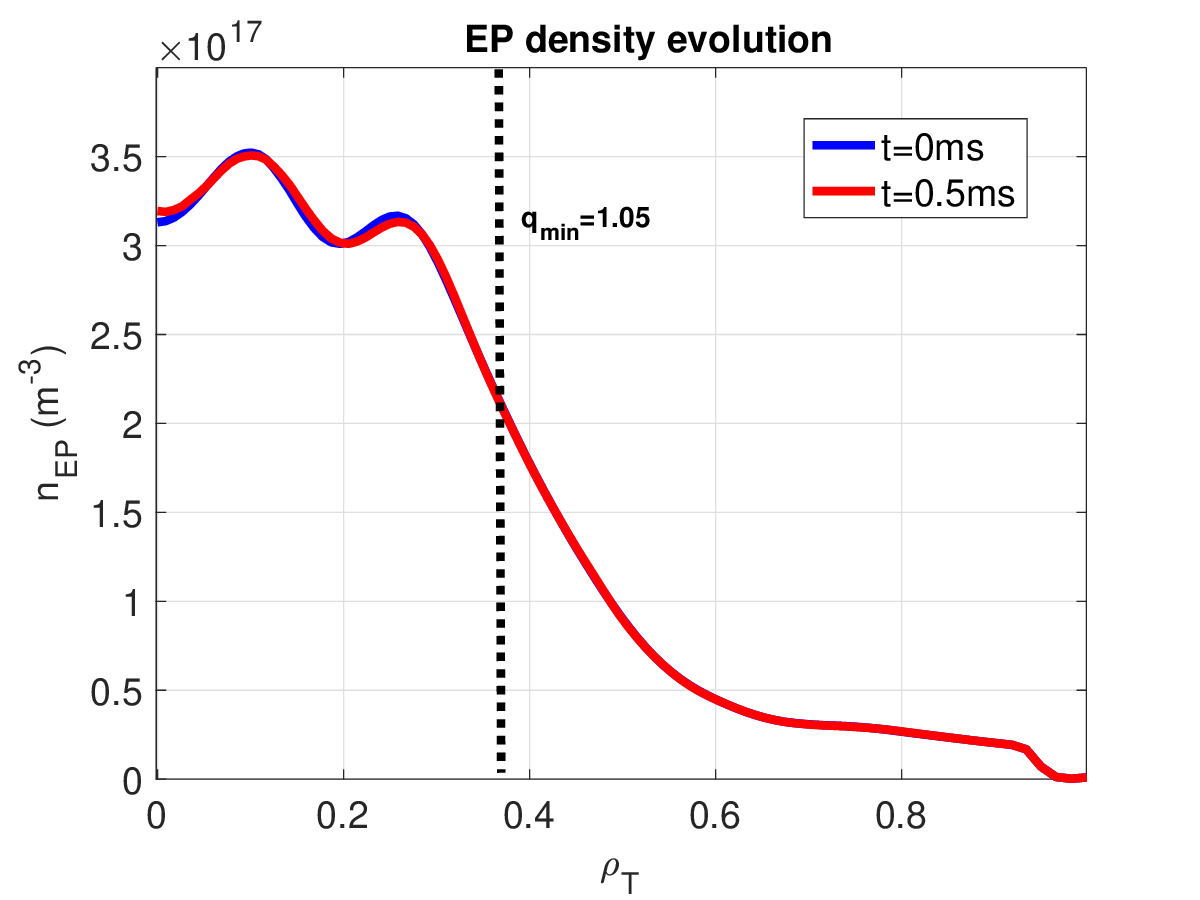}
   \caption{}
\end{subfigure}     
\begin{subfigure}{.5\textwidth} 
   \centering
   \includegraphics[scale=0.4]{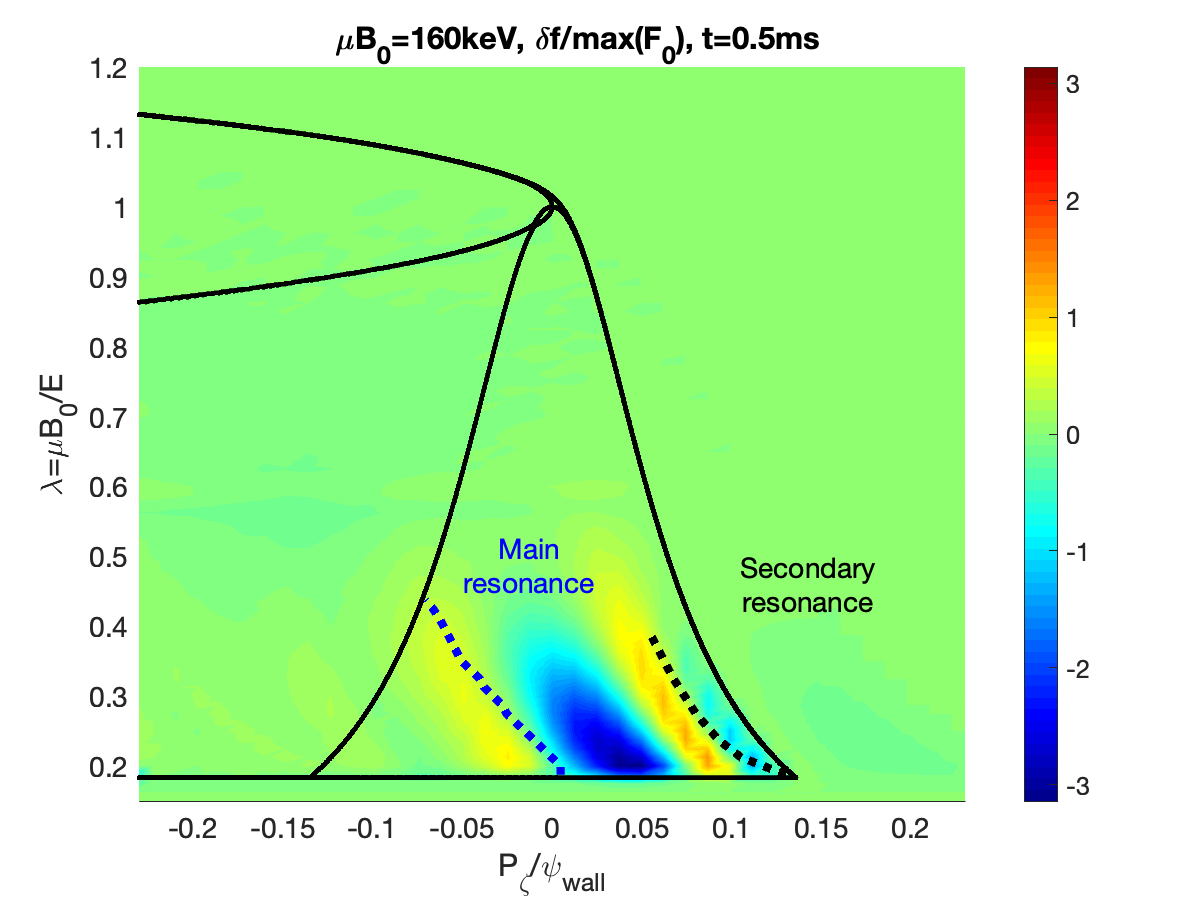}
   \caption{}
\end{subfigure}
\caption{(a) EP density profile before and after fishbone burst without zonal flows. (b) Perturbed distribution function in $(P_{\zeta},\lambda)$ CoM phase space at $\mu B_0=$160keV.)}
\label{ITER_EP}
\end{figure}
\\ The fishbone-induced EP transport is analysed with the GTC simulation without zonal flows, the one with zonal flows not lasting long enough to quantify such a transport. The EP transport levels reported here therefore represent the upper bound of the transport expected in this ITER plasma, as zonal flows reduce the EP transport by lowering the fishbone saturation amplitude. The EP density profiles before and after the fishbone burst in GTC simulations are shown in Fig. \ref{ITER_EP} (a). It can be observed that the EP redistribution only takes place within the $q_{min}$ surface at $\rho\sim0.4$. Only 2\% of the EP population is redistributed by the fishbone, with both inward and outward EP fluxes due to presence of negative and positive EP pressure gradients within the $q_{min}$ volume. These low redistributions levels are confirmed by looking at $\delta f_{EP}$ in CoM space, displayed in Fig. \ref{ITER_EP}. Hole and clump structures characterizing an outward EP transport form around each resonant position. However, their amplitude only correspond to a few percents of the initial EP distribution, which imply that the EP redistribution will be marginal in this ITER plasma, and should therefore not impact significantly the efficiency of the beam heating. This conclusion is similar to what was reported for the alpha-fishbone in a ITER 15 MA baseline DT scenario \cite{Brochard2020b}, where the amount of redistributed alpha particles is too low to affect significantly the burning plasma self-heating. 
\section{Conclusion and perspectives}
In this paper, the fishbone-zonal flows interplay and its impact on the EP redistribution has been studied in DIII-D and ITER prefusion baseline plasmas. The DIII-D discharge has been selected as a matching case for the considered ITER scenario, in order to first validate nonlinear first-principle codes using DIII-D experimental measurements, before applying them to predict the fishbone dynamics in ITER. The gyrokinetic code GTC and the kinetic-MHD codes M3D-C1 and XTOR-K were used in this modelling analysis. A fishbone mode driven by both precessional and drift-transit resonances was found unstable for the DIII-D configuration. Zonal flows were observed to be generated by the fishbone mode, and to dominate the fishbone saturation in GTC simulations. These results imply that the fishbone saturation mechanism is more complex than the conventional picture of EP distribution flattening through resonant wave-particle interactions. Saturation levels for both the $\delta T_e$ envelope and the neutron drop in GTC simulation were found to be in quantitative agreement with ECE and neutron flux measurements in DIII-D, thus supporting this novel saturation mechanism of fishbone instability by self-generated zonal flows. The underlying mechanisms of the two-way fishbone-zonal flows interplay were then discussed in details. The zonal flows generation was identified self-consistently with gyrokinetic simulations to be due to the fishbone-induced EP redistribution, which creates a gyrocenter charge separation leading to the emergence of a radial zonal electric field. This dominant contribution was demonstrated by successful comparisons with analytical theory only taking EP redistribution into account for the zonal flow generation. The mechanism for the fishbone saturation by self-generated zonal flows was identified in phase space, where a zonal Doppler-shift affects the nonlinear dynamics of phase space zonal structures by modifying the position of the resonances. The zonal flows are therefore able to reduce the fishbone EP avalanche by preventing linearly non-resonant particles to resonate with the mode, through a locking of the precessional resonance in its linear position and a detuning of the drift-transit resonance. These effects therefore lead to lower saturation levels for the fishbone instability, by reducing the extent of the EP distribution flattening through wave-particle interactions. The down-chirping of the fishbone frequency was then shown using analytical theory to be entirely due to mode locking, with quantitative agreement between GTC and analytical chirping rates. These results imply that mode-locking may be a universal mechanism through which waves destabilised by wave-particles interaction undergo non-adiabatic frequency chirping in both laboratory and astrophysical plasmas, with similar results for whistler chorus waves \cite{Zonca2022}. Moreover, the fishbone-induced zonal flows were found likely responsible for an ion-ITB formation in the DIII-D discharge, since these zonal flows can suppress turbulent transport as their shearing rate is larger than the growth rate of the most unstable drift-wave mode. Finally, GTC simulations were performed on the ITER prefusion baseline scenario. A fishbone mode was observed to be excited by a drift-transit resonance. Zonal flows were also found to be generated by the fishbone mode and to dominate its nonlinear saturation. The zonal flows shearing rate at the drift-wave location is also large enough to suppress microturbulence in this ITER plasma, and can lead to ITB formation. The fishbone-induced EP transport is observed to be marginal in the limit without zonal flows, confirming previous findings for the alpha-fishbone in ITER 15 MA DT scenarios \cite{Brochard2020b}.\\\\
Global EP-driven instabilities such as the fishbone instability have been considered since their identification as modes to be avoided in burning plasmas such as those of ITER, as they can degrade plasma self-heating and damage the first wall through EP transport. However, since benign fishbones lead to negligible EP transport and can create strongly sheared zonal flows that may suppress turbulent transport, it could therefore be of great interest to trigger fishbone modes on purpose in ITER plasmas to increase fusion performances, rather than avoiding them. This could be done by optimizing the NBI and ICRH depositions, as well as the alpha pressure profile, to excite fishbone resonances. Nonetheless, the relevant experimental actuators that lead to strongly sheared fishbone-induced zonal flows first have not yet been identified theoretically nor experimentally. Additional first-principles simulations and tokamak experiments are therefore required to identify the optimal regimes in which fishbones generate such flows, without inducing a large EP loss. Furthermore, the causality between fishbone modes and ITB formation also needs to be established. Other physical mechanisms could explain ITB formation in tokamak discharge featuring fishbone modes, such as weakly reversed magnetic shear configurations and equilibrium flows \cite{Wolf2003}. The different mechanisms need to be disentangled to clearly establish that fishbone bursts are the dominant mechanism in the ITB formation observed in multiple tokamak experiments. \\ To address these aspects, new DIII-D experiments have been proposed to quantify the impact of the q profile, the EP pressure profile and the beam deposition on the nonlinear interplay between fishbones and ITBs, with a specific emphasis on the effects of fishbones driven by monotonic q profile on bulk confinement. Cross-scale gyrokinetic simulations self-consistently coupling drift-waves, Alfv\'en eigenmodes and fishbones will also be performed, with the overall zonal flow levels determined by each of these instabilities. The integrated simulation of a transport barrier through gyrokinetic simulations is essential to confirm microturbulence suppression by fishbone instabilities.
\section*{Acknowledgements}
The views and opinions expressed herein do not necessarily reflect those of the ITER organization. This work was supported by DOE SciDAC ISEP, INCITE, and the ITPA-EP group. This material is based upon work supported by the U.S. DOE, Office of Science, Office of Fusion Energy Sciences, using the DIII-D National Fusion Facility, a DOE Office of Science user facility, under Award(s) DE-FC02-04ER54698, as well as computing resources from ORNL (DOE Contract DE-AC05-00OR22725), NERSC (DOE Contract DE- AC02-05CH11231), and PPPL (DOE Contract DE-AC02-09CH11466). This report was prepared as an account of work sponsored by an agency of the United States Government. Neither the United States Government nor any agency thereof, nor any of their employees, makes any warranty, express or implied, or assumes any legal liability or responsibility for the accuracy, completeness, or usefulness of any information, apparatus, product, or process disclosed, or represents that its use would not infringe privately owned rights. Reference herein to any specific commercial product, process, or service by trade name, trademark, manufacturer, or otherwise does not necessarily constitute or imply its endorsement, recommendation, or favoring by the United States Government or any agency thereof. The views and opinions of authors expressed herein do not necessarily state or reflect those of the United States Government or any agency thereof.
\appendix
\section{Derivation of the gyrokinetic ion weight equation with anisotropic slowing down distribution}
As previously discussed in section 3, the following anisotropic slowing-down distribution is considered in GTC, taking into account one injection energy for simplicity
\begin{equation}
F_{SD,ani}(\psi,v,\lambda) = \frac{n_f(\psi)}{C}\frac{H(v_0-v)}{v^3+v_c^3(\psi)}\exp\bigg[-\bigg(\frac{\lambda-\lambda_0}{\Delta\lambda}\bigg)^2\bigg]
\end{equation}
the normalization constant $C$ is given by
\begin{equation}
\fl
C = \frac{2\pi}{3}\ln\Big[1+\Big(\frac{v_0}{v_c}\Big)^3\Big]\int_{-1}^{1}d\theta \exp\bigg[-\bigg(\frac{\sin^2\theta B_0/B-\lambda_0}{\Delta\lambda}\bigg)\bigg]\sin\theta,\ \ \ \sin\theta=\frac{v_{\perp}}{v}
\end{equation}
the critical velocity is defined in general as
\begin{equation}
v_c(\psi) = \bigg(\frac{3\sqrt{\pi}m_e}{4m_f}\bigg)^{1/3}\sqrt{\frac{T_e(\psi)}{m_e}}
\end{equation}
and $v_0$ stands for the birth velocity, $n_f$ the fast ion density profile, $\lambda_0$ the peak pitch angle of the distribution and $\Delta\lambda$ its width along $\lambda$. The critical velocity can also be taken as a constant to fit experimental distributions, as discussed in section 3.\\\\
When using such distributions, the ion weight equation needs to be modified since the terms $\partial_{v_{\parallel}}F|_{\mu,\textbf{R}}/F$ and $\nabla F|_{\mu,v_{\parallel}}/F$ are explicitly required. Following \cite{Dong2017}, the ion weight equation can indeed be expressed as
\begin{eqnarray}
\fl
\frac{dw_f}{dt} = (1-w_f)\bigg[\bigg(v_{\parallel}\frac{\delta\textbf{B}_{\perp}}{B_{\parallel}^*} +\frac{\textbf{b}_0\times\nabla\phi_{\delta B_{\parallel}}}{B_{\parallel}^*}\bigg)\cdot\frac{\nabla F_{SD,ani}|_{\mu,v_{\parallel}}}{F_{SD,ani}} +  \nonumber\\ \bigg(\frac{\mu\delta\textbf{B}_{\perp}\cdot\nabla B_0}{B_{\parallel}^*} + Z_f\frac{(\textbf{B}_0^*+\delta\textbf{B}_{\perp})}{B_{\parallel}^*}\cdot\nabla\phi_{\delta B_{\parallel}}+ Z_f\frac{\partial \delta A_{\parallel}}{\partial t}\bigg)\frac{\partial_{v_{\parallel}}F_{SD,ani}|_{\mu,\textbf{R}}}{m_fF_{SD,ani}}\bigg]
\end{eqnarray}
where $\phi_{\delta B_{\parallel}} = \phi + \mu\delta B_{\parallel}/Z_f$. Using the set of variables $(\textbf{R},\mu,v_{\parallel})$, the anistropic slowing down distribution expands as
\begin{equation}
F_{SD,ani}(\textbf{R},v_{\parallel},\mu) = \frac{n_f(\psi)}{C}\frac{ \exp\bigg[-\bigg(\bigg(\frac{\mu B_0}{m_fv_{\parallel}^2/2+\mu B_0} - \lambda_0\bigg)/\Delta \lambda\bigg)^2\bigg]}{(v_{\parallel}^2+2\mu B_0/m_f)^{3/2} + v_c^3(\psi)}
\end{equation}
Since $B_0$ depends on $\textbf{R}$, $\nabla F|_{\mu,v_{\parallel}}/F$ can be expressed as
\begin{equation}
\frac{\partial F_{SD,ani}}{\partial \textbf{R}}\bigg|_{\mu,v_{\parallel}} = \frac{\partial F_{SD,ani}}{\partial \textbf{R}}\bigg|_{\mu B_0,v_{\parallel}} + \frac{\partial F_{SD,ani}}{\partial (\mu B_0)}\mu\nabla B_0
\end{equation}
The derivatives of the anistropic slowing-down distribution required in the ion weight equation therefore read
\begin{equation}\label{der_1}
\frac{\partial_{v_{\parallel}}F_{SD,ani}|_{\mu,\textbf{R}}}{F_{SD,ani}} = 4\bigg(\frac{\lambda - \lambda_0}{\Delta\lambda^2}\bigg)\frac{v_{\parallel}\lambda}{v^2} - 3\frac{vv_{\parallel}}{v^3+v_c^3}
\end{equation}
\begin{equation}\label{der_2}
\frac{\nabla F_{SD,ani}|_{\mu,v_{\parallel}}}{F_{SD,ani}} = \frac{\nabla n_f}{n_f} - 3\frac{v_c^2\nabla v_c}{v^3+v_c^3}
\end{equation}
\begin{equation}\label{der_3}
\fl
\frac{\partial_{\mu B_0}F_{SD,ani}|_{v_{\parallel},\textbf{R}}}{F_{SD,ani}} = \frac{1}{m_f}\bigg[ 4\bigg(\frac{\lambda - \lambda_0}{\Delta\lambda^2}\bigg)\frac{(\lambda-1)}{v^2}-3 \frac{v}{v^3+v_c^3} \bigg] = \frac{v_{\parallel}}{m_f}\frac{\partial_{v_{\parallel}}F_{SD,ani}|_{\mu,\textbf{R}}}{F_{SD,ani}} - \frac{4}{m_f}\frac{(\lambda-\lambda_0)}{v^2\Delta\lambda^2}
\end{equation}
Considering $\delta \textbf{B}_{\perp} \approx \textbf{b}_0\times\nabla\delta A_{\parallel}$ and $\textbf{B}_0^* = \textbf{B}_0 + B_0\frac{v_{\parallel}}{\Omega_f}\nabla\times\textbf{b}_0$, Eq. (4) can be expanded as
\begin{eqnarray}\label{exp_ion_eq}
\fl
\frac{dw_f}{dt} = (1-w_f)\bigg[-\frac{\textbf{b}_0}{B_{\parallel}^*}\times\nabla(\phi_{\delta B_{\parallel}}-v_{\parallel}\delta A_{\parallel})\cdot\nabla\ln F_{SD,ani}|_{\mu B_0,v_{\parallel}} + \frac{Z_f}{m_f} \bigg(\frac{\textbf{v}_c\cdot\nabla\phi_{\delta B_{\parallel}}}{v_{\parallel}}  \nonumber\\ \fl - \frac{\textbf{b}_0\times\nabla\delta A_{\parallel}\cdot\nabla\phi_{\delta B_{\parallel}} }{B_{\parallel}^*}-  E_{\parallel} + \frac{\mu\textbf{B}_0}{Z_fB_{\parallel}^*}\cdot\nabla\delta B_{\parallel}\bigg)\frac{\partial \ln F_{SD,ani}}{\partial v_{\parallel}}\bigg|_{\mu,\textbf{R}}  + Z_f\textbf{v}_g\cdot\nabla\phi_{\delta B_{\parallel}}\frac{\partial\ln F_{SD,ani}}{\partial(\mu B_0)}\bigg|_{v_{\parallel},\textbf{R}}   \nonumber\\ \fl + \frac{\mu\textbf{b}_0\times\nabla\delta A_{\parallel}}{B_{\parallel}^*}\cdot\nabla B_0\bigg(v_{\parallel} \frac{\partial \ln F_{SD,ani}}{\partial(\mu B_0)}\bigg|_{v_{\parallel},\textbf{R}} - \frac{1}{m_f}\frac{\partial \ln F_{SD,ani}}{\partial v_{\parallel}}\bigg|_{\mu,\textbf{R}}\bigg) \bigg]
\end{eqnarray}
Using Eqs. (\ref{der_1}-\ref{der_3}), Eq. (\ref{exp_ion_eq}) reduces to
\begin{eqnarray}\label{final_ion}
\fl
\frac{dw_f}{dt} = (1-w_f)\bigg[-\frac{\textbf{b}_0}{B_{\parallel}^*}\times\nabla(\phi_{\delta B_{\parallel}}-v_{\parallel}\delta A_{\parallel})\cdot \bigg(\frac{\nabla n_f}{n_f} - \nabla v_c\Big[\frac{3v_c^2}{v^3+v_c^3}\Big]\bigg)  + \frac{Z_f}{m_f}\bigg(\textbf{v}_d\cdot\nabla\phi_{\delta B_{\parallel}} \nonumber\\ \fl - \frac{v_{\parallel}}{B_{\parallel}^*}\textbf{b}_0\times\nabla\delta A_{\parallel}\cdot\nabla\phi_{\delta B_{\parallel}}  -v_{\parallel} E_{\parallel} + \frac{v_{\parallel}\mu\textbf{B}_0}{Z_f B_{\parallel}^*}\cdot\nabla\delta B_{\parallel} \bigg)\bigg(4\Big[\frac{\lambda-\lambda_0}{\Delta\lambda^2}\Big]\frac{\lambda}{v^2} - 3\frac{v}{v^3+v^3_c}\bigg) \nonumber\\ - \frac{4\mu}{B_{\parallel}^*m_f}\frac{(\lambda-\lambda_0)}{v^2\Delta\lambda^2}\textbf{b}_0\times\nabla B_0\cdot\nabla(\phi_{\delta B_{\parallel}} - v_{\parallel}\delta A_{\parallel})\bigg]
\end{eqnarray}
The third line of Eqs. (\ref{exp_ion_eq},\ref{final_ion}) cancels out for maxwellian distributions since $\partial_{v_{\parallel}} F_M|_{\mu,\textbf{R}} = -(m_f/v_{\parallel})\partial_{\mu B_0} F_M|_{v_{\parallel},\textbf{R}}$.\\\\
\section*{References}
\bibliography{fishbone_zonal_prl}{}

\begin{thebibliography}{10}

\bibitem{ITER}
{ITER Physics Expert Group on Energe Drive and ITER Physics Basis Editors}.
\newblock {\em Nuclear Fusion}, 39(12):2471--2495, dec 1999.

\bibitem{Heidbrink2008}
W.~W. Heidbrink.
\newblock Basic physics of alfvén instabilities driven by energetic particles
  in toroidally confined plasmas.
\newblock {\em Physics of Plasmas}, 15(5), February 2008.

\bibitem{McGuire1983}
K.~McGuire and al.
\newblock {\em Physical Review Letters}, 51(20):1925--1925, nov 1983.

\bibitem{Chen1984}
L.~Chen, R.~B. White, and M.~N. Rosenbluth.
\newblock {\em Physical Review Letters}, 52(13):1122--1125, mar 1984.

\bibitem{Fu2006}
G.~Y.~Fu et~al.
\newblock {\em Physics of Plasmas}, 13(5):052517, may 2006.

\bibitem{Brochard2020b}
G.~Brochard et~al.
\newblock {\em Nuclear Fusion}, 60(12):126019, oct 2020.

\bibitem{Diamond2005}
P~H Diamond, S-I Itoh, K~Itoh, and T~S Hahm.
\newblock {\em Plasma Physics and Controlled Fusion}, 47(5):R35--R161, apr
  2005.

\bibitem{Lin1998}
Z.~Lin, T.~S. Hahm, W.~W. Lee, W.~M. Tang, and R.~B. White.
\newblock {\em Science}, 281(5384):1835--1837, sep 1998.

\bibitem{Spong1994}
D.~A. Spong, B.~A. Carreras, and C.~L. Hedrick.
\newblock {\em Physics of Plasmas}, 1(5):1503--1510, may 1994.

\bibitem{Bass2010}
E.~M. Bass and R.~E. Waltz.
\newblock Gyrokinetic simulations of mesoscale energetic particle-driven
  alfvénic turbulent transport embedded in microturbulence.
\newblock {\em Physics of Plasmas}, 17(11), November 2010.

\bibitem{Todo2012}
Y.~Todo, H.L. Berk, and B.N. Breizman.
\newblock {\em Nuclear Fusion}, 52(9):094018, sep 2012.

\bibitem{Zhang2013}
Huasen Zhang and Zhihong Lin.
\newblock Nonlinear generation of zonal fields by the beta-induced alfv{\'{e}}n
  eigenmode in tokamak.
\newblock {\em Plasma Science and Technology}, 15(10):969--973, oct 2013.

\bibitem{Shen2015}
Wei Shen, G.~Y. Fu, Benjamin Tobias, Michael Van~Zeeland, Feng Wang, and
  Zheng-Mao Sheng.
\newblock Nonlinear hybrid simulation of internal kink with beam ion effects in
  diii-d.
\newblock {\em Physics of Plasmas}, 22(4), April 2015.

\bibitem{Ge2022}
W.~Ge, Z.~X. Wang, F.~Wang, Z.~Liu, and L.~Xu.
\newblock {\em Nuclear Fusion}, 63(1):016007, dec 2022.

\bibitem{Rosenbluth1998}
M.~N. Rosenbluth and F.~L. Hinton.
\newblock {\em Physical Review Letters}, 80(4):724--727, jan 1998.

\bibitem{Guenter2001}
S~G\"unter et~al.
\newblock {\em Nuclear Fusion}, 41(9):1283--1290, sep 2001.

\bibitem{Pinches2001}
S.~D.~Pinches et~al.
\newblock {\em 28th EPS Conference on Contr. Fusion and Plasma Phys. (Funchal)
  Vol 25A p 57}, 2001.

\bibitem{Liu2023}
Zhaoyang Liu and Guoyong Fu.
\newblock A simple model for internal transport barrier induced by fishbone in
  tokamak plasmas.
\newblock {\em Journal of Plasma Physics}, 89(6), December 2023.

\bibitem{Hahm1995}
T.~S. Hahm and K.~H. Burrell.
\newblock {\em Physics of Plasmas}, 2(5):1648--1651, may 1995.

\bibitem{Liu2022a}
P.~Liu et~al.
\newblock {\em Physical Review Letters}, 128(18):185001, may 2022.

\bibitem{Giacalone1999}
J.~Giacalone and J.~R. Jokipii.
\newblock {\em The Astrophysical Journal}, 520(1):204--214, jul 1999.

\bibitem{Wang2016}
Feng Wang, G.Y. Fu, and Wei Shen.
\newblock Nonlinear fishbone dynamics in spherical tokamaks.
\newblock {\em Nuclear Fusion}, 57(1):016034, November 2016.

\bibitem{Vlad2016}
G~Vlad, V~Fusco, S~Briguglio, G~Fogaccia, F~Zonca, and X~Wang.
\newblock Theory and modeling of electron fishbones.
\newblock {\em New Journal of Physics}, 18(10):105004, October 2016.

\bibitem{Wang2023}
X~Wang, S~Briguglio, A~Bottino, M~Falessi, T~Hayward-Schneider, Ph~Lauber,
  A~Mishchenko, L~Villard, and F~Zonca.
\newblock Nonlinear dynamics of nonadiabatic chirping-frequency alfvén modes
  in tokamak plasmas.
\newblock {\em Plasma Physics and Controlled Fusion}, 65(7):074001, May 2023.

\bibitem{Tao2021}
Xin Tao, Fulvio Zonca, and Liu Chen.
\newblock A trap release amplify model of chorus waves.
\newblock {\em Journal of Geophysical Research: Space Physics}, 126(9), August
  2021.

\bibitem{Zonca2022}
F.~Zonca, X.~Tao, and L.~Chen.
\newblock A theoretical framework of chorus wave excitation.
\newblock {\em Journal of Geophysical Research: Space Physics}, 127(2),
  February 2022.

\bibitem{Zonca2015}
F~Zonca et~al.
\newblock {\em New Journal of Physics}, 17(1):013052, jan 2015.

\bibitem{Chen2016}
Liu Chen and Fulvio Zonca.
\newblock {\em Reviews of Modern Physics}, 88(1):015008, mar 2016.

\bibitem{Heidbrink2020}
W.W. Heidbrink, M.A. Van~Zeeland, M.E. Austin, A.~Bierwage, Liu Chen, G.J.
  Choi, P.~Lauber, Z.~Lin, G.R. McKee, and D.A. Spong.
\newblock ‘baae’ instabilities observed without fast ion drive.
\newblock {\em Nuclear Fusion}, 61(1):016029, December 2020.

\bibitem{Polevoi2021}
A.R.~Polevoi et~al.
\newblock {\em Nuclear Fusion}, 61(7):076008, jun 2021.

\bibitem{Deng2012}
W.~Deng, Z.~Lin, and I.~Holod.
\newblock Gyrokinetic simulation model for kinetic magnetohydrodynamic
  processes in magnetized plasmas.
\newblock {\em Nuclear Fusion}, 52(2):023005, jan 2012.

\bibitem{Dong2017}
G.~Dong, J.~Bao, A.~Bhattacharjee, A.~Brizard, Z.~Lin, and P.~Porazik.
\newblock {\em Physics of Plasmas}, 24(8):081205, aug 2017.

\bibitem{Field2011}
A.R.~Field et~al.
\newblock {\em Nuclear Fusion}, 51(6):063006, apr 2011.

\bibitem{Chen2016a}
W.~Chen et~al.
\newblock {\em Nuclear Fusion}, 56(4):044001, mar 2016.

\bibitem{Gao2018}
X.~Gao et~al.
\newblock {\em Physics Letters A}, 382(18):1242--1246, may 2018.

\bibitem{Lin2023}
Z.~Lin, E.~Bass, G.~Brochard, Y.~Ghai, N.~Gorelenkov, M.~Idouakass, C.~Liu,
  P.~Liu, M.~Podesta, D.~Spong, X.~Wei, W.~Heidbrink, G.~McKee, R.~Waltz,
  J.~Bao, B.~Cornille, V.~Duarte, R.~Falgout, M.~Gorelenkova,
  T.~Hayward-Schneider, S.~H. Kim, W.~Joubert, S.~Klasky, I.~Lyngaas, K.~Mehta,
  J.~Nicolau, S.~D. Pinches, A.~Polevoi, M.~Schneider, G.~Sitaraman, W.~Tang,
  P.~Wang, and S.~Williams.
\newblock Prediction of energetic particle confinement in iter operation
  scenarios.
\newblock {\em Proceedings of the 29th International Conference on Plasma
  Physics and Controlled Nuclear Fusion Research (London, 2023) (International
  Atomic Energy Agency, Vienna, Austria, 2023), Paper IAEA-CN/TH-2}, 2023.

\bibitem{Pankin2004}
Alexei~Pankin et~al.
\newblock {\em Computer Physics Communications}, 159(3):157--184, jun 2004.

\bibitem{Lao1985}
L.L. Lao, H.~St. John, R.D. Stambaugh, A.G. Kellman, and W.~Pfeiffer.
\newblock Reconstruction of current profile parameters and plasma shapes in
  tokamaks.
\newblock {\em Nuclear Fusion}, 25(11):1611--1622, nov 1985.

\bibitem{Rice1995}
B.~W. Rice, D.~G. Nilson, and D.~Wróblewski.
\newblock Motional stark effect upgrades on diii-d.
\newblock {\em Review of Scientific Instruments}, 66(1):373--375, January 1995.

\bibitem{Jardin2012}
S~C Jardin, N~Ferraro, J~Breslau, and J~Chen.
\newblock {\em Computational Science {\&} Discovery}, 5(1):014002, may 2012.

\bibitem{Liu2022}
C.~Liu et~al.
\newblock {\em Computer Physics Communications}, 275:108313, jun 2022.

\bibitem{Luetjens2010}
H.~L\"utjens and J.~F. Luciani.
\newblock {\em Journal of Computational Physics}, 229(21):8130--8143, oct 2010.

\bibitem{Brochard2020a}
G.~Brochard, R.~Dumont, H.~L\"utjens, and X.~Garbet.
\newblock {\em Nuclear Fusion}, 60(8):086002, jul 2020.

\bibitem{Brochard2022}
G.~Brochard et~al.
\newblock {\em Nuclear Fusion}, 62(3):036021, jan 2022.

\bibitem{Grierson2018}
B.~A.~Grierson et~al.
\newblock {\em Fusion Science and Technology}, 74(1-2):101--115, feb 2018.

\bibitem{Bussac1975}
M.~N. Bussac, R.~Pellat, D.~Edery, and J.~L. Soule.
\newblock Internal kink modes in toroidal plasmas with circular cross sections.
\newblock {\em Physical Review Letters}, 35(24):1638--1641, December 1975.

\bibitem{Moseev2019}
D.~Moseev and M.~Salewski.
\newblock {\em Physics of Plasmas}, 26(2):020901, feb 2019.

\bibitem{Hirvijoki2014}
E.~Hirvijoki, O.~Asunta, T.~Koskela, T.~Kurki-Suonio, J.~Miettunen, S.~Sipilä,
  A.~Snicker, and S.~Äkäslompolo.
\newblock Ascot: Solving the kinetic equation of minority particle species in
  tokamak plasmas.
\newblock {\em Computer Physics Communications}, 185(4):1310--1321, April 2014.

\bibitem{Schmidt2023}
B.S. Schmidt, M.~Salewski, D.~Moseev, M.~Baquero-Ruiz, P.C. Hansen,
  J.~Eriksson, O.~Ford, G.~Gorini, H.~Järleblad, Ye~O. Kazakov, D.~Kulla,
  S.~Lazerson, J.E. Mencke, D.~Mykytchuk, M.~Nocente, P.~Poloskei, M.~Rud,
  A.~Snicker, L.~Stagner, and S.~Äkäslompolo.
\newblock 4d and 5d phase-space tomography using slowing-down physics
  regularization.
\newblock {\em Nuclear Fusion}, 63(7):076016, May 2023.

\bibitem{Bierwage2022}
Andreas Bierwage, Michael Fitzgerald, Philipp Lauber, Mirko Salewski, Yevgen
  Kazakov, and Žiga Štancar.
\newblock Representation and modeling of charged particle distributions in
  tokamaks.
\newblock {\em Computer Physics Communications}, 275:108305, June 2022.

\bibitem{Fitzgerald2013}
M.~Fitzgerald, L.C. Appel, and M.J. Hole.
\newblock Efit tokamak equilibria with toroidal flow and anisotropic pressure
  using the two-temperature guiding-centre plasma.
\newblock {\em Nuclear Fusion}, 53(11):113040, October 2013.

\bibitem{Qu2014}
Z~S Qu, M~Fitzgerald, and M~J Hole.
\newblock Analysing the impact of anisotropy pressure on tokamak equilibria.
\newblock {\em Plasma Physics and Controlled Fusion}, 56(7):075007, May 2014.

\bibitem{Pereverzev2002}
G.~V. Pereverzev and P.~N. Yushmanov.
\newblock Astra auto- mated system for transport analysis in a tokamak.
\newblock {\em Max- Planck IPP Report vol 5/98}, 2002.

\bibitem{Polevoi2019}
A.~R.~Polevoi et~al.
\newblock Reassessment of steady state operation in iter with nbi and ec
  heating and current drive.
\newblock {\em 46th EPS Conf. on Plasma Physics}, 2019.

\bibitem{White2006}
Roscoe~B White.
\newblock {\em The Theory of Toroidally Confined Plasmas: Revised Second
  Edition}.
\newblock PUBLISHED BY IMPERIAL COLLEGE PRESS AND DISTRIBUTED BY WORLD
  SCIENTIFIC PUBLISHING CO., April 2006.

\bibitem{Porcelli1994}
F.~Porcelli, R.~Stankiewicz, W.~Kerner, and H.~L. Berk.
\newblock Solution of the drift-kinetic equation for global plasma modes and
  finite particle orbit widths.
\newblock {\em Physics of Plasmas}, 1(3):470--480, March 1994.

\bibitem{Ohshima2007}
S~Ohshima, A~Fujisawa, A~Shimizu, H~Nakano, H~Iguchi, Y~Yoshimura, K~Nagaoka,
  T~Minami, M~Isobe, S~Nishimura, C~Suzuki, T~Akiyama, C~Takahashi, M~Takeuchi,
  T~Ito, T~Watari, R~Kumazawa, S-I Itoh, K~Itoh, K~Matsuoka, and S~Okamura.
\newblock Zonal flow driven by energetic particle during magneto-hydro-dynamic
  burst in a toroidal plasma.
\newblock {\em Plasma Physics and Controlled Fusion}, 49(11):1945--1952,
  October 2007.

\bibitem{Brochard2024}
G.~Brochard et~al.
\newblock Saturation of fishbone instability by self-generated zonal flows in
  tokamak plasmas.
\newblock {\em Physical Review Letters, in press}, 2024.

\bibitem{Chen2012}
L.~Chen and F.~Zonca.
\newblock Nonlinear excitations of zonal structures by toroidal alfv\'en
  eigenmodes.
\newblock {\em Physical Review Letters}, 109(14):145002, oct 2012.

\bibitem{Chen2000}
L.~Chen, Z.~Lin, and R.~White.
\newblock {\em Physics of Plasmas}, 7(8):3129--3132, aug 2000.

\bibitem{Zonca2014}
F~Zonca, L~Chen, S~Briguglio, G~Fogaccia, A~V Milovanov, Z~Qiu, G~Vlad, and
  X~Wang.
\newblock Energetic particles and multi-scale dynamics in fusion plasmas.
\newblock {\em Plasma Physics and Controlled Fusion}, 57(1):014024, November
  2014.

\bibitem{Liu2024}
P.Liu, X.~Wei, Z.~Lin, W.~W. Heidbrink, G.~Brochard, G.~J. Choi, J.~H. Nicolau,
  and W.~Zhang.
\newblock Cross-scale interaction between microturbulence and meso-scale
  reversed shear alfven eigenmodes in diii-d plasmas.
\newblock {\em Submitted to Nuclear Fusion}, 2024.

\bibitem{Luetjens1996}
H.~Lütjens, A.~Bondeson, and O.~Sauter.
\newblock The chease code for toroidal mhd equilibria.
\newblock {\em Computer Physics Communications}, 97(3):219--260, September
  1996.

\bibitem{Rosenbluth1996}
M.N Rosenbluth and F.L Hinton.
\newblock Plasma rotation driven by alpha particles in a tokamak reactor.
\newblock {\em Nuclear Fusion}, 36(1):55--67, January 1996.

\bibitem{Peeters1998}
A.~G. Peeters.
\newblock Equations for the evolution of the radial electric field and poloidal
  rotation in toroidally symmetric geometry.
\newblock {\em Physics of Plasmas}, 5(3):763--767, March 1998.

\bibitem{Xiao2015}
Y.~Xiao, I.~Holod, Z.~Wang, Z.~Lin, and T.~Zhang.
\newblock {\em Physics of Plasmas}, 22(2):022516, feb 2015.

\bibitem{Fang2019}
K.~Fang, J.~Bao, and Z.~Lin.
\newblock Gyrokinetic simulations of nonlinear interactions between magnetic
  islands and microturbulence.
\newblock {\em Plasma Science and Technology}, 21(11):115102, September 2019.

\bibitem{Falessi2023}
Matteo~Valerio Falessi, Liu Chen, Zhiyong Qiu, and Fulvio Zonca.
\newblock Nonlinear equilibria and transport processes in burning plasmas.
\newblock {\em New Journal of Physics}, 25(12):123035, December 2023.

\bibitem{Falessi2019}
Matteo~Valerio Falessi and Fulvio Zonca.
\newblock Transport theory of phase space zonal structures.
\newblock {\em Physics of Plasmas}, 26(2), February 2019.

\bibitem{Cowley1996}
S.~Cowley et~al.
\newblock {\em Phys. Plasmas}, 3(5):1848--1852, may 1996.

\bibitem{Berk1999}
H.~L. Berk, B.~N. Breizman, J.~Candy, M.~Pekker, and N.~V. Petviashvili.
\newblock Spontaneous hole–clump pair creation.
\newblock {\em Physics of Plasmas}, 6(8):3102--3113, August 1999.

\bibitem{Guo2009}
Z.~Guo et~al.
\newblock {\em Physical Review Letters}, 103(5):055002, jul 2009.

\bibitem{Teng2023}
Shangchun Teng, Yifan Wu, Yuki Harada, Jacob Bortnik, Fulvio Zonca, Liu Chen,
  and Xin Tao.
\newblock Whistler-mode chorus waves at mars.
\newblock {\em Nature Communications}, 14(1), June 2023.

\bibitem{Chen1994}
Liu Chen.
\newblock Theory of magnetohydrodynamic instabilities excited by energetic
  particles in tokamaks*.
\newblock {\em Physics of Plasmas}, 1(5):1519--1522, May 1994.

\bibitem{Zonca2000}
F.~Zonca and L.~Chen.
\newblock Destabilization of energetic particle modes by icrf induced fast
  minority ion tails on tftr.
\newblock {\em Proceedings of the 6th IAEA TCM on Energetic Particles in
  Magnetic Confinement Systems, JAERI- Conf. 2000-004 p. 52}, 2000.

\bibitem{Zhang2012}
H.~S. Zhang, Z.~Lin, and I.~Holod.
\newblock Nonlinear frequency oscillation of alfvén eigenmodes in fusion
  plasmas.
\newblock {\em Physical Review Letters}, 109(2):025001, July 2012.

\bibitem{Yu2022}
L.~M. Yu, F.~Zonca, Z.~Y. Qiu, L.~Chen, W.~Chen, X.~T. Ding, X.~Q. Ji, T.~Wang,
  T.~B. Wang, R.~R. Ma, B.~S. Yuan, P.~W. Shi, Y.~G. Li, L.~Liu, Z.~B. Shi,
  J.~Y. Cao, J.~Q. Dong, Yi~Liu, Q.~W. Yang, and M.~Xu.
\newblock Experimental evidence of nonlinear avalanche dynamics of energetic
  particle modes.
\newblock {\em Europhysics Letters}, 138(5):54002, June 2022.

\bibitem{Zonca2023}
F.~Zonca, L.~Chen, M.~V. Falessi, and Z.~Qiu.
\newblock On the nonlinear dynamics of fishbone and energetic particle modes.
\newblock {\em Proceeding of the 28th IAEA- Fusion Energy Conference (FEC
  2023)}, 2023.

\bibitem{White1983}
R.~B. White, R.~J. Goldston, K.~McGuire, Allen~H. Boozer, D.~A. Monticello, and
  W.~Park.
\newblock Theory of mode-induced beam particle loss in tokamaks.
\newblock {\em The Physics of Fluids}, 26(10):2958--2965, October 1983.

\bibitem{Biancalani2018}
A.~Biancalani, N.~Carlevaro, A.~Bottino, G.~Montani, and Z.~Qiu.
\newblock Nonlinear velocity redistribution caused by energetic-particle-driven
  geodesic acoustic modes, mapped with the beam-plasma system.
\newblock {\em Journal of Plasma Physics}, 84(6), December 2018.

\bibitem{Yang2017}
Y~Yang, X~Gao, H~Q Liu, G~Q Li, T~Zhang, L~Zeng, Y~K Liu, M~Q Wu, D~F Kong, T~F
  Ming, X~Han, Y~M Wang, Q~Zang, B~Lyu, Y~Y Li, Y~M Duan, F~B Zhong, K~Li, L~Q
  Xu, X~Z Gong, Y~W Sun, J~P Qian, B~J Ding, Z~X Liu, F~K Liu, C~D Hu, N~Xiang,
  Y~F Liang, X~D Zhang, B~N Wan, J~G Li, and Y~X~Wan and.
\newblock Observation of internal transport barrier in {ELMy} h-mode plasmas on
  the {EAST} tokamak.
\newblock {\em Plasma Physics and Controlled Fusion}, 59(8):085003, jun 2017.

\bibitem{Adam1976}
J.~C. Adam, W.~M. Tang, and P.~H. Rutherford.
\newblock Destabilization of the trapped-electron mode by magnetic curvature
  drift resonances.
\newblock {\em The Physics of Fluids}, 19(4):561--566, April 1976.

\bibitem{Wolf1999}
R~C Wolf, O~Gruber, M~Maraschek, R~Dux, C~Fuchs, S~Günter, A~Herrmann,
  A~Kallenbach, K~Lackner, P~J McCarthy, H~Meister, G~Pereverzev, J~Schweinzer,
  U~Seidel, and the ASDEX Upgrade~Team.
\newblock Stationary advanced scenarios with internal transport barrier on
  {ASDEX} upgrade.
\newblock {\em Plasma Physics and Controlled Fusion}, 41(12B):B93--B107, dec
  1999.

\bibitem{Zhang2023}
Xuexi Zhang, M.Q. Wu, Gongshun Li, Guoqiang Li, Tengfei Tang, Y.~Yang, F.B.
  Zhong, F.F. Long, M.F. Wu, T.~Zhang, T.F. Ming, X.~Zhu, K.N. Geng, Haiqing
  Liu, and Xiang Gao.
\newblock Investigation of key factors for itb formation and maintenance in
  east high $\beta$ discharges.
\newblock {\em Physics Letters A}, 462:128646, February 2023.

\bibitem{Wolf2003}
R.~C. Wolf.
\newblock {\em Plasma Physics and Controlled Fusion}, 45(1):R1--R91, nov 2003.

\end{thebibliography}
\end{document}